\documentclass[fleqn,usenatbib]{mnras}

\usepackage{newtxtext,newtxmath}

\usepackage[T1]{fontenc}
\usepackage{ae,aecompl}

\usepackage{graphicx}	
\usepackage{amsmath}	
\usepackage{amssymb}	
\usepackage{enumitem}

\usepackage{bm}
\usepackage[usenames,dvipsnames,svgnames,hyperref]{xcolor}

\newcommand{\myri}{\mathrm{M}_\odot\ \mathrm{yr}^{-1}}

\newcommand{\msuni}{\mathrm{M}_\odot}
\newcommand{\msuno}{$\mathrm{M}_\odot$}

\title[Robust identification of galaxies in simulations]
      {Introducing a new, robust galaxy finder algorithm for simulations}

\author[R. Ca\~nas et al.]{Rodrigo Ca\~nas$^{1,2}$\thanks{E-mail: rodrigo.canas@icrar.org (RC)},
Pascal J. Elahi$^{1,2}$,
Charlotte Welker$^{1,2}$,
Claudia del P. Lagos$^{1,2}$,
\newauthor
Chris Power$^{1,2}$,
Yohan Dubois$^{3,4,5}$ 
and
Christophe Pichon$^{3,4,5}$
\\
$^{1}$International Centre for Radio Astronomy Research, University of Western Australia, 
      35 Stirling Highway, Crawley, WA 6009, Australia\\
$^{2}$ARC Centre of Excellence for All Sky Astrophysics in 3 Dimensions (ASTRO 3D)\\
$^{3}$CNRS and UPMC Univ. Paris 06, UMR 7095, Institut d'Astrophysique de Paris, 98 bis Boulevard Arago, F-75014 Paris, France\\
$^{4}$Institute for Astronomy, University of Edinburgh, Royal Observatory, Blackford Hill, Edinburgh, EH9 3HJ, United Kingdom\\
$^{5}$Korea Institute of Advanced Studies (KIAS) 85 Hoegiro, Dongdaemun-gu, Seoul, 02455, Republic of Korea
}

\date{Accepted XXX. Received YYY; in original form ZZZ}

\pubyear{2018}

\begin{document}
\label{firstpage}
\pagerange{\pageref{firstpage}--\pageref{lastpage}}
\maketitle

\begin{abstract}
	  Identifying galaxies in hydrodynamical simulations is a difficult 
      task, particularly in regions of high density such as galaxy groups
      and clusters.
      We present a new scale-free shape-independent algorithm to robustly
      and accurately identify galaxies in simulation, implemented within
      the phase-space halo-finder code {\sc VELOCIraptor}.
      This is achieved by using the full phase-space dispersion tensor
      for particle assignment and an iterative adjustment of search
      parameters, which help us overcome common structure finding problems.
      We apply our improved method to the Horizon-AGN simulation and
      compare galaxy stellar masses ($M_*$), star formation rates (SFR)
      and sizes with the elaborate configuration-space halo finder,
      {\sc HaloMaker}.
      Galaxies living in halos with $>1$ galaxy are the most affected by
      the shortcomings of real-space finders, with their mass, SFR, and
      sizes being $> 2$ times larger (smaller) in the case of host
      (satellite) galaxies.
      Thus, our ability to measure minor/major merger rates and
      disentangle environmental effects in simulations can be generally
      hindered if the identification of galaxies is not treated carefully.
      Though large systematic differences are obtained on a one-to-one
      basis, the overall Galaxy Stellar Mass Function, the Star
      Formation Rate Function and mass-size relations are not
      greatly affected.
      This is due to isolated galaxies being the most abundant population, 
      dominating broad statistics.
\end{abstract}

\begin{keywords}
    methods: numerical -- galaxies: evolution -- cosmology: theory-dark matter
\end{keywords}



          
	\section{Introduction}
      Galaxies are the result of a wide variety of physical processes.
      Their evolution and properties are determined by both their hierarchical 
      assembly and the complex interplay between many multi-scale non-linear 
      processes such as star formation, radiative cooling, and feedback
      loops \citep[see][for a recent review]{Somerville2015}.
      Cosmological hydro-dynamical simulations of galaxy formation are ideal 
      laboratories to explore and isolate the effects of these physical 
      processes on the evolution of galaxies in realistic 
      environments \citep[][]{Dubois2014,Vogelsberger2014,Schaye2015}.
      The advantage of these simulations over other numerical methods, such as 
      abundance matching \citep[e.g.][]{Berlind2002,Berlind2001}
      and semi-analytic models of galaxy formation 
      \citep[][]{Lacey1993,Cole2000,Kauffmann1998}
      is the ability to predict the internal structure of galaxies,
      as the hydrodynamics that give rise to it is
      resolved through direct resolution of the equations
      of physics down to sub-galactic scales.
      %
      
      In recent years a major breakthrough in the capability of 
      cosmological hydrodynamical simulations to produce realistic
      galaxy populations has taken place. 
      This has been achieved thanks to the combined results of major
      improvements in numerical algorithms, availability of computing
      resources, improved subgrid models for unresolved feedback
      processes, and the calibration of subgrid feedback parameters
      to match key observables. 
      Examples of this new generation of simulations include
      Horizon-AGN \citep{Dubois2014}, 
      EAGLE \citep{Schaye2015}, 
      Illustris \citep{Vogelsberger2014} 
      and IllustrisTNG \citep{Pillepich2018}.
      Simulated boxes of $\sim(100\rm \ {\mathrm cMpc})^3$ with 
      sub-kpc resolution are becoming common.
      These simulations reproduce observables beyond those they
      were tuned for, with various degrees of success.
      For example, these simulations produce reasonable
      morphological diversity of galaxies, the colour bimodality
      of galaxies, the 
      SFR-stellar mass relation, the stellar mass function and
      the cosmic SFR density evolution 
      \citep[e.g.][]{Furlong2015,Genel2014,Trayford2015,
      Trayford2016,Snyder2015,Dubois2016,Nelson2018}.
      %

      In order to understand the physics involved in the 
      formation of galaxies through simulations, we first need
      to understand and test the extent to which such results
      depend on numerical effects rather than on the physics
      \citep[e.g.][]{Klypin1999}.
      This issue has been pointed out over the years by several
      studies which have shown that properties of galaxies and
      galaxy populations sensitively depend on the specific code
      used, the implemented subgrid physics and their respective
      tuning, as well as numerical resolution
      \citep[see e.g.][]{Frenk1999,Kim2014,Power2014,Knebe2015,
      Scannapieco2012,Schaye2015,Elahi2016,Sembolini2016a,
      Sembolini2016b}.
      %
      
      Often overlooked is the issue of the robustness with which
      we can measure galaxy properties in these simulations that
      can affect the conclusions reached.
      The latter ultimately depends on how well we identify
      structures in the simulations
      \citep[][]{Knebe2011,Knebe2013b}.
      These issues are of particular interest for the new and
      coming generation of hydrodynamical simulations, which
      have taken the route of fine tuning the free parameters
      of the subgrid physics modules (i.e. which describe the 
      processes that are expected to take place at scales below
      the resolution limit) against a desired observable (e.g. 
      the galaxy stellar mass function, GSMF, and the size-mass
      relation, \citealt{Crain2015}).
      Robustly measuring the desired galaxy property to perform
      the tuning in simulations is therefore crucial.
      %

      In the first studies of hierarchical formation, simple
      structure finding algorithms, such as spherical
      over-density \citep[SO,][]{Press1974}
      and Friends-of-Friends \citep[FOF,][]{Davis1985}, were able
      to give a reasonable estimation of ``condensed'' structures
      in simulations.
      However, with the ever increasing size of simulations
      and the need of higher accuracy in measurements, such
      simple approaches are not necessarily optimal, and a
      large number of codes have appeared in the literature
      addressing the finding of structures in simulations 
      \citep[see][and references therein]{Knebe2011,Knebe2013b}.
      Early approaches have been characterised by using solely
      configuration-space information (e.g.
      {\sc bdm}, \citealt{Klypin1997}; 
      {\sc hop}, \citealt{Eisenstein1998};
      {\sc skid}, \citealt{Stadel2001};
      {\sc subfind}, \citealt{Springel2001};
      {\sc ahf}, \citealt{Gill2004}),
      while more recent sophisticated algorithms have addressed
      the problem adding the velocity-space information (e.g.
      {\sc 6dfof}, \citealt{Diemand2006};
      {\sc hsf}, \citealt{Maciejewski2009};
      {\sc VELOCIraptor}, \citealt{Elahi2011};
      {\sc rockstar}, \citealt{Behroozi2013}).
      Although all these algorithms attempt to solve the same
      problem, the specific details of each implementation
      can introduce artifacts in the final results.
      Other approaches tackle the problem by using temporal
      information by following (sub)haloes' bound particles
      through simulation snapshots to identify structures and
      \emph{de-blend} systems in interaction, which can be done
      either from late to earlier times (e.g. {\sc surv}
      \citealt{Tormen2004,Giocoli2008,Giocoli2010}) or
      vice-versa (e.g. {\sc hbt, hbt+} \citealt{Han2012,Han2018}).
      Though powerful in principle, these method rely heavily
      on identification at sufficiently early times and having
      at hand snapshots at a high cadence.
      %

      It is essential that we understand the reliability of
      measurements and the associated systematic uncertainties.
      This has been addressed by many comparison projects
      in which structure finding codes are tested against
      the same data to study the similarities and differences
      on the measurements of the properties of dark matter
      haloes \citep{Knebe2011},
      subhaloes \citep{Onions2012}, 
      galaxies \citep{Knebe2013a} and tidal
      structures \citep{Elahi2013}.
      Such studies have found overall agreement when
      analysing dark matter halo populations
      \citep[][]{Knebe2011}. 
      However, large differences are obtained on the overall
      mass recovered for dark matter subhaloes, satellite
      galaxies and tidal streams      
      \citep{Knebe2013b,Onions2012,Knebe2013a,Elahi2013}.
      While the identification of substructures depends
      on the identification of density peaks, the major
      challenge is to assign the ``background'' particles
      to statistically significant density peaks
      which can affect drastically the properties of
      the structures.
      For this reason, algorithms that only use
      configuration space information, although fast, struggle
      to identify appropriately subhaloes in dense environments
      (e.g. galaxy groups and clusters, and merging systems),  
      while finders that include also include velocity-space
      information obtain better results in these regimes
      \citep{Knebe2011}.
      %
      
      This paper presents a new galaxy finding algorithm which
      makes use of the full configuration and velocity space
      information, and presents a thorough study of the effects
      that the identification method has on the properties of
      individual galaxies and galaxy populations.
      This implementation is an extension of the halo-finder
      code {\sc VELOCIraptor}
      \citep[][Elahi et al. in prep]{Elahi2011}. 
      We pay special attention to two regimes that have been
      traditionally challenging for galaxy finding algorithms:
      (i) mergers and interactions and (ii) identification of
      substructures in high density environments.
      The main problem in both of these regimes is that
      the outskirts of hosts and satellite structures 
      can have similar densities, making it difficult to
      distinguish to which structure they belong.
      This is even harder if only configuration space
      information is taken into account.
      These problems are equally valid for dark matter haloes
      and galaxies, while there is a plethora of literature
      that addresses the former
      \citep[see for reference][]{Knebe2013b}, the latter has
      not yet been thoroughly addressed.
      Galaxies have a range of morphologies which during
      interactions produce complex stellar structures that
      form on an already significant density peak.
      Thus, the problem of identifying galaxies cannot be 
      solved using dark matter halo finding tools.
      We show that the undesirable consequences of poor
      identification affect radial mass profiles, sizes
      and total masses.
      We apply our new galaxy finding algorithm
      to the state-of-the-art cosmological hydrodynamical
      simulation Horizon-AGN \citep{Dubois2014}
      and compare our results with the original galaxy
      catalog, which was obtained by applying the
      configuration space finder {\sc HaloMaker} 
      \citep{Aubert2004,Tweed2009}.
      %

      This work is organised as follows:
      In Section~\ref{sec:NumericalMethods} we provide a general and
      brief description of the code {\sc VELOCIraptor} and the
      Horizon-AGN simulation.
      In Section~\ref{sec:Algorithm} we describe in detail the 
      improved algorithm to identify galaxies in simulations and 
      implemented in {\sc VELOCIraptor}. 
      In Section~\ref{sec:casestudy} we present examples of the
      performance of our new algorithm on strongly interacting
      scenarios.
      In Section~\ref{sec:results} we compare results obtained with 
      {\sc VELOCIraptor} and the original Horizon-AGN galaxy catalog
      on a galaxy-to-galaxy basis, as well as comparing the entire 
      galaxy populations. 
      Discussion is presented in Section~\ref{sec:discussion}, and 
      summary and conclusions are presented in 
      Section~\ref{sec:conclusions}.
      Lastly, in Appendix~\ref{appndx:ll} we show how configuration
      space linking length affect galaxy delimitation, and in
      Appendix~\ref{appndx:weight} we show different weights
      affect particle assignment.
      %
      
            
  \section{Numerical Methods}
  \label{sec:NumericalMethods}
      In this section, we briefly describe the Horizon-AGN simulation,
      and the structure finding code {\sc VELOCIraptor}. For further details the 
      interested reader is referred to \citet{Dubois2014}, where the 
      Horizon-AGN simulation was presented, and to \citet{Elahi2011} 
      for a detailed description of {\sc VELOCIraptor}.

      \subsection{Horizon-AGN Simulation}
      \label{sec:HAGN}
      Horizon-AGN is a state-of-the-art hydrodynamical simulation, presented
      in \citet{Dubois2014}. It follows the formation and evolution 
      of galaxies in a standard $\Lambda$ cold dark matter ($\Lambda$CDM) 
      cosmology, adopting values of a total matter density $\Omega_\mathrm{m} = 0.272$, 
      dark energy density $\Omega_\Lambda = 0.728$, amplitude of the linear power
      spectrum $\sigma_8 = 0.81$, baryon density $\Omega_\mathrm{b} = 0.045$,
      Hubble constant $H_0 =70.4$ km\,s$^{-1}$\, Mpc $^{-1}$, and spectral 
      index $n_s = 0.967$, in concordance to results from the 
      \emph{Wilkinson Microwave Anisotropy Probe 7} \citep[WMAP7,][]
      {Komatsu2011}.
      
      The simulation was run using the adaptive mesh refinement (AMR) code  
      {\sc ramses} \citep{Teyssier2002}, and it has a comoving box size
      of $L_\mathrm{box} = 100\,h^{-1}$ Mpc, a total of 1024$^3$ dark matter 
      particles with mass $M_\mathrm{dm} = 8 \times 10^7$ 
      \msuno; and an initial number of 1024$^3$ gas cells, which are refined
      up to seven times to a maximum physical resolution of $\Delta x = 1$ kpc.
      
      Implemented subgrid physics include: gas cooling, heating from a uniform 
      redshift-dependent UV background, star formation, stellar feedback driven 
      by supernovae (SNe) Type Ia, II and stellar winds, and
      black hole (BH) accretion and its associated active galactic nucleus (AGN) 
      feedback.
      Following \citet{Dubois2012}, BHs are created with a seed mass of 
      $M_\mathrm{BH} = 10^5 \ \msuni$, and grow according to a Bondi-Hoyle-Lyttleon
      accretion scheme capped at Eddington accretion rate. A two-mode AGN feedback is
      explicitly implemented as a bipolar outflow at accretion rates smaller than 
      1\% the Eddington accretion \citep{Dubois2010}, and as an isotropic 
      thermal energy injection otherwise \citep[see][for further details]
      {Dubois2014,Volonteri2016}.

      Galaxies in Horizon-AGN were originally identified with the code
      {\sc HaloMaker} \citep{Tweed2009}.
      {\sc HaloMaker} uses {\sc AdaptaHOP} \citep{Aubert2004} algorithm
      (which is itself based on {\sc HOP} \citealt{Eisenstein1998}) to
      identify structures and their corresponding substructures.
      The algorithm identifies high-density regions and the particles
      associated to those.
      This is done by estimating the density of all particles
      from $N_\mathrm{SPH}$ neighbours using an Smoothed Particle
      Hydrodynamics (SPH) kernel.
      Then, starting at a reference particle, the density field
      gradient is followed by linking it to the densest
      particle within $N_\mathrm{HOP}$ neighbours, and 
      hopping it as the new reference particle.
      This process is iteratively done until the reference particle is
      the densest within its $N_\mathrm{HOP}$ neighbours.
      Particles with density above a density threshold $\rho_\mathrm{t}$
      linked to the same peak constitute groups.
      Hierarchy is established by looking for saddle points in 
      the density field between peaks, and using them as boundaries
      to define hierarchy levels.
      Groups whose saddle point is above $\rho_\mathrm{t}$ are linked
      as members of the same branch.
      The process is repeated iteratively for each level until no
      saddle points are found.
      The main structure (either dark matter halo or galaxy) is defined
      by following the branch to which the most massive or densest peak
      belongs.
      Groups from other branches will then become substructures, while 
      those in branches within branches will be sub-substructures, and so on.
      Galaxies are identified using star particles information only,
      the local density for each particle is calculated using 
      $N_\mathrm{SPH} = 20$ neighbours, and a local threshold of 
      $\rho_\mathrm{t} = 178$ times the average total matter density is
      applied to select relevant densities.
      A minimum physical size above which substructures 
      are considered relevant of $\sim$2 kpc is also applied.
      Only galactic structures with more
      than 50 star particles were considered.
      %
      
      Horizon-AGN has been used to study the alignments between the spin of
      galaxies and the cosmic web filaments, and how mergers change the
      spin orientation of galaxies \citep{Dubois2014,Welker2014}.
      Its BH growth and AGN feedback implementations have succeeded in producing
      a BH population whose overall properties agree with observations
      \citep{Volonteri2016}, and have shown the importance of AGN feedback 
      in helping the simulation to reproduce the observed morphology and kinematic
      properties of massive galaxies \citep{Dubois2016}.
      Additionally, the simulation has also been used to study the evolution
      of the galaxy luminosity and stellar mass functions, star formation
      main sequence and galaxy colours 
      \citep{Kaviraj2017}\footnote{Further research projects
      and publications can be found in the Horizon-AGN simulation website 
      (\href{http://www.horizon-simulation.org}
      {http://www.horizon-simulation.org}).}.
      %
      
      
    \subsection{{\sc VELOCIraptor}}
    \label{sec:velociraptor}
      {\sc VELOCIraptor} \citep[also known as {\sc stf},][]{Elahi2011} 
      is a structure finding algorithm capable of identifying dark matter haloes,
      galaxies and substructures such as satellite subhaloes and streams in 
      simulations. Here we briefly summarize the algorithm presented in 
      \citet{Elahi2011}. 
      
      The standard {\sc VELOCIraptor}'s algorithm is based on the 
      assumption that the velocity distribution of a system composed 
      by many objects can be split into a smooth background component
      with overdense features in it. The former would correspond to 
      the main halo, and the latter to the substructures embedded in it.
      Hence, substructures are found by identifying the particles whose 
      local velocity density $f_\mathrm{l}(\bm{v})$ stands out from the 
      expected background velocity density $f_\mathrm{bg}(\bm{v})$,
      effectively looking for clustering in orbit space.

      In order to calculate these quantities for each particle, the main
      halo is split in volume cells using the KD-tree algorithm 
      \citep{Bentley1975}. 
      This is done such that each cell
      contains enough particles to minimize statistical errors, but not
      too many to avoid variations in the gravitational potential and
      velocity density in each cell.
      The expected background velocity density, $f_\mathrm{bg}$, is
      estimated as a multivariate Gaussian. Hence, for a particle $i$
      with velocity $\bm{v}_i$
      
      \begin{equation}
        f_\mathrm{bg}(\bm{v}_i) = \frac{\exp[-\frac12 (\bm{v}_i - \bm{\bar{v}}(\bm{x}_i) \ \Sigma_v(\bm{x}_i)^{-1}\ 
                                   (\bm{v}_i - \bm{\bar{v}}(\bm{x}_i)]}{(2\pi^{3/2}|\Sigma_v(\bm{x}_i)|^{1/2})} \ ,
      \end{equation}
      
      \noindent where $\bm{\bar{v}}(\bm{x}_i)$ and $\Sigma_v(\bm{x}_i)$ are
      respectively the local average velocity and velocity dispersion
      tensor about $\bm{\bar{v}}(\bm{x}_i)$, at the $i^\mathrm{th}$ particle's position $\bm{x}_i$.
      To accurately determine $\bm{\bar{v}}(\bm{x}_i)$ and $\Sigma_v(\bm{x}_i)$,
      the $\bar{\bm{v}}_k$ and $\Sigma_{v,k}$ of each cell $k$ are calculated, 
      and these quantities are linearly interpolated to the $i^\mathrm{th}$ particle's
      position using the cell containing the particle and six neighbouring cells.
      For each cell $k$ 
      
      \begin{equation}
        \bar{\bm{v}}_k = \frac{1}{M_k} \sum_j^{N_k} m_j \bm{v}_j \ ,
      \end{equation}

      \noindent and
      
      \begin{equation}
        \Sigma_{v,k} = \frac{1}{M_k}  \sum_j^{N_k} m_j (\bm{v}_j - \bar{\bm{v}}) (\bm{v}_j - \bar{\bm{v}})^T \ ,
      \end{equation}
      
      \noindent where $m_j$ and $\bm{v}_j$ are the particle $j$'s mass and velocity 
      respectively, and $N_k$\footnote{
      {\sc VELOCIraptor}
      constructs KD-trees at several stages to 
      calculate velocity density distribution, FOF searches and estimate 
      gravitational potentials.
      The number of particles inside each cell $N_k$ will vary depending on
      the purpose of the tree.
      To estimate $f_\mathrm{bg}(\bm{v})$ a $N_k = 16$ is used when the
      $f(\bm{v})$ is estimated using 32 velocity-space nearest neighbours.
      For efficient FOF searches $N_k$ is selected to be similar to the
      minimum number of particles threshold to define a structure.
      Finally, to calculate the gravitational potential $N_k = 8$ is used.}      
      and $M_k$ are the number of particles and mass of
      the cell $k$, respectively.
      Finally, the local velocity density $f_\mathrm{l}(\bm{v}_i)$ is calculated
      using a smoothing kernel scheme from velocity-space nearest neighbours.
      %

      For each particle $i$, the logarithmic ratio of the local and
      background velocity distributions 
      
      \begin{equation}
        \mathcal{R}_i = \ln \frac{f_\mathrm{l}(\bm{v}_i)}{f_\mathrm{bg}(\bm{v}_i)}\ ,
      \end{equation}
      
      \noindent is calculated.
      Particles with $\mathcal{R}_\mathrm{th}$ above a 
      $\mathcal{R}_\mathrm{th}$ threshold are kept and
      classified as potential substructures.
      %
      
      Once the outlying particles are found, they are
      clustered into groups using a Friends-of-Friends
      \citep[FOF,][]{Davis1985} motivated algorithm.
      Particles $i$ and $j$ are grouped if:
      
      \begin{equation}
        \label{eq:3dfof}
        \frac{(\bm{x}_i-\bm{x}_j)^2}{l^2_\mathrm{x}} < 1 \ ,
      \end{equation}
      \begin{equation}
        1/\mathcal{V}_\mathrm{r} \leq v_i/v_j \leq \mathcal{V}_\mathrm{r} \ ,
      \end{equation}
      \begin{equation}
        \cos\Theta_\mathrm{op} \leq \frac{\bm{v}_i\cdot\bm{v}_j}{v_i v_j} \ ,
      \end{equation}
      
      \noindent where $\bm{x}$ and $\bm{v}$ are a
      particle's position and velocity respectively,
      $l_\mathrm{x}$ is the configuration-space linking
      length, $\mathcal{V}_\mathrm{r}$ is the velocity
      ratio threshold determining the range in which the
      norm of the particles' velocities are considered
      to be similar, and  $\Theta_\mathrm{op}$ is an
      opening angle threshold within which directions of
      the particles' velocity vector must align.
      This effectively means that particles in a group
      need not only to be physically close, but they
      also need to be close in orbital space.
      %

      {\sc VELOCIraptor} has been employed in several
      comparison projects that have confirmed its
      versatility and ability to accurately find 
      structures and substructures in $N$-body and
      hydrodynamical simulations
      \citep[e.g.][]{Knebe2013a,Elahi2013,
      Behroozi2015,Onions2012}.
      An updated version of the code along with new
      features and tools will be presented in
      Elahi et al. in prep. 
      %
      

  \section{Robust Identification of Galaxies}
  \label{sec:Algorithm}
    
      {\sc VELOCIraptor} was originally designed to
      find dark matter structures in simulations,
      including haloes, subhaloes and dark matter
      streams. 
      While it has also been used to identify
      galaxies in hydrodynamical simulations
      \citep{Knebe2013a}, the treatment of the 
      baryonic component was limited to first
      identifying dark matter (sub-)haloes, and
      then linking gas and stellar particles to the 
      nearest dark matter particle in phase-space.
      Though this procedure in principle provides a
      phase-space assignment of baryons to dark matter
      haloes, there were two key aspects that needed
      improvement.
      First, the metric used for baryon assignment
      was quite simple, which could cause incorrect
      assignment of particles especially for
      non-spherical or complex geometries, which are
      particularly present in interacting galaxies.
      Secondly, for some interacting galaxies, the
      dark matter haloes might be indistinguishable,
      assigning the merging galaxies to the a single
      halo.
      %
      
      These problems could be solved by running
      {\sc VELOCIraptor} independently over stellar
      particles to identify galaxies.
      However, the original {\sc VELOCIraptor} 
      algorithm assumes the existence of a smooth,
      semi-virialised background.
      The code was not optimised to find substructures
      in any system where the background is sparsely
      sampled.
      %
      
      Here, we describe a new algorithm that uses the
      tools already implemented in {\sc VELOCIraptor}
      to perform fast and efficient phase-space FOF
      searches, but modifying several search and
      assignment criteria to get the desired robustness
      in the identification of galaxies.

      \begin{figure*}
        \centering
        \includegraphics[width=\textwidth]{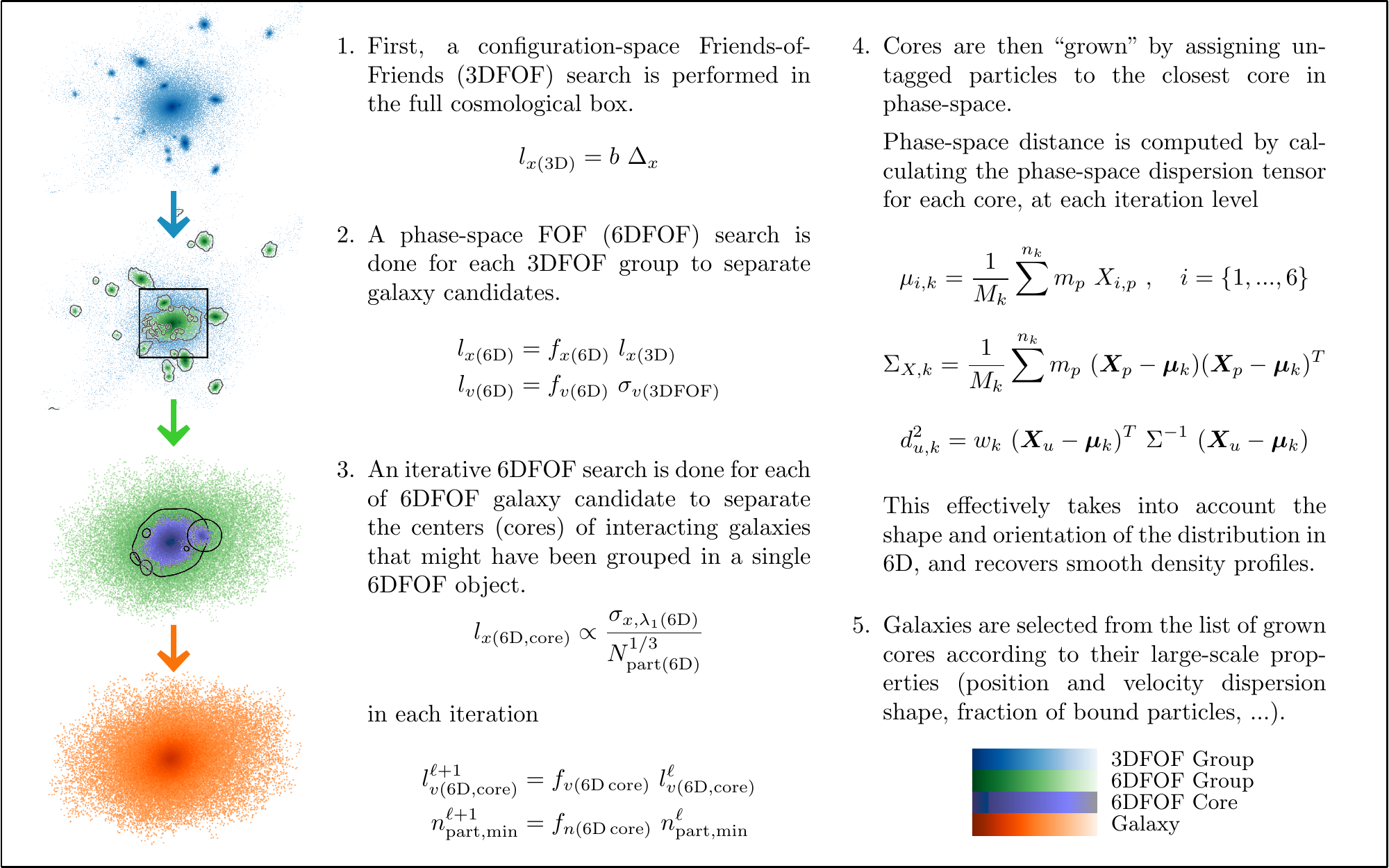}
        \caption{Summary of the algorithm to find galaxies with {\sc VELOCIraptor}
                introduced in Section~\ref{sec:Algorithm}.
                Structures are search by separating particles in the simulation 
                in 3DFOF objects (1), and posteriorly 
                doing a 6DFOF search (2).
                Then, an iterative 6DFOF search is done in each of these objects
                to look for dense cores of galaxies in close interactions
                or mergers (3).
                Once cores are found they are grown by assigning
                particles in the original 6DFOF object (4).
                Finally, properties of all objects found are calculated and
                galaxies are selected according to these properties (5).
                A key aspect of this algorithm is the particle assignment 
                procedure (core growth), as even in the presence of
                satellites close to the host centre (6DFOF Core objects in 
                purple), we can obtain smooth profiles (see central galaxy
                in orange).
                See text for further details.
        }
        \label{fig:algorithm}
      \end{figure*}

	\subsection{An Improved Algorithm to Identify Galaxies}
      The exact definition of what a galaxy is non-trivial in
      both simulations and observations.
      For hydrodynamical simulations a commonly adopted definition
      of a galaxy is all the baryonic mass bound to dark matter
      (sub)haloes.
      Hence, the identification of a galaxy  relies on how well
      (sub)haloes are identified.
      Instead, our aim here is to be able to identify galaxies robustly
      independently of dark matter by using star particle
      information only.
      This is done first by identifying the regions where galaxies
      are expected to be and then separating kinematically distinct
      phase-space overdense structures.
      In this section we describe in detail the algorithm; a schematic
      representation is shown in Fig.~\ref{fig:algorithm}.

      \subsubsection{Step 1 - 3DFOF}
      \label{sec:step1}
      In the dark matter cosmological framework, galaxies
      reside inside large virialised dark 
      matter haloes.
      Our `first guess' of where galaxies are located
      will be the region delimited by the 
      extent of its host dark matter halo.
      This is done by grouping particles that are close
      in physical space using a configuration-space FOF
      search (3DFOF), described by equation~(\ref{eq:3dfof}),
      on the star particles.
      Since its introduction in \citet{Davis1985}, this
      first step is commonly used by many finding algorithms
      \citep[e.g. {\sc Subfind}, {\sc HaloMaker}, 
      {\sc Rockstar},][]{Springel2001,Aubert2004,Tweed2009,
      Behroozi2013} due to its simplicity and versatility.
      For cosmological simulations, a widely adopted scheme is
      
      \begin{equation}
        \label{eq:lx3d}
        l_{x(\mathrm{3D})} = b\ \Delta x \ ,
      \end{equation}

      \noindent where $l_{x(\mathrm{3D})}$ is the 
      configuration-space linking length, $\Delta x$ is the
      simulation's mean inter-particle spacing,
      and $0 < b < 1$.
      We adopt the commonly used value of $b = 0.2$ 
      \citep[e.g.][]{BoylanKolchin2009,Schaye2015,
      Vogelsberger2014} which will group
      star particles inside the dark matter halo.

      \subsubsection{Step 2 - 6DFOF}
      \label{sec:step2}
      Galaxies are centrally concentrated distributions
      of stars in configuration and velocity space.
      In simulations, the positions and velocities of the 
      constituent particles are expected to be found
      close in phase-space.
      Galaxies are identified by performing a phase-space
      FOF (6DFOF) search separating each 3DFOF
      object into kinematically distinct substructures.
      Particles $i$ and $j$ are linked into 6DFOF groups
      if and only if
      
      \begin{equation}
      \label{eq:6dfof}
      \frac{(\bm{x}_i-\bm{x}_j)^2}{l_{x\mathrm{(6D)}}^2} + \frac{(\bm{v}_i - \bm{v}_j)^2}{l_{v\mathrm{(6D)}}^2} \leq 1 ,
      \end{equation}
                 
      \noindent where $l_{x\mathrm{(6D)}}$ and
      $l_{v\mathrm{(6D)}}$ are the configuration
      space and velocity linking lengths, respectively. 
      %
      
      We stress that appropriate values of 
      $l_{x\mathrm{(6D)}}$ and $l_{v\mathrm{(6D)}}$
      have to be chosen in 6DFOF searches. 
      If a very large value of $l_{x\mathrm{(6D)}}$
      is adopted, this would result in a velocity-only
      FOF search and vice-versa; while very small
      values of linking lengths would result in either
      splitting single structures into multiple
      components, or missing structures.
      %
      
      At this point we are interested in separating
      structures that have been found in a common
      3DFOF envelope.
      For this purpose $l_{x(\mathrm{6D})}$ is chosen
      to be a function of $l_{x(\mathrm{3D})}$ and 
      $l_{v(\mathrm{6D})}$ is estimated from the
      velocity dispersion of the full 3DFOF object
 
      \begin{equation}
        \label{eq:lx6d} 
        l_{x(\mathrm{6D})} = f_{x\mathrm{(6D)}} \ l_{x(\mathrm{3D})} \ ,
      \end{equation}      
 
      \noindent and
            
      \begin{equation}
        \label{eq:lv6d}
        l_{v(\mathrm{6D})} = f_{v\mathrm{(6D)}}\ \sigma_v
                           = f_{v\mathrm{(6D)}}
                             \sqrt{\sigma_{v,x}^2 
                             + \sigma_{v,y}^2 
                             + \sigma_{v,z}^2} \ .
      \end{equation}
      
      \noindent Here $0 < f_{x\mathrm{(6D)}} < 1$, 
      $\sigma_{v,j}$ is the velocity dispersion in
      the $j$ direction, and $f_{v\mathrm{(6D)}}$ is
      a user defined parameter which should be of
      order unity.
      As local properties of each 3DFOF object are
      used for its 6DFOF search, we are effectively
      performing a `tailored' 6DFOF search
      \footnote{Consider trying to link particles
      belonging to a Gaussian distribution. Its
      dispersion, $\sigma$, provides a good starting
      point for linking length.}.
      The above choice of parameters is motivated 
      by the fact that galaxies reside in (sub)haloes
      centres, hence their overdensities are expected
      to be much higher than that of the dark matter
      halo.
      This condition is imposed by shrinking the 
      configuration space linking length.
      The velocity space linking effectively removes
      particle bridges in configuration space, 
      resulting in the identification of kinematically 
      distinct structures.
      %
      
      Intuitively it would be more consistent to
      compute $l_{x\mathrm{(6D)}}$ using similar
      arguments as for $l_{v\mathrm{(6D)}}$. 
      However, due to the complexity of the environment
      in which some galaxies reside, measurements of
      position dispersion of the particles would actually
      result in very large values of $l_{x\mathrm(6D)}$. 
      This is especially the case for galaxy groups
      and clusters where particle bridges between
      galaxies make 3DFOF structures too extended.
      A similar argument can be stated against using 
      equation~(\ref{eq:lv6d}), as large 3DFOF objects
      are expected to have very large velocity dispersion,
      and consequently very large values of 
      $l_{v\mathrm{(6D)}}$.
      However, in this case we do not have \emph{a priori} 
      knowledge of what the scale of the velocity linking
      length should be, as this is the first 6DFOF search,
      $\sigma_v$ provides a good first estimation of
      $l_{v\mathrm{(6D)}}$.

      \subsubsection{Step 3 - Iterative 6DFOF core search}
      \label{sec:step3}
      Although the 6DFOF search should already have
      separated galaxies with distinct phase-space
      distributions, multiple galaxies can still be
      found in single 6DFOF groups.
      This is the case of merging galaxies whose 
      outskirts have phase-mixed to some degree but
      whose cores (dense kinematically cold galactic
      centres) have not yet fully merged, or satellites
      that orbit close to the centre of a much bigger
      galaxy.
      Instead of trying to recover a group in its
      entirety, we adopt a different approach and
      attempt to isolate their cores.
      In order to separate galaxies in these
      structures we perform an iterative 6DFOF core
      search for each preliminary 6DFOF group.
      For this iterative 6DFOF core search we use
      the same criteria as equation~(\ref{eq:6dfof})
      to link particles, but using a different choice
      of linking lengths, which for clarity will be
      identified with the subscript $\mathrm{(6D,core)}$.
      These linking lengths scale with the dispersion
      of the system being searched.
      %
      
      FOF algorithms, particularly when used in an
      iterative fashion, are sensitive to the choice
      of linking parameters: too large and separate
      structures can be joined; too small and structures
      can be fragmented. 
      {\sc Rockstar} \citep{Behroozi2013}, which uses
      a 6DFOF to recover groups in full, addresses the
      latter problem by merging groups if their centres
      are closer than a phase-space distance threshold
      to clean for false positives.
      Although useful, our approach is oriented towards
      a robust search of the  \emph{densest portions of
      groups}, followed by carefully growing candidate
      cores, and does not solely rely on the effectiveness
      of cleaning procedures.
      Therefore, we first set appropriately the search
      parameters, which are then modified in each iteration.
      %

      For the initial velocity space linking length we
      adopt
      
      \begin{equation}
        \label{eq:lv6dcore}
         l_{v,\mathrm{(6D,core)}} = \sigma_{v,\lambda_1}\ ,
      \end{equation}
      
      \noindent where $\sigma_{v,\lambda_1}^2$ is the
      length of the largest principal axis of the velocity
      dispersion tensor, $\Sigma_v$.
      As for the first 6DFOF search,
      equation~(\ref{eq:lv6dcore})
      sets the scale for the initial
      velocity space linking length.
      For the following iterations 
      $l_{v\mathrm{(6D,core)}}$ is iteratively
      shrunk, i.e.
      
      \begin{equation}
        \label{eq:itlv6dcore}
          l^{\ell+1}_{v\mathrm{(6D,core)}} = f_{v\mathrm{(6D,core)}}\ 
                                             l^\ell_{v\mathrm{(6D,core)}} \ ,
      \end{equation}
      
      \noindent where the super-script indicates
      the iteration level, and
      $0 < f_{v\mathrm{(6D,core)}} < 1$ is a
      user-defined shrinking factor.
      By shrinking the velocity space linking
      lengths this way, we remove the wings and
      bridges in the distribution, because in each
      iteration we truncate the original
      distribution towards the coldest regions,
      separating cores.
      For this study we adopt 
      $f_{v\mathrm{(6D,core)}} = 0.8$.
      %

      The adopted configuration space linking
      length here is
      
      \begin{equation}
        \label{eq:lx6dcore}
        l_{x\mathrm{(6D,core)}} = 3\ \sigma_{x,\lambda_1} \left(\frac{4\pi}{3}
        \frac{1}{N_\mathrm{part(6D)}}\right)^{1/3} \ ,
      \end{equation}
      
      \noindent where $\sigma_{x,\lambda_1}^2$ is
      the length of the largest principal axis of
      the configuration dispersion tensor,
      $\Sigma_x$, and $N_\mathrm{part(6D)}$ is the
      number of particles in the 6DFOF group.
      Equation~(\ref{eq:lx6dcore}) is then the mean
      inter-particle spacing in a 
      $3\sigma_{x,\lambda_1\mathrm{(6D)}}$ radius
      sphere.
      This linking length scales with 
      configuration-space dispersion and 
      the extent to which the distribution
      is well sampled.
      The logic of including a scaling that
      decreases the linking length with increasing
      number of particles is as follows. 
      With a well-sampled distribution, the 3$\sigma$
      scaling used will link not only the central
      region but the outskirts as well, possibly
      joining this distribution with neighbouring
      ones.
      Decreasing the linking length, if well sampled,
      reduces the likelihood of artificially joining
      structures.
      Conversely, if poorly sampled, the measured
      dispersion will underestimate the true one. 
      Therefore, relative to a well sampled system,
      we scale up the linking length.
      %
      
      Although at this stage the iterative 6DFOF
      search is done to separate structures,
      configuration space linking length is kept
      fixed through iterations.
      We could in principle modify
      $l_{x\mathrm{(6D,core)}}$ by some factor
      $f_{x,\mathrm{(6D,core)}}$ at each iteration
      as is done for $l_{v\mathrm{(6D,core)}}$.
      However, equation~(\ref{eq:lx6dcore})
      already includes the information on how
      concentrated the distribution (6DFOF object)
      is in configuration space.
      Reducing $l_{x\mathrm{(6D,core)}}$ value
      will likely cause that we either miss or
      fragment structures.
      Our approach requires a fixed
      $l_{x,\mathrm{(6D,core)}}$ short enough
      to separate structures in configuration space,
      and a $l_{v\mathrm{(6D,core)}}$ long enough
      to gather statistically significant groups
      of particles.
      In each iteration $l_{v\mathrm{(6D,core)}}$
      is shrunk to separate structures that
      might be linked by their velocity-space
      outskirts.
      %

      %
      %
      \begin{table}
        \centering
        \caption{Suggested values for the parameters used for galaxy 
                 identification with {\sc VELOCIraptor}.}
        \label{tab:values}
        \begin{tabular}{lrcrll} 
          \hline
          Parameter                      & & Value     & & \multicolumn{2}{c}{Reference}    \\
          \hline
          $b$                            & & 0.2       & & Equation  &~\ref{eq:lx3d}        \\
          $f_{x\mathrm{(6D)}}$           & & 0.2       & & Equation  &~\ref{eq:lx6d}        \\
          $f_{v\mathrm{(6D)}}$           & & 1.0       & & Equation  &~\ref{eq:lv6d}        \\
          $f_{v\mathrm{(6D,core)}}$      & & 0.8       & & Equation  &~\ref{eq:itlv6dcore}  \\ 
          $f_{n\mathrm{(6D,core)}}$      & & 1.5       & & Equation  &~\ref{eq:itnpartmin}  \\
          $N_\mathrm{iter}^\mathrm{max}$ & & 8         & & Section   &~\ref{sec:step3}      \\
          $n_\mathrm{part,min}$          & & $\geq$ 50 & & Section   &~\ref{sec:step3}      \\
          $\alpha$                       & & 0.5       & & Equation  &~\ref{eq:weight}      \\
          \hline
        \end{tabular}
      \end{table}

      For each FOF search, a minimum particle
      number, $n_\mathrm{part,min}$ has to be
      set to define statistically significant
      structures.
      For steps 1 and 2 (Sections~\ref{sec:step1}
      and~\ref{sec:step2}) we suggest a 
      $n_\mathrm{part,min} = 50$.
      For the iterative search, however,
      $n_\mathrm{part,min}$ is updated after
      each iteration as
      
      \begin{equation}
        \label{eq:itnpartmin}
          n^{\ell+1}_{\mathrm{part,min}} = f_{n\mathrm{(6D,core)}}\
                                           n^\ell_\mathrm{part,min} \ ,
      \end{equation}
      
      \noindent where $n_{\mathrm{part,min}}$
      is the minimum number of particles,
      $f_{n\mathrm{(6D,core)}} > 1$ and superscript 
      $\ell$ indicates the iteration level.
      Increasing the minimum number of particles
      while shrinking linking lengths may sound
      non-intuitive at first as we expect to
      link \emph{fewer} particles per group in
      each iteration.
      However, as the linking length 
      $l_{v,\mathrm{(6D,core)}}$ becomes smaller,
      it also becomes easier to identify small
      phase-space overdense (noisy) patches in
      the distribution, which can result in
      finding multiple spurious structures.
      Iteratively increasing $n_\mathrm{part,min}$
      reduces the likelihood of finding noisy patches.
      For this study we adopt 
      $f_{n\mathrm{(6D,core)}} \sim 1.5$.
      %
      
      A more intuitive choice of 
      $f_{n\mathrm{(6D,core)}}$ would be one that
      scales with the number of particles in a
      given group or iteration level, instead of
      choosing a fixed $f_{n\mathrm{(6D,core)}}$
      for all searches.
      However, bearing in mind that the number of
      particles can differ by orders of magnitude
      between galaxies in the same system, even
      using a logarithmic scale of the number
      of particles can lead to 
      $f_{n\mathrm{(6D,core)}} \gg 1$, and
      consequently to very large
      $n_{\mathrm{part,min}}$ in a couple of
      iterations.
      %
      
      This iterative 6DFOF search starts with
      the entire 6DFOF object.
      For subsequent iterations the 6DFOF search
      is done only for the largest core.
      This prevents the loss of an already found
      structure due to the increment of
      $n_\mathrm{part,min}$.
      These cores are kept for particle
      assignment (core growth, 
      Section~\ref{sec:step4}) and are revisited
      later to look for possible mergers or
      close interactions.
      Iterations on the largest core stop when a 
      user-defined maximum number of iterations,
      $N_\mathrm{iter}^\mathrm{max}$, has been
      reached, or when no more structures are
      found with the current iteration
      level search parameters.

      \subsubsection{Step 4 - Core growth}
      \label{sec:step4}
      The critical step once cores are identified
      is assigning particles to these cores,
      reconstructing the galaxies.
      We assign particles that belong to the
      original 6DFOF structure (step 2, 
      Section~\ref{sec:step2}) that are
      not member of a core. 
      This process is crucial as the final
      product of structure searches (either
      galaxies or dark matter halos) can be
      severely affected by how this is done.
      %
      
      Given the phase-space nature of the
      6DFOF searches, the obvious criteria
      would be to assign a given particle
      to the closest core in phase-space.  
      This concept has been previously used
      by other algorithms, but several
      implementations can exist.
      A naive 6D phase-space distance as
      implied by \citet{Behroozi2013},
      implicitly assumes a spherical
      morphology. 
      This might work well for dark matter
      haloes but can lead to systematic
      effects due to the complex morphologies
      of galaxies.
      %
      
      Instead, starting at level $\ell$, we
      characterize the phase-space distribution
      of each core $k$, by calculating its
      mean $\bm{\mu}$ (phase-space centre-of-mass
      vector), and phase-space dispersion tensor
      $\bm\Sigma_X$ (distribution's covariance
      matrix),
      
      \begin{equation}
          \bm{\mu}_{k} = \{\mu_{i,k}\}\ , \quad i=\{1,...,6\}  \ ,
      \end{equation}
      \begin{equation}
          \mu_{i,k} = \frac{1}{M_{k}} \sum^{n_k} m_{p}\ X_{i,p} \ ,
      \end{equation}
      \begin{equation}
          \bm\Sigma_{X,k}  = \frac{1}{M_{k}}\ \sum^{n_k} \ m_{p} \ 
                         (\bm{X}_p - \bm{\mu}_k )
                         (\bm{X}_p - \bm{\mu}_k )^T  \ .
      \end{equation}
      
      \noindent Here, $M_k$ and $n_k$ are the
      total mass and the total number of
      particles in the core $k$, respectively;
      $X_{i,p}$ is the $i^\mathrm{th}$ coordinate
      of the phase-space coordinate vector
      $\bm{X}$ of particle $p$ with mass $m_p$,
      that belongs to core $k$.
      Then, for all the particles at $\ell-1$ that
      were not assigned to any core at level
      $\ell$, we calculate
      
      \begin{equation}
         \label{eq:coredistance}
          d^2_{u,k} = w_k \ (\bm{X}_u - \bm{\mu}_k )^T
                     \bm\Sigma_X^{-1} \ 
                     (\bm{X}_u - \bm{\mu}_k ) \ . 
      \end{equation}
      
      \noindent Here $d_{u,k}$ is the phase-space
      distance from untagged particle $u$ to core
      $k$ and $w_k$ is a weighting constant. 
      A weighting scheme is necessary to avoid
      assigning too many particles to tidal streams
      and shells.
      Without a weighting, this could happen as
      these structures can be quite extended and
      have large position and velocity dispersion
      compared to those of galaxies (compact centrally
      concentrated distributions).
      To compensate for this, we adopt
      
      \begin{equation}
        \label{eq:weight}
        w_k = \frac{1}{M_k^{\alpha}}\ ,
      \end{equation}
      
      \noindent with $\alpha$ a free parameter.
      Taking $\alpha = 1$ can cause all particles
      to be assigned to the largest object, again, 
      as galaxy masses in the same system can differ
      by orders of magnitude.
      Values of $1/3 \leq \alpha \leq 2/3$ give a
      $w$ that scales with tidal radius.
      We have found that $\alpha = 0.5$ leads to
      good results; we justify this choice of 
      $\alpha$ in Appendix~\ref{appndx:weight}.
      %
            
      After calculating these distances, particles
      are assigned to the closest core in phase-space.
      When a single core is found at level $\ell$,
      all untagged particles at the previous level,
      $\ell-1$ are assigned to that single core.
      Then, $\bm{\mu}$ and $\bm{\Sigma}$ are
      recalculated for all the cores in the following
      levels and the process is repeated until all
      particles in the original 6DFOF group
      have been assigned to a core.
      %
            
      This approach is particularly powerful for
      many reasons: 
      (i) it effectively takes into account the
      shape and orientation of the distribution; 
      (ii) it allows the shape of the distribution
      to change from the inner to the outer parts;
      (iii) this produces smooth density profiles
      for the galaxies even when galaxies are
      passing through the inner radii of larger
      galaxies. Hence, galaxies will not have
      missing holes or bubble-like structures 
      (see Figs ~\ref{fig:algorithm} 
      and~\ref{fig:clusters} for some examples).
      This is essential when measuring galaxy
      properties' radial profiles.
      %
      
      For each 6DFOF object (step 2, 
      Section~\ref{sec:step2}) the algorithm
      continues as follows.
      After performing step 3
      (Section~\ref{sec:step3}) on the largest
      core, particles are assigned to all
      cores inside following step 4.
      The top hierarchy level, $i$, is assigned
      to the largest core (candidate central
      galaxy).
      The rest of the cores will have hierarchy
      level $i + 1$.
      Steps 3 and 4 are then repeated for all
      $i + 1$ substructures.
      If any sub-substructures are found they
      are assigned a hierarchy level $i+2$, and
      so on.
      The algorithm finishes when all 
      (sub)structures have been iteratively
      searched.

\subsubsection{Step 5 - Selecting galaxies}
      Once all (sub)structures have passed
      through the iterative core search and
      their respective core growth, bulk
      properties of the structures are 
      calculated to determine if they are
      galaxies or not.
      This is necessary because the versatility 
      of the algorithm allows us to identify not
      only galaxies but also tidal features such
      as streams and shells.
      This catalogue can be cleaned if only
      galaxies are desired. 
      %
      
      We classify objects as galaxies or streams
      following \citet{Elahi2013}.
      We calculate the ratios $q \equiv \lambda_2/\lambda_1$
      and $s \equiv \lambda_3/\lambda_1$ of
      the eigenvalues, $\lambda_i$,  of the position and
      velocity dispersion tensors for all the
      structures, as well as the bound fraction of
      particles $f_\mathrm{b}$.
      A structure is not considered as a galaxy if 
      
      \begin{equation}
        \label{eq:isgalaxy}
        \begin{split}
          &(f_\mathrm{b} < 0.01)\quad \cup                                 \\
          &((q_x < 0.3 \cap s_x < 0.2) \cup (q_v < 0.5 \cap s_v < 0.2))\ \cup  \\
          &(f_\mathrm{b} < 0.2 \cap ((q_x < 0.6 \cap s_x < 0.5) \cup (q_v < 0.5 \cap s_v < 0.4)))\ ,\\
        \end{split}
      \end{equation}
      
      \noindent that is, galaxies are expected
      to be bound ellipsoidal distributions of
      stars.
      Structure with less than 1\% of bound
      particles are unlikely to be galaxies.
      Highly elongated structures either in
      configuration or velocity space (i.e.
      low values of $q_x$, $s_x$, $q_v$, and
      $s_v$), which can be bound to some
      degree, are likely to be streams or
      shells.
      The fraction of bound particles is kept
      to such low thresholds, as neither gas
      nor dark matter information is taken
      into account when computing the
      gravitational potential.
      Parameters and thresholds used in 
      equation~(\ref{eq:isgalaxy}) are suggested values
      that were derived from calibration tests to
      give desired results.
      At $z=0$ this selection discards
      $\sim 30$\% of structures with 
      $10^8 < M_* / \msuni < 10^9$, $\sim 1.5$\%
      for $10^9 < M_* / \msuni < 10^{10}$, and 
      $\sim 0.2$\% for 
      $10^{10} < M_* / \msuni < 10^{11}$.
      If desired, other selection criteria could be
      used.
      %
      
      It is important to note that, equation~(\ref{eq:isgalaxy})
      was only tested for {\sc VELOCIraptor}
      outputs. 
      Comparisons throughout this study between
      {\sc VELOCIraptor} and {\sc HaloMaker} are
      done using \emph{raw} catalogues.
      We argue that selection of galaxies 
      using equation~(\ref{eq:isgalaxy}) does
      not impact on the results of this study
      as we focused on \emph{well resolved}
      structures with $M_* > 10^9 \msuni$.

      \subsubsection{Intra-halo stellar component}
      \label{sec:IHSC}
      Once galaxies have been identified inside
      a 3DFOF object, the remaining stellar
      particles are kept and labelled as
      Intra-Halo Stellar Component
      (IHSC).
      The extent, distribution and shape of this
      component relies on the definition itself
      of galaxies (see Appendix~\ref{appndx:ll}
      and Fig.~\ref{fig:ll}).
      The IHSC is therefore all the material that
      is kinematically different enough from the
      distribution of any structure in the 3DFOF
      object.
      This diffuse component can be associated to
      either extended stellar haloes on Milky Way
      like systems, up to Intra-Cluster Light in 
      densely populated environments.
      In-depth analysis of the IHSC is beyond the
      scope of this work; thus, we address this
      in upcoming studies (Ca\~nas et al, in prep).

      \subsection{Adjustable parameters}
      \label{sec:parameters}
      Our new algorithm introduces a few tunable parameters, 
      which determine key aspects of how the search is done.
      We show in Table~\ref{tab:values} the values of
      the parameters used in this work.
      These values are, however, not fixed and can be
      modified to achieve different desired results.
      Here, we briefly describe how modifications to 
      these values can change the identification.
      %
      
     \begin{itemize}
       \item $b$ - Step 1 (Section~\ref{sec:step1}):
             As mentioned above our choice of $b$ is the widely adopted
             $b = 0.2$, which is a good reference to define the extent
             of dark matter haloes in which we are interested in finding
             galaxies.
             This parameter can be changed if a different definition of
             the extent of FOF dark matter halo is adopted 
             \citep[e.g. $b=0.28$][]{Behroozi2013}.

       \item $f_{x\mathrm{(6D)}}$ - Step 2 (Section~\ref{sec:step2}):
             This parameter shrinks $b$ in order to identify higher
             overdensities than those of dark matter haloes.
             From the tests and calibrations  we have performed (Appendix~\ref{appndx:ll}), we found that 
             $f_{x\mathrm{(6D)}} = 0.2$ separates most of the galaxies
             and satellites, leaving only strongly interacting
             systems linked as a single 6DFOF object.
             We further discuss the impact of $f_{x\mathrm{(6D)}}$
             in Appendix~\ref{appndx:ll}.

       \item $f_{v\mathrm{(6D)}}$ - Step 2 (Section~\ref{sec:step2}):
             The velocity dispersion $\sigma_v$ of a 3DFOF object can have
             different meanings for isolated and highly interacting
             systems due to the large dynamical range that is covered
             in cosmological simulations.
             As our aim is to have an automated algorithm to identify
             all the galaxies in such simulations, we suggest to keep it
             $\sigma_v$ unchanged with $f\_v(6D) = 1$. 
             However, $f_{v\mathrm{(6D)}} = 1$ is left as free parameter
             for the possibility of tuning the initial 6DFOF for specific
             cases such as zoom simulations or non-cosmological models.

       \item $f_{v\mathrm{(6D,core)}}$ - Step 3 (Section~\ref{sec:step3}):
             This parameter sets how the velocity linking length 
             scales in each iteration and  can impact on how many
             iterations are performed.
             Small values of $f_{v\mathrm{(6D,core)}}$ will lead to
             fewer iterations, therefore less use of computational
             resources; however, the identification of cores can be
             missed as aggressively shrinking $l_{v\mathrm{6D,core}}$
             can cause particles not to be linked.
             A conservative choice would be values of
             $f_{v\mathrm{(6D,core)}} \sim 1$, which in principle 
             would be able to find all cores; however, this 
             can lead to a very large number of iterations to
             separate cores, and consequently more use of
             computational resources, especially for
             major mergers; for such values a successful separation
             of all cores will then depend on
             $N^\mathrm{max}_\mathrm{iter}$.
             From calibration tests we found that values of 
             $0.7 \leq f_{v\mathrm{(6D,core)}} \leq 0.8$
             successfully separate structures and minimize
             the total number of iterations.

       \item $f_{n\mathrm{(6D,core)}}$ - Step 3 (Section~\ref{sec:step3}):
             This parameters dictates how the minimum number of particles 
             threshold is modified between iterations.
             The purpose of this parameter is to avoid identifying small
             spurious structures, due to shrinking of $l_{v\mathrm{6D,core}}$,
             which happens to be overdense patches in phase-space.
             This parameter is particularly important for galaxy groups and
             clusters due the amount of particle bridges caused by the
             large number of particles in the system and their interactions, and the large
             dynamical range of galaxy masses within them.
             The threshold $n^\ell_\mathrm{part,min}$ changes 
             $\propto n_\mathrm{part,min} \times f_{n\mathrm{(6D,core)}}^m$
             $m^\mathrm{th}$ iteration.
             Values of $f_{n\mathrm{(6D,core)}} \sim 1$ practically  do not 
             change $n^\ell_\mathrm{part,min}$, contradicting the purpose
             of this parameter.
             Values of $f_{n\mathrm{(6D,core)}} \gg 1$ can lead to missing the
             identification of cores of small galaxies specially for
             systems composed of a large number of particles. 
             For example, for $f_{n\mathrm{(6D,core)}} = 3$ and starting with
             $n_\mathrm{part,min} = 50$, would  require a \emph{core} to
             have at least 4,050 particles at a fourth iteration to identified.
             From calibration tests we found $f_{n\mathrm{(6D,core)}} = 1.5$
             to give the desired results in a large simulation, such as Horizon-AGN.
             Deviations of $\pm 0.1$ from the suggested value and starting with 
             $n_\mathrm{part,min} = 50$, lead to differences of 40\% with particle thresholds of 192 for
             $f_{n\mathrm{(6D,core)}} = 1.4$ and 327 for 
             $f_{n\mathrm{(6D,core)}} = 1.6$  at a \emph{fifth} iteration, which are reasonable 
             thresholds for the purpose of this parameter.

       \item $N^\mathrm{max}_\mathrm{iter}$ - Step 3 (Section~\ref{sec:step3}):
             The iterative core search stops when no further cores are
             found with the parameters at a given an iteration.
             Depending on the choices of $f_{v\mathrm{(6D,core)}}$ and 
             $f_{n,\mathrm{(6D,core)}}$, it is possible that a large number
             of iterations are needed before the loop stops.
             This parameter sets the maximum number of iterations in case the
             iterative core search has not stopped.
             Using the values in Table~\ref{tab:values}, the algorithm 
             stops at the $6^\mathrm{th}$ iteration for the largest galaxy cluster
             in Horizon-AGN at $z = 0$.
             Choosing $N^\mathrm{max}_\mathrm{iter} = 8$ sets a reasonable
             threshold in case more iterations are needed.
       \item $n_\mathrm{partmin}$ - Steps 1, 2 and 3 
             (Sections~\ref{sec:step1},~\ref{sec:step2} and~\ref{sec:step3}):
             This parameters sets a threshold over which structures are
             considered as relevant.
             This limit can be adjusted depending on the galaxies of
             interest. In our study we adopt a value of $50$. 

       \item $\alpha$ - Step 4 (Section~\ref{sec:step4}):
             This parameter sets the strength of the mass-dependent
             weight to scale phase-space distances from untagged
             particles to cores.
             The purpose of this parameter is to compensate between
             tidal features with large dispersions and compact 
             dense cores with small ones.
             The value of $\alpha$ can be adjusted depending on 
             the scientific question to be addressed.
             For identification of galaxies and from our calibrations tests, 
             we found $\alpha = 0.5$ to give the best results.
             A thorough discussion and comparison of different
             values of $\alpha$ as well as other choices of $w$
             for the core growth can be found in Appendix~\ref{appndx:weight}.
     \end{itemize}

      \subsection{Comments}
      \label{sec:comments}
      This core growth method has been also implemented
      in the {\sc VELOCIraptor} algorithm to find merging
      dark matter haloes.
      %
      
      We note that none of the finding
      algorithms is exempt from finding undesired
      (spurious) structures. Although for this study
      most of such structures are removed from our
      galaxy catalogue with the criteria described
      in equation~(\ref{eq:isgalaxy}), some spurious
      structures can still be present if they
      happen to be not very elongated in phase-space
      and are marginally bound.
      We leave methods and discussions on this
      matter for the upcoming {\sc VELOCIraptor} 
      paper (Elahi et al., in prep).
      %

      Many structure identification codes implement
      particle unbinding procedures to `clean'
      substructures from particles that  likely belong to a parent structure.
      This means that algorithms are generally 
      focused on finding density peaks (either in
      configuration, velocity or phase-space), 
      while the assignment of particles to these
      peaks is not well addressed and is
      generally overlooked \citep[][]{Knebe2011}.
      {\sc VELOCIraptor} performs unbinding
      procedures for dark matter (sub)halo
      identification.
      In the present study we only use the stellar
      particles information to identify galaxies; that is, 
      we do not take into account any information 
      either from the gas or dark matter distributions.
      Therefore we cannot estimate accurately the
      true gravitational potential at each particle
      position to determine whether it is bound
      or not to a given (sub)structure.
      The latter is also true for the galaxy catalogues
      generated by {\sc HaloMaker} for the
      Horizon-AGN simulation \citep[][]{Dubois2014}.
      We argue that for {\sc VELOCIraptor}, binding 
      information is included to a certain degree by
      requiring that particles belonging to the same
      structure are close in phase-space
      \footnote{This can also be thought
      the other way around.
      The way in which configuration space based finders
      include velocity space information is by 
      including unbinding procedures. 
      Estimating the kinetic energy of each 
      particle takes into account the information of the
      relative velocity of a particle with
      respect to the bulk velocity of a structure
      (either centred on mass, deepest potential
      or highest density), bound particles
      would then need to be those which are
      close in phase space
      }.      
      We stress though that it is crucial how
      particles with lower densities than the peaks
      are assigned to them.
      Even if particle unbinding is fully implemented,
      if the first guess of what a (sub)structure
      is is wrong, no unbinding procedures will fix the 
      problems, as particles would 
      be assigned automatically to its direct host.
      In the structure finding codes found in the literature 
      (to the knowledge of the authors), 
      particles are never re-assigned from hosts to
      substructures, unless using temporal information (tracing)
      to decide where to re-assign particles  
      \citep[e.g. {\sc hbt,hbt+}][]{Han2012,Han2018}.
      %

      As it is shown in the following sections,
      this algorithm is quite efficient and powerful
      at finding galaxies at all simulation-resolved
      mass scales, in all environments.
      We note, however, that this is not the definitive
      method for finding simulated galaxies
      because we do not include
      baryons in the form of gas.
      Hence, we may miss gas-dominated dwarf galaxies,
      which would have very few stellar particles or
      with a bound fraction of particles below our
      adopted threshold.
      This is anyway solved by applying conservative
      particle number thresholds when selecting
      galaxies.
      In the future we plan to link gas to galaxies
      in a similar fashion as we do in the core growth,
      but to do this properly we need to take into
      account the thermal energy of the gas.
      This needs to be carefully implemented to include
      both particle-based and mesh-based algorithms.
      Further discussion on this matter is beyond the
      scope of this paper and is left for future studies.
      %
 
      %
      %
      \begin{figure*}
        \centering
        \includegraphics[width=\textwidth]{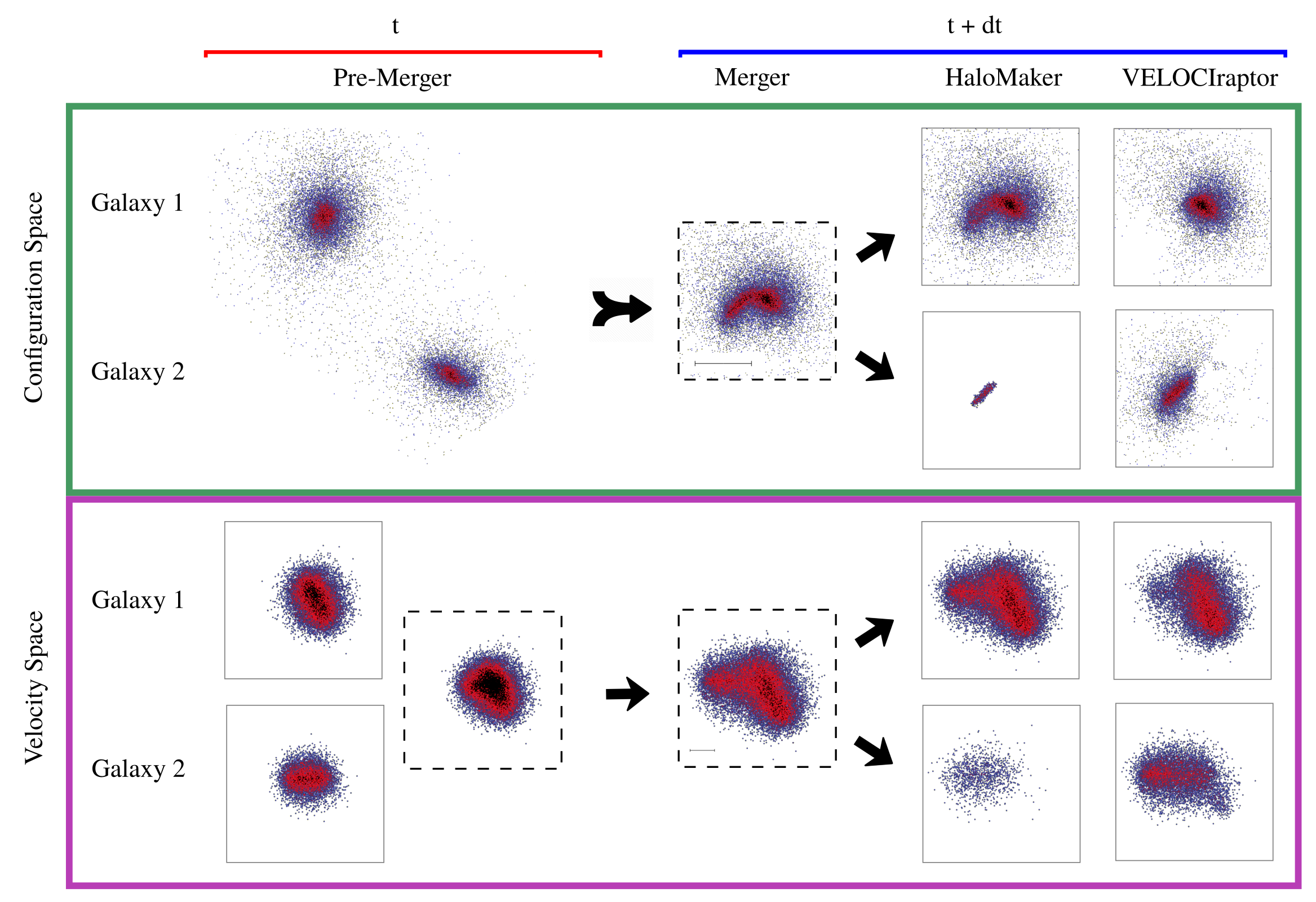}
        \caption{Projected stellar density for a major merger
                 in Horizon-AGN simulation in configuration space
                 and velocity space, as labelled.
                 $t$ shows the galaxies before the merging occurs,
                 and $t + dt$ during the merger.
                 For clarity, for the velocity space visualization
                 at $t$, galaxies are shown both individually and
                 as part of the same velocity space.
                 Galaxies identified with {\sc HaloMaker} and
                 {\sc VELOCIraptor} at $t + dt$ are shown. 
                 We show for each space the projection in which
                 particle distributions are most distinguishable.
                 It can be seen that although {\sc HaloMaker} is
                 able to identify two galaxies, it appears that for
                 the small galaxy only the core of it is identified
                 as an individual one, while its outer regions are
                 assigned to its companion.
                 Due to its phase-space implementation for search and
                 core growth, {\sc VELOCIraptor} is able to find both
                 galaxies and provide a better estimate of their mass
                 and size.
                 The horizontal line in the merger inset shows a
                 length of 20 kpc (200 km s$^{-1}$), which is the
                 same for all the configuration (velocity) space insets.
                 }
        \label{fig:merger}
      \end{figure*}
  
      \section{Case Studies}
      \label{sec:casestudy}
      Here we present two case studies in which we
      compare the results of the improved algorithm
      of {\sc VELOCIraptor} and the galaxies from
      the original catalogue identified with
      {\sc HaloMaker}. 
      With these case studies we address the most
      challenging cases for galaxy identifications,
      which our new algorithm solves well:
      (i) strongly interacting and merging galaxies
      and (ii) robust identification and particle
      assignment in dense environments, such as
      galaxy groups and clusters.

      \subsection{Close interactions}
      \label{sec:closeinteractions}
      Structure finding algorithms have been known
      to struggle to produce robust results when
      trying to separate dark matter haloes and
      galaxies in the process of merging
      \citep{Knebe2011,Behroozi2013,Behroozi2015}.
      The reason behind this problem is that as
      structures start to get closer, the particle
      distributions that describe them start to mix,
      and separating them becomes a complicated task.
      For FOF finders, particle mixing creates
      bridges between the centres of the structures
      that link them together;
      while for density threshold algorithms, the
      mixture of the distribution reduces the contrast
      between peaks and saddle points in the density
      field, making it more difficult to identify
      correctly the components.
      As particle distributions also mix in
      phase-space, even iterative procedures can
      struggle to find peaks, and to assign particles
      correctly to structures, hence host and
      substructure identities can be swapped between
      snapshots \citep[see e.g.][]{Behroozi2015,
      Poole2017}.
      Here we show how our improved galaxy finding
      algorithm performs in such cases.
      %
      
      We show an example of a close merger in
      Fig.~\ref{fig:merger}.
      At a given snapshot, $t$, the galaxies are
      still separated, and have masses of
      $5.61 \, \times \, 10^{10} \, \msuni$ and
      $2.88 \, \times \, 10^{10} \, \msuni$
      respectively, giving a merger ratio of
      $1:1.9$.
      In a subsequent snapshot, $t+dt$, 
      {\sc HaloMaker} identifies two galaxies
      with very different masses of 
      $8.25 \times 10^{10} \, \msuni$ and 
      $4.49 \, \times 10^9 \, \msuni$ respectively,
      corresponding to a merger ratio of $1:18$.
      During a merger, we expect some of the mass
      of one galaxy to be accreted by the other.
      However, from visual inspection we can tell
      that the galaxies have not been well
      separated by {\sc HaloMaker}, as it seems
      that only the core of one of them is
      identified as an individual galaxy, while
      its outer parts have been assigned to its
      companion.
      Although two galaxies are identified, the
      mass of the smallest galaxy is underestimated,
      while the mass of the larger one is
      overestimated.
      %

      We ran {\sc VELOCIraptor} on the same
      merger and it can be seen from simple visual
      inspection that a better result is obtained,
      despite the complexity of the interaction.
      The recovered masses of the galaxies are
      $5.4 \, \times \, 10^{10}\ \msuni$ and
      $3.08 \, \times \, 10^{10}\ \msuni$, giving
      merger ratio of $1:1.75$.
      This is in much better agreement with what
      is measured at $t$, when galaxies were far
      enough as to be easily identified by a 3DFOF
      algorithm.
      It can be seen that not only both the galaxy
      centres are found, but also the shapes of the
      galaxies are well recovered thanks to the
      improved particle assignment (core growth)
      implementation.
      In order to confirm the latter, we analysed different 
      projections of the stellar mass maps of the galaxies, 
      together with the velocity maps and found that prior to
      the merger both galaxies have clear rotation-dominated
      kinematics, and flattened stellar disks, while during
      the merger the primary galaxy continues to have
      rotation-supported kinematics, while the secondary 
      galaxy becomes more disturbed.
      Correctly assigning particles to galaxy
      centres is crucial for an accurate estimation
      of the overall properties of galaxies.
      It affects the ratio of the merger, which in
      turn can affect the overall minor and major
      merger rate estimates, especially when only
      single snapshots are taken into account.
      %

      The (in)capability of disentangling structures in such
      complex interactions might not be considered as a relevant
      problem for finders, as it is easier to look for the
      progenitor structures at earlier times when they are still
      well separated, which ends up not affect the merger ratio
      estimation in a major way.
      However, in general there is not always data available at
      high enough cadence to identify the galaxies at a mass that
      represents best the merger (e.g. maximum mass as is done by
      \citealt{RodriguezGomez2015}) or simply snapshots may not be
      available.
      The capability of identifying robustly galaxies in these
      cases will become more important with the advent of even 
      larger simulations for which storage of a large number of
      snapshots becomes undesirable and even implausible.
      The fact that {\sc VELOCIraptor} succeeds in this
      task without using any temporal information is a major success of our algorithm. 

      \subsection{Groups and Clusters of Galaxies}
      \label{sec:clusters}
      %
      %
      \begin{figure}
            \centering
            \includegraphics[width=\columnwidth, clip=true, trim=0cm 0cm 0cm 0cm]
                            {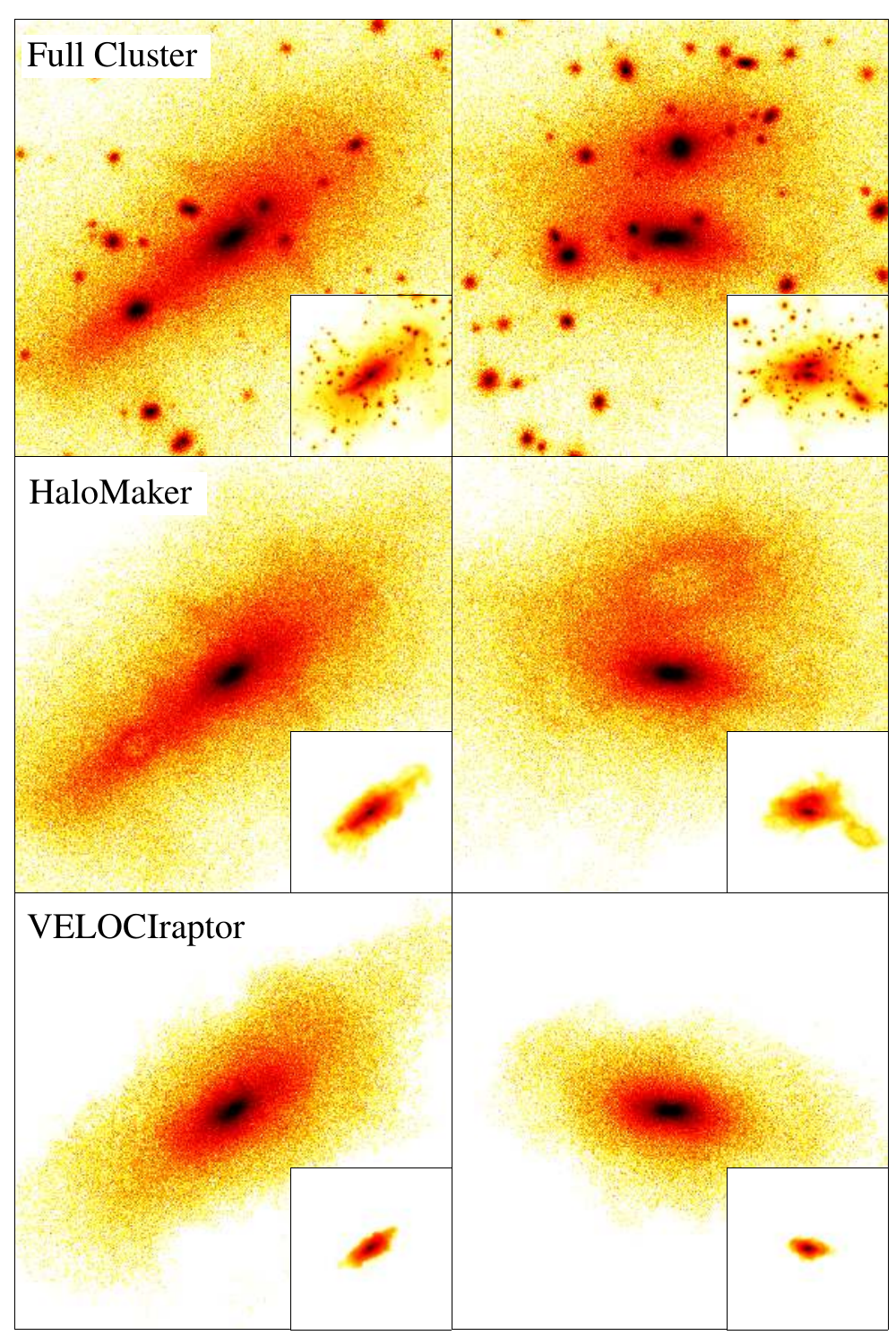}
            \caption{Projected stellar density of the two most massive galaxies 
                     found by {\sc HaloMaker} (middle row), their respective galaxy 
                     cluster (top row), and their {\sc VELOCIraptor} counterparts
                     (bottom row). 
                     Although both codes are able to identify the central
                     galaxy, {\sc HaloMaker} fails to separate stellar content
                     that belongs to other galaxies.
                     To emphasize the full extension of the galaxies, insets show
                     a zoomed-out visualization of the same objects.
                     Panels (insets) have a box size of 600 (2000) kpc.
                     }
        \label{fig:clusters}
      \end{figure}
      Galaxy identification can be a complex task 
      in galaxy groups and clusters.
      Stripped material from multiple interactions 
      generates particle bridges and decreases the
      contrast in the density field, causing similar
      problems as the ones discussed in 
      Section~\ref{sec:closeinteractions}.
      Robust identification of galaxies in 
      such systems is crucial as it can affect
      a very large number of galaxies.
      This can in principle affect environmental
      studies, as well as impact on galaxy
      population measurements.
      %
        
      We show in Fig.~\ref{fig:clusters} two
      galaxy clusters in Horizon-AGN that host
      the two most massive galaxies identified
      by {\sc HaloMaker}.
      We show projected stellar density of the
      full 3DFOF structure (step 1,
      Section~\ref{sec:step1}), the central galaxy
      identified with {\sc HaloMaker}, and the
      {\sc VELOCIraptor} counterpart; a zoomed-out
      visualization of the objects is shown in
      the insets.
      %

      We can see that both codes are able to
      identify correctly a single peak in the
      central galaxy, meaning that there is no
      contamination from undetected satellite
      galaxies.
      In the zoomed-out images it can be seen
      that {\sc HaloMaker} tends to assign a
      large number of particles to the central
      galaxy that belong to other galaxies in
      the cluster. 
      This leads to the odd bubble shapes
      observed for the second {\sc HaloMaker}
      galaxy on its top, and bottom-right
      in the zoom-out inset.
      This problem causes the mass and size of
      the central galaxy to be overestimated.
      On the other hand, because it searches
      for structure in phase space,
      {\sc VELOCIraptor} is able to identify
      kinematically distinct structures, 
      resulting in a better
      delimitation of the galaxy's boundaries.
      %
      
      This example also demonstrates how
      crucial the particle assignment is for
      the robust identification of structures.
      Although both finders are capable of
      identifying the cores of the central and
      satellite galaxies, galaxies can be 
      greatly different due to particle
      assignment procedures.
      This occurs for the galaxy in the first
      column of Fig.~\ref{fig:clusters}, where
      {\sc HaloMaker} assigns particles from 
      an orbiting satellite to the central galaxy.
      Similarly as above, this is seen as a 
      bubble-shaped feature corresponding to 
      the outskirts of the satellite.
      On the other hand, due to the improvements
      of particle assignment using phase-space
      dispersion tensors, {\sc VELOCIraptor} is
      able to separate distinct components
      even if their distributions overlap.
      This produces not only a better estimation
      of the masses of the galaxies, but also
      allows us to recover smooth density
      profiles of the galaxies, which is important
      if we are interested in studying radial
      profiles of galaxy properties.
      %
      
      This problem is not unique of 
      {\sc HaloMaker}, but of structure finding
      codes in general.
      This could in principle be tackled
      by re-assigning procedures, for which 
      particles from central galaxies could
      be returned to any of the other substructures
      identified.
      However, as mentioned in Section~\ref{sec:comments}
      particles are never returned to substructures
      as particles that are not originally part of
      a substructure are expected to be bound to the central 
      halo-galaxy system.
      {\sc VELOCIraptor} attempts to minimize this
      issue by carefully assigning particles to cores
      at each iteration level (Step 4 Section~\ref{sec:step4})
      without any prior assumption on whether cores
      will become central or satellite galaxies.

	\subsection{Temporal evolution of galaxy properties}
    \label{sec:evol}
     We have shown how our new implementation to find galaxies with
     {\sc VELOCIraptor} is capable of identifying galaxies in complex
     environments.
     However, a robust algorithm requires that structures are identified
     consistently over time.
     This is necessary to ensure that studies focused on the evolution
     of single galaxies or systems, are not affected by the finder.
     Temporal evolution of structures is either tracked by linking
     structures across catalogues using merger trees
     \citep[see for reference][]{Srisawat2013,Avila2014,Wang2016,Poole2017},
     or is done \emph{on the fly} during structure identification by 
     tracing algorithms \citep[e.g. {\sc hbt,hbt+}][]{Han2012,Han2018}.
     It is well known that the evolution traced by merger trees can be
     severely affected by the specific implementation of structure finding
     algorithms \citep[e.g.][]{Avila2014,Poole2017}.
     Though the goal of this study is not focused on testing the consistency
     of merger trees for our galaxy catalogues, we show in this section how
     large-scale properties of galaxies as well as their radial distribution
     evolve for our catalogues.
     %

     For this purpose we generate galaxy catalogues on high-cadence
     snapshots produced for stellar particles only in Horizon-AGN.
     These catalogues are temporally spaced every $\sim 25$ Myrs, 
     with a total of 778 snapshots being  available.
     To test time consistency in the properties of the galaxies identified
     by our algorithm and {\sc HaloMaker}, we selected from the
     most massive galaxy cluster at $z = 0$ the four most massive
     galaxies, hereafter referred to as \texttt{Galaxy 1-4}, respectively,
     and follow their evolution backwards in time for $40$ of the above
     mentioned snapshots, corresponding to $\sim 1$ Gyr of evolution.
     %
     
     We trace galaxies between snapshots using 
     {\sc TreeFrog} \citep[Elahi et al. in prep,][]{Poulton2018}
     a tool associated to the {\sc VELOCIraptor} repository to
     construct merger trees for simulations.
     Galaxies in a reference snapshot are matched by finding
     the structure that shares the most particles in a
     subsequent snapshot.
     This is done by looking at the individual particle IDs
     that belong to the galaxies and computing a merit function
     \begin{equation}
     \label{eq:merit}
     \mathcal{M}_{ij} = \frac{N^2_\mathrm{sh}}{N_i \ N_j} \ .
     \end{equation}
     \noindent Here, $N_i$ and $N_j$ are the total number
     of particles in structures $i$ and $j$ respectively,
     and $N_\mathrm{sh}$ is the number of shared particles,
     i.e. that exist both in $i$ and $j$.
     This method ensures that galaxies in one snapshot
     are matched to the galaxy in the subsequent snapshot
     that is most similar in particle members and that shares
     a large fraction of those.
     %
     
      %
      %
      \begin{figure}
            \centering
            \includegraphics[width=\columnwidth, clip=true, trim=0cm 0cm 0cm 0cm]
                            {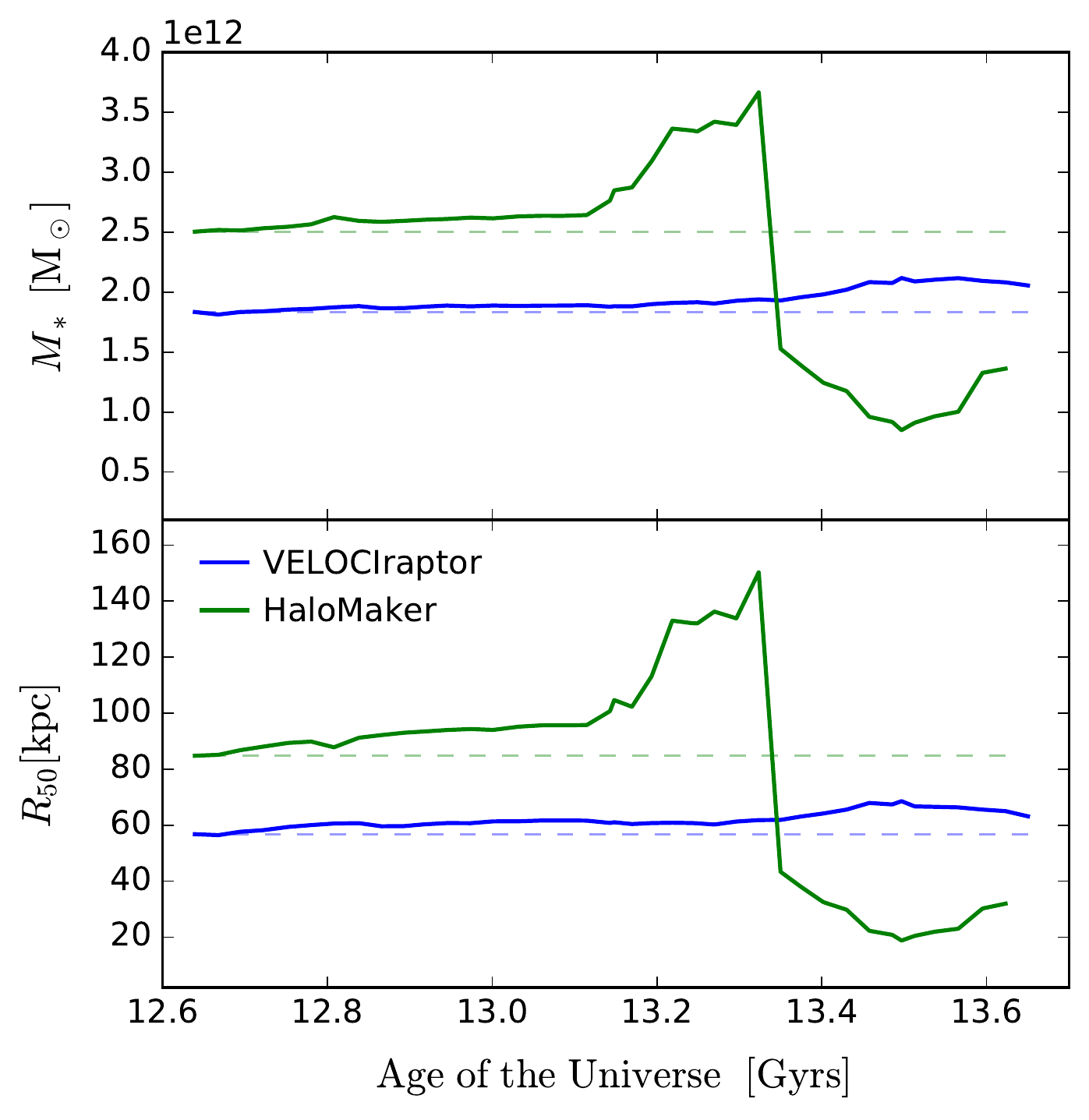}
            \caption{Evolution of the total stellar mass ($M_*$, upper panel),
                     and spherical half-mass radius ($R_{50}$, bottom panel), for
                     the central galaxy of the most massive cluster in 
                     Horizon-AGN at $z=0$, as estimated by {\sc VELOCIraptor} 
                     (blue) and {\sc HaloMaker} (green). A dashed line is shown
                     as reference for the initial estimated value of each 
                     property. {\sc VELOCIraptor} is capable to following 
                     consistently the evolution of the bulk galaxy properties
                     in complex environments without applying any temporal
                     corrections. Further details for other three massive
                     members of the same cluster can be found in 
                     Appendix~\ref{appndx:evol}.
                    }
        \label{fig:gal1props}
      \end{figure}

     The upper panel of Fig.~\ref{fig:gal1props} shows 
     the evolution of $M_*$ for the most massive galaxy in the
     cluster, \texttt{Galaxy 1}, found by {\sc VELOCIraptor}
     (blue) and {\sc HaloMaker} (green).
     We calculate $M_*$ simply by adding the stellar mass
     of all the particles in the galaxy.
     The bottom panel shows the evolution of $R_{50}$, which
     is the spherical radius which encloses     half of $M_*$.
     Solid lines show the evolution of each quantity, and
     a dashed line shows, as reference, the initial amplitude
     of each quantity for each finder.
     %
     
     We demonstrate that for {\sc VELOCIraptor} the evolution
     of $M_*$ and $R_{50}$ is stable through time.
     Slight increments and decrements are expected due to the
     evolution of the galaxy through mergers and interactions.
     For {\sc HaloMaker} it is seen  that the evolution for the first 
     $\sim 500$ Myrs is quite stable; however,
     past that point there is huge increment of both $M_*$
     and $R_{50}$ for $\sim 200$ Myrs; then a sudden drop, which
     decreases to a minimum at $t = 13.5$ Gyrs. During the last $100$~Myrs,
     the magnitude of the properties increase steadily with time.
     The sudden increment at $t = 13.1$ Gyrs is consistent with
     the case studies presented above, which  show that central galaxies
     identified by {\sc HaloMaker} include other galaxies' outskirts.
     In this case, as \texttt{Galaxy 1} gets closer to the other
     massive galaxies, {\sc HaloMaker} for a short period of time
     adds their outskirts as part of \texttt{Galaxy 1};
     the abrupt decrement happens when \texttt{Galaxy 1} is not
     considered to be the central anymore, and galaxies' outskirts are
     assigned to a companion galaxy, \texttt{Galaxy 2}.
     A visualization of the evolution, as well as the evolution of the 
     properties of the other 3 massive galaxies in the cluster can 
     be found in Appendix~\ref{appndx:evol}.
     %
     
\begin{figure}
            \centering
            \includegraphics[width=\columnwidth, clip=true, trim=0cm 0cm 0cm 0cm]
                            {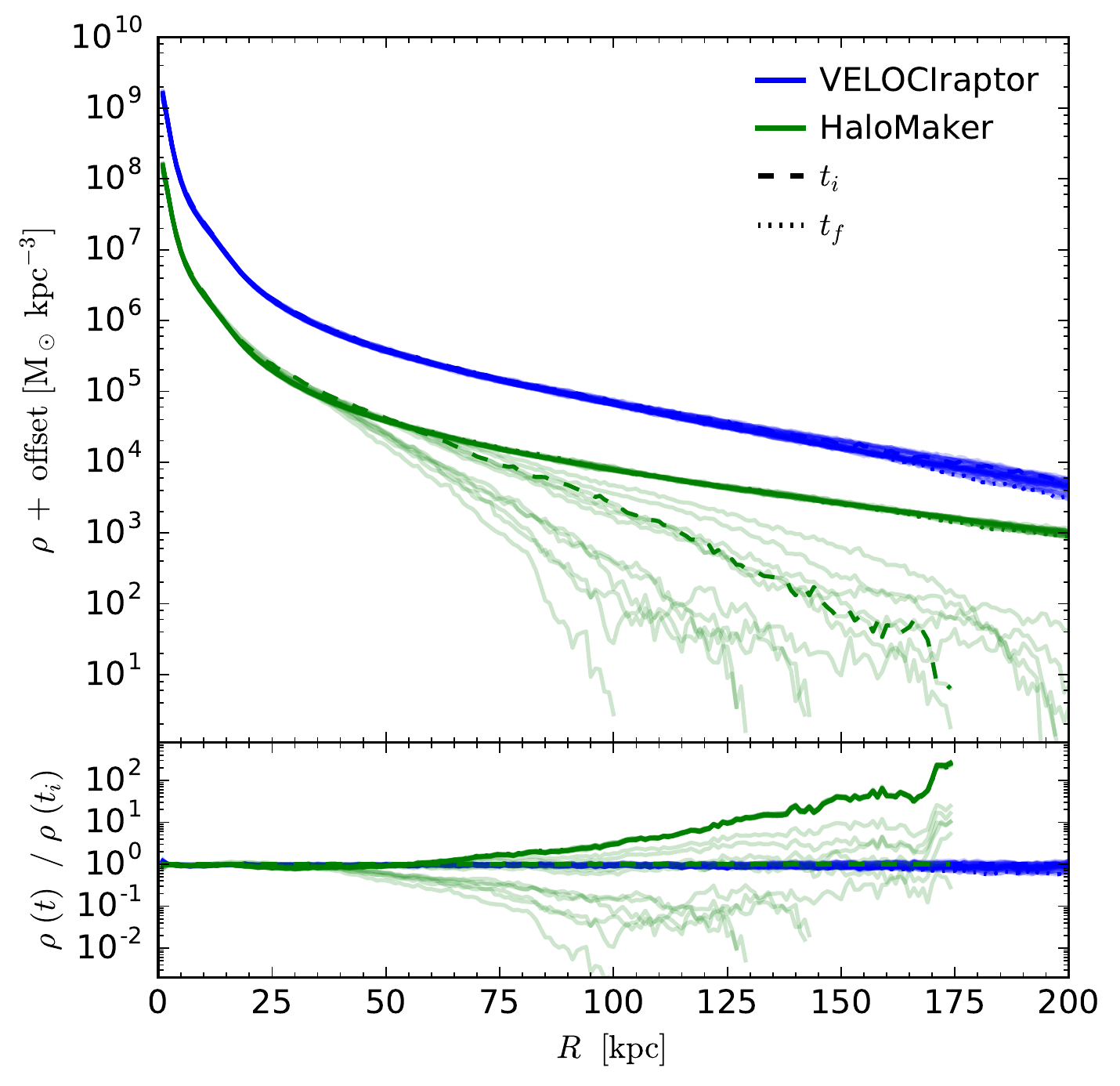}
            \caption{Evolution of the stellar density profile $\rho$ for the 
                     same galaxy of Fig.~\ref{fig:gal1props} for $\sim1$ Gyr. 
                     The profile estimated
                     by {\sc VELOCIraptor} and {\sc HaloMaker} is shown in solid blue and
                     green lines, respectively. {\sc HaloMaker}'s profile is shifted
                     by -1 dex for clarity. Dashed and dotted lines show the
                     profile measured at the snapshot when we start ($t_i$) and stop
                     ($t_f$) tracking the galaxies, respectively. This implementation of
                     {\sc VELOCIraptor} is capable also of obtaining consistent density 
                     profiles to very large radii (>100 kpc) even for massive galaxies
                     with multiple orbiting satellites interacting in a complex 
                     galaxy cluster. The evolution of
                     the density profile for the other 3 most massive galaxies in the same
                     cluster is shown in Appendix~\ref{appndx:evol}.
                    }
        \label{fig:gal1profile}
      \end{figure}

     We further test temporal consistency by measuring the radial 
     stellar mass distribution of the galaxies.
     The top panel of Fig.~\ref{fig:gal1profile} shows the 
     stellar volume density profile $\rho$ of \texttt{Galaxy 1} produced by
     {\sc VELOCIraptor} (blue lines) and {\sc HaloMaker}
     (green lines), from the snapshot where galaxies were first
     identified $t_i$ ($z = 0$) to the last snapshot used
     $t_f$; for {\sc HaloMaker} we have offset the profile by
     $-1$ dex for clarity.
     The profile is calculated by adding the mass of all stellar
     particles inside fixed 1 kpc bins describing concentric 
     spherical shells around the centre of mass of the galaxy, and
     dividing over the volume of the shell.
     The bottom panel shows the ratio of the density at each bin
     at a time, $t$, with respect to the density of the same bin
     at time $t_i$, as solid lines.
     The density profiles at $t_i$ and $t_f$ for each finder are
     shown as dashed and dotted lines, respectively.
     We show that {\sc VELOCIraptor} does not only produce stable
     large-scale properties, but also the mass profile of  
      \texttt{Galaxy 1}.
     On the other hand, the {\sc HaloMaker} stellar mass profile is 
     only stable for the inner $40$ kpc, fluctuating by up to
     two orders of magnitude at large radii.
     This is also due to particle assignment , which truncates the outskirts of 
     \texttt{Galaxy 1} when is identified as a satellite rather than the central
     by {\sc HaloMaker}; this is seen as the decrement 
     in the density profile, which corresponds to the `valley'
     observed for $M_*$ and $R_{50}$ at $t \gtrsim 13.3$ Gyrs.
     Between $12.6 \leq t/{\rm Gyr} \leq 13.1$, {\sc HaloMaker}'s profile
     seems stable and smooth for three reasons: (i) the outskirts 
     are not truncated, (ii) even if other galaxies' outskirts are
     added (producing the bumps in $M_*$ and $R_{50}$), those particle have radii
     much greater than 200 kpc, 
     and (iii) even if those particles are asymmetrically distributed
     with respect to the \texttt{Galaxy 1}'s centre-of-mass, the profile
     looks smooth, as calculating $\rho$ spherically averages that
     added outskirts (see Appendix~\ref{appndx:evol} for further details).
     %

      %
      %


    \section{Results}
    \label{sec:results}
      In this section we study the differences
      between the {\sc HaloMaker} and 
      {\sc VELOCIraptor}.
      For this purpose we generate a new galaxy
      catalogue for the Horizon-AGN simulation
      using our improved algorithm.
      In Section~\ref{sec:galaxytogalaxy} we
      compare the catalogues on a
      galaxy-to-galaxy basis to study how much
      galaxy properties can be affected by the
      identification method.
      In Section~\ref{sec:popstatistics} we
      investigate how differences in the
      identification can affect measurements
      of galaxy population properties.
      %
      

      \subsection{Galaxy-to-Galaxy Comparison}
      \label{sec:galaxytogalaxy}
          We investigate differences between the
          finders by performing a galaxy-to-galaxy
          comparison.
          Matching structures between catalogues
          is a similar process as building merger
          trees.
          The best match of a galaxy is found 
          by looking at the particles IDs information
          only.
          Therefore we use {\sc TreeFrog} as a 
          catalogue correlator, and equation~(\ref{eq:merit})
          to select the best match.
          %

          We compare the total stellar mass $M_*$, SFR,
          and sizes $R_{50,90}$ of matched galaxies by
          computing
          \begin{equation}
              f_Y = Y_{\mathrm{{\sc HaloMaker}}}/
                    Y_{\mathrm{{\sc VELOCIraptor}}} \ ,
          \end{equation}
          \noindent which is the ratio between the above
          mentioned quantities, $Y$, as measured for the
          {\sc HaloMaker} galaxy, over the one measured
          for its {\sc VELOCIraptor} counterpart.
          In order to make a proper comparison and avoid
          resolution effects, we only show $f_Y$ of
          galaxies whose total stellar mass is greater
          than $10^9 \ \msuni$ in both catalogues, and
          only matches with $\mathcal{M} > 0.1$ (galaxies
          sharing $\gtrsim 30$\% of  particles) are shown.
          %

          %
          %
          \begin{figure*}
            \centering
            \includegraphics[width=0.98\textwidth]{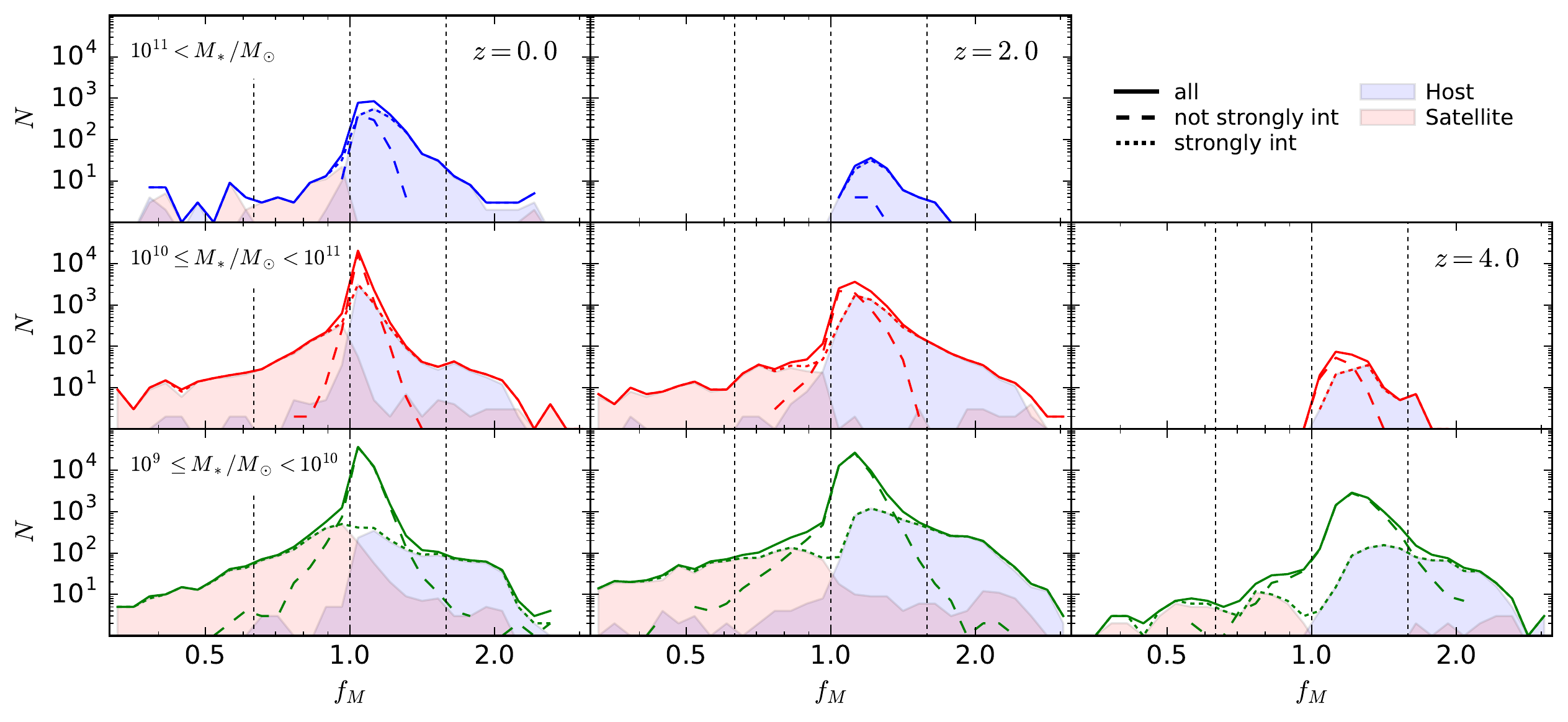}
            \caption{Distributions of the mass ratio 
                     $f_M = M_{*\mathrm{\sc HaloMaker}}
                     / M_{*\mathrm{\sc VELOCIraptor}}$ 
                     at $z=0$ (left panels), $z = 2.0$ 
                     (middle panels), and $z = 4.0$ 
                     (right panels) for different 
                     stellar mass ranges, as labelled.
                     The contribution from isolated galaxies
                     and loosely interacting galaxies
                     is shown as a dashed line, from
                     interacting galaxies as a dotted line,
                     and the combined distribution as a solid
                     line.
                     For strongly interacting galaxies the
                     contribution from hosts is shown
                     as shaded blue region, while for
                     satellites as shaded red region.
                     Vertical dashed lines are shown as
                     reference at $\pm 0.2$ dex
                     from an exact match ($f_M \equiv 1$).
                    }
            \label{fig:fmdist}
          \end{figure*}
          
          To properly account for the cases shown in 
          Section~\ref{sec:casestudy}, we labelled galaxies
          depending on their degree of interaction as:
          
          \begin{itemize}
            \item \textbf{Isolated} - The galaxy is the
                  only structure found in the initial
                  3DFOF envelope.
            \item \textbf{Loosely interacting} - The galaxy
                  belongs to a 3DFOF object with multiple
                  structures, and \emph{no} structures were
                  found in its iterative search, i.e. a
                  single structure in a 6DFOF object.
            \item \textbf{Strongly interacting} - The galaxy
                  belongs to a 3DFOF object with multiple
                  structures, and one or more additional
                  structures were found in the
                  6DFOF object iterative search.
                  The most massive galaxy in the 6DFOF object
                  will be referred to as \emph{host}, otherwise
                  as \emph{satellite}.
                  Note that strongly interacting hosts are
                  not necessarily central galaxies of the
                  3DFOF object (e.g. most massive galaxy of
                  an interacting pair falling into a galaxy
                  cluster).
          \end{itemize}

         A total of 81,583 matches fulfil the above criteria
         at $z = 0$.
         Of those, a total of 54,113 are isolated; from the
         remaining 27,470 interacting galaxies, 16,851 (10,619)
         are loosely (strongly) interacting.
         Approximately 3\% of all the structures in each of
         the catalogues do not have a counterpart in the 
         other catalogue.
         %
          
        
        \subsubsection{Total Stellar Mass}
        \label{sec:totmass}
          We measure the impact of identification on
          one of the most fundamental properties of a 
          galaxy, the total stellar mass, $M_*$.
          We show in Fig.~\ref{fig:fmdist} the 
          distributions of the ratio $f_M$ for galaxies
          in {\sc VELOCIraptor} in the mass ranges of
          $10^{9} \leq M_*/\msuni \leq 10^{10}$,
          $10^{10} \leq M_*/\msuni \leq 10^{11}$, and
          $M_*/\msuni \geq 10^{11}$, which we will
          refer to as M09, M10, and M11 respectively.
          The $f_M$ distribution of all galaxies is
          shown as a solid line; dashed lines show
          the contribution from both isolated and
          loosely interacting galaxies (i.e. not strongly
          interacting); dotted lines show the
          contribution from strongly interacting galaxies.
          For the latter we show the contribution
          from host and satellite galaxies as shaded
          blue and red regions, respectively.
          Vertical dashed lines are shown as reference
          at $\pm 0.2$ dex with respect to an identical
          estimated mass, i.e. $f_M \equiv 1$.
          %

\begin{figure}
            \centering
            \includegraphics[width=\columnwidth]{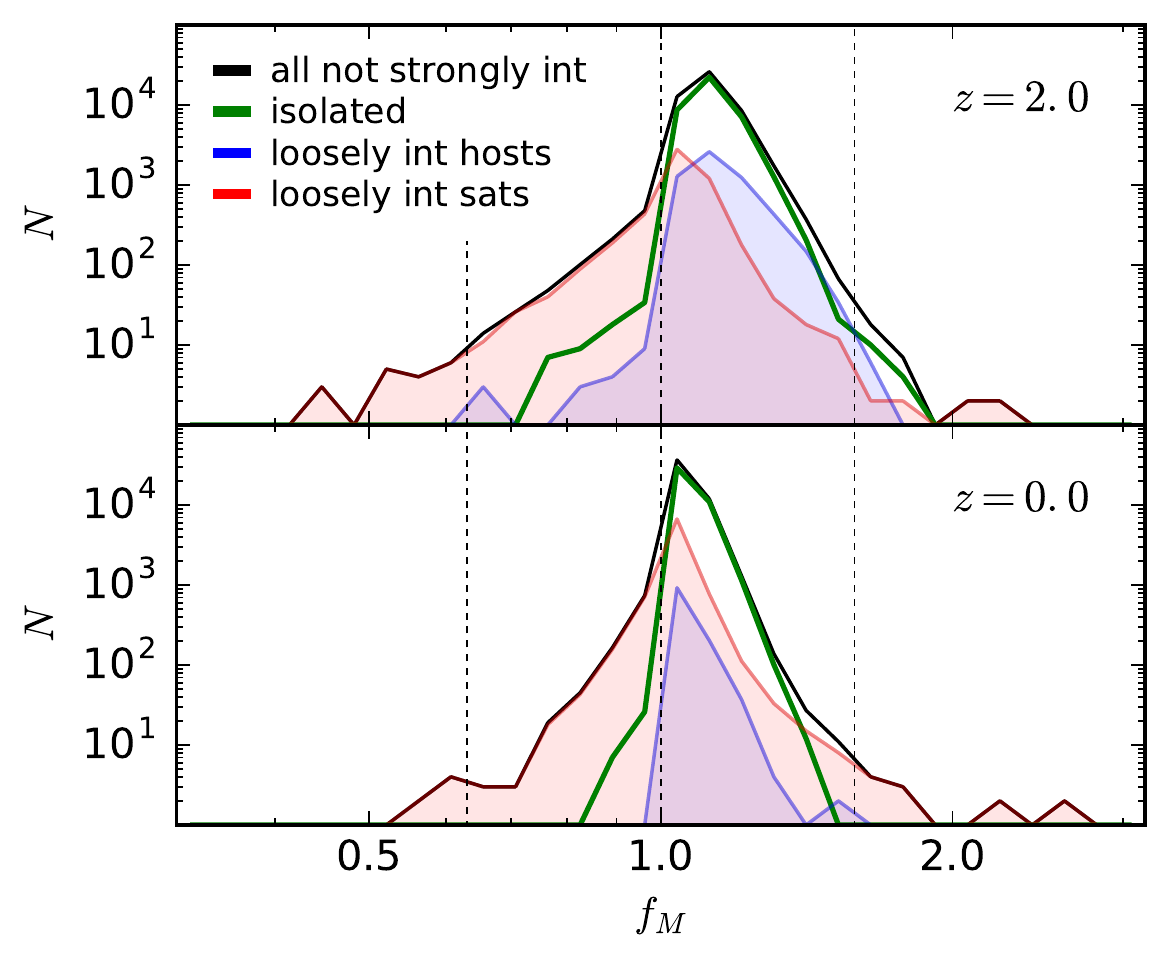}
            \caption{Distributions of the mass ratio $f_M $ for 
                     not strongly interacting galaxies
                     with $10^9 \leq M_*/\msuni \leq 10^{10}$
                     at $z = 2$ (top panel) and $z = 0$
                     (bottom panel).
                     The contribution from isolated galaxies is
                     shown as a solid green line, and from loosely
                     interacting hosts and satellites as shaded blue
                     and red regions, respectively.
                     Vertical dashed lines are shown as reference
                     at $\pm 0.2$ dex from an exact match ($f_M \equiv 1$).
                    }
            \label{fig:fmdistnotint}
          \end{figure}

          At $z = 0$, the $f_M$ distributions of the M09,
          M10 and M11 samples show that galaxies that 
          are not strongly interacting are overall distributed
          around $f_M = 1$.
          This is something we would expect because any
          finder in the literature should not have any
          problem with the identification of isolated
          density peaks or particle distributions.
          Looking closely we see that the peak is slightly
          skewed towards $f_M > 1$, as a result of 
          differences in the initial galaxy identification
          steps taken by the codes.
          {\sc HaloMaker} assigns all particles inside
          a 3DFOF object to identified structures, but
          {\sc VELOCIraptor} performs an additional 6DFOF
          search which delimits galaxies by grouping
          only phase-space close particles.
          This procedure effectively gets rid of the
          furthest phase-space particles reducing the
          `available' mass to distribute between
          galaxies inside the original 3DFOF object; 
          the mass excess observed for {\sc HaloMaker}
          galaxies is the mass we consider to be part
          of the IHSC (Section~\ref{sec:IHSC}).
          The reason why the M09 $f_M$ distribution is
          not as narrow as the other sample is likely
          due to all galaxies in M10 and M11 being
          well resolved, while some galaxies in M09
          could still be affected by resolution
          effects.
          In Fig.~\ref{fig:fmdistnotint} the $f_M$
          distributions for all not strongly interacting
          galaxies in the M09 sample are shown.
          The majority of
          isolated and loosely interacting hosts
          have $f_M > 1$, consistent with
          the above description.
          On the other hand, loosely interacting
          satellites describe a symmetric $f_M$
          distribution centred at $f_M = 1$, suggesting
          that their mass can either
          be over- or underestimated by {\sc HaloMaker}
          compared to {\sc VELOCIraptor}.

          The greatest difference between the catalogues is
          seen in the strongly interacting population
          (dotted lines in Fig.~\ref{fig:fmdist}).
          Their $f_M$ distributions at all stellar masses 
          are broader than for the rest of the population.
          This shows that there is a non-negligible number
          of \emph{resolved} galaxies that are affected by
          the artificial transfer of mass in interacting systems
          due to the particle assignment criteria of the finder
          (see Fig.~\ref{fig:merger} for an example).
          Host galaxies display a significant preference
          for $f_M > 1$, while the $f_M < 1$ part of the
          distribution is predominantly dominated by satellite
          galaxies, which is more evident for the M11
          sample.
          This picture is consistent with the examples shown
          in Section~\ref{sec:casestudy} (Figs.~\ref{fig:merger}
          and~\ref{fig:clusters}), meaning that the behaviour
          observed in those examples are not simple
          {\sc HaloMaker} outliers, but are a recurrent
          phenomenon in the simulation.
          %
      
          For a considerable amount of strongly interacting
          satellites {\sc HaloMaker} \emph{overestimates}
          their mass compared to {\sc VELOCIraptor}, i.e. 
          they have $f_M > 1$.
          As these are interacting systems, a fraction of
          these cases can be explained by host-satellite
          swapping, where a galaxy that is considered a
          satellite by {\sc VELOCIraptor}
          is in fact the host galaxy in {\sc HaloMaker}.
          This artificially increases their mass
          compared to what {\sc VELOCIraptor} estimates.
          It is not uncommon to see this phenomenon across
          catalogues from different finders, and it is even
          present for the same finder between
          different snapshots, especially in the case of major
          mergers \citep[see e.g.][]{Behroozi2015,Poole2017}.
          %

          At higher redshifts, the $f_M$ distributions
          display an amplified version of the behaviour
          observed at $z = 0$.
          Distributions for all galaxies at all masses
          become wider, and peak further from $f_M > 1$.
          The $f_M$ distributions at $z = 2$, shown in the
          middle panels of Fig.~\ref{fig:fmdist}, are wider,
          with more prominent wings compared to $z = 0$.
          In addition, the M09 $f_M$ at $z = 2$ peaks at
          $f_M \approx 1.2$ for not strongly interacting
          galaxies.
          Strongly interacting galaxies show a similar 
          behaviour to the $z = 0$ ones, with the $f_M > 1$
          region being dominated by host galaxies, and the 
          $f_M < 1$ region by satellites, with some
          fraction of satellites also having $f_M > 1$,
          likely due to the host-satellite swapping.
          The $f_M$ distributions at $z = 4$ (right panels
          in Fig.~\ref{fig:fmdist}) show an even wider
          distribution than that at $z = 2$ with a less
          prominent peak in the case of isolated galaxies.
          %
          
          In order to understand some of the differences
          at $z = 2$ and $z = 4$ we have to bear in mind
          the nature of the AMR calculation,
          and the properties of high-redshift galaxies.
          Horizon-AGN was run using an AMR code for
          which the grid cells used to compute gravity and
          hydrodynamics change as the simulation evolves
          depending on local density, affecting
          the effective resolution of the simulation.
          Cells are allowed to be refined when the universe
          has an expansion factor of $a = 0.1,\ 0.2,\ 0.4,\ 0.8$
          ($z = 0.25,\ 1.5,\ 4.0,\ 9.0$), in order to keep
          the physical size of the cell somewhat constant.
          This affects the spatial and mass scales at which
          gas forms stars, hence the scales on which galaxies
          are resolved, impacting also on their identification.
          Additionally, such large differences for the same 
          galaxy can be explained by the fact that at high
          redshifts, galaxies are clumpier and more compact
          than at the present time.
          Bursts of star formation within the same galaxy
          could be easily identified as separate structures,
          by either of the finders.
          However, we expect that {\sc VELOCIraptor} is capable
          of joining structures that are kinematically similar
          that might appear as separate structures in
          configuration space.
          Lastly, at high redshifts we also expect the number
          of mergers to increase \citep[e.g.][]{Fakhouri2010},
          hence we expect that the example analysed in
          Section~\ref{sec:casestudy} becomes more frequent.
          %


        \subsubsection{Star Formation Rate}
        \label{sec:sfr}
          Another fundamental quantity measured for galaxies is
          their star formation rate (SFR).
          We calculate the SFR for each galaxy by adding up the
          mass of all stellar particles with age smaller than a 
          given $\Delta t$ threshold, and dividing the sum over
          that period of time.
          Results presented here were obtained adopting 
          $\Delta t = 50\ \mathrm{Myr}$, and were corrected for
          a recycling fraction of 0.44 for a Kroupa initial 
          mass function following \citet{Courteau2014},
          implicitly assuming instantaneous recycling.
          We have also calculated SFRs using $\Delta t$ windows
          of $\Delta t = 20$ and $100$ Myrs, find that results
          are robust.
          %

          \begin{figure}
            \centering
            \includegraphics[width=\columnwidth]
                     {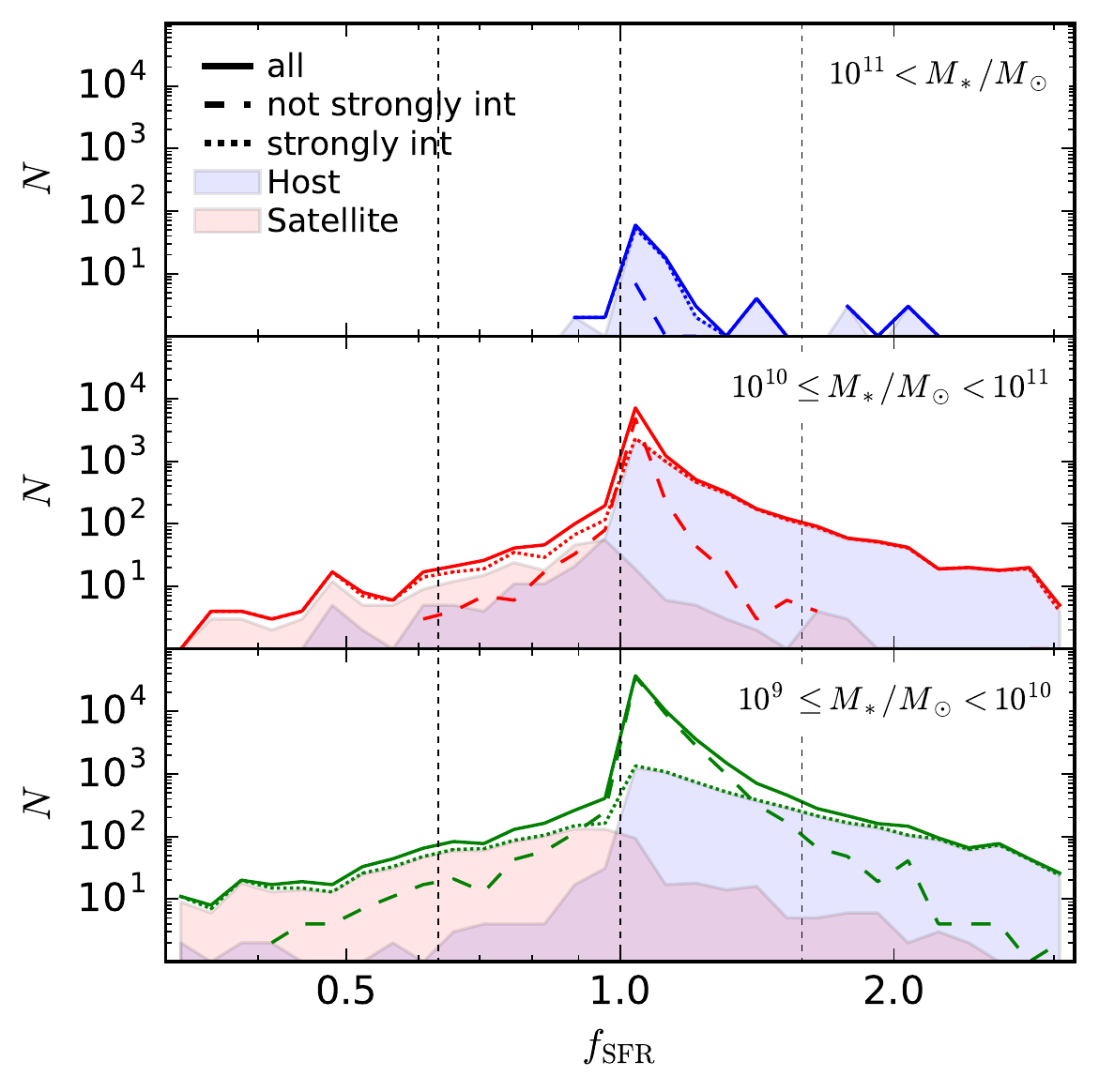}
            \caption{Distribution of the SFR ratio
                     $f_\mathrm{SFR}$ at $z = 2$ for different
                     mass ranges, as labelled. The contribution
                     from isolated galaxies is shown as a dashed
                     line, interacting galaxies as a dotted line,
                     and the combined distribution as a solid line.
                     For interacting galaxies the contribution
                     from host and satellite galaxies is shown as
                     shaded blue and red region, respectively.
                     For reference vertical dashed lines are shown
                     at $\pm 0.2$ dex from an exact match
                     $f_\mathrm{SFR} \equiv 1$.
                     }
            \label{fig:fsfrdist}
          \end{figure}
          
          We measure 
          $f_\mathrm{SFR} = \mathrm{SFR}_\mathrm{HaloMaker} /
          \mathrm{SFR}_\mathrm{VELOCIraptor}$ to quantify the 
          galaxy-to-galaxy difference in the estimated SFR.
          Figure~\ref{fig:fsfrdist} shows the $f_\mathrm{SFR}$
          distribution at $z = 2$ for samples M09, M10, and M11
          as green, red and blue lines, respectively.
          Similarly to Fig.~\ref{fig:fmdist}, we show the total
          $f_\mathrm{SFR}$ distribution, and the contribution
          from not strongly interacting galaxies, and strongly 
          interacting hosts and satellites, as labelled.
          $f_\mathrm{SFR}$ peaks close to $f_\mathrm{SFR} = 1$
          for not strongly interacting galaxies in all mass ranges.
          %
          
          In this case the $f_\mathrm{SFR}$ distribution's 
          peak is slightly more narrow compared to the one
          displayed by the $f_M$ distribution (Fig.~\ref{fig:fmdist}),
          and is also slightly shifted towards $f_\mathrm{SFR} > 1$
          due to the 6DFOF search of {\sc VELOCIraptor} that
          `removes' the phase-space outermost particles of the
          initial 3DFOF object, as discussed
          in Section~\ref{sec:totmass}.
          The spread in the $f_\mathrm{SFR}$ distribution comes
          mostly from strongly interacting galaxies, with
          $f_\mathrm{SFR} > 1$ ($f_\mathrm{SFR} < 1$) 
          corresponding to host (satellite) galaxies.
          It is interesting though that for host galaxies the
          $f_\mathrm{SFR}$ tail is quite prominent and extends
          to $f_\mathrm{SFR} > 2$ at all stellar masses.
          For satellites, on the other hand, 
          $f_\mathrm{SFR} < 1$ tails are more prominent at lower
          stellar masses.
          Galaxies whose mass is overestimated via the 
          spurious acquisition of outer material of an 
          orbiting satellite, will for instance
          increase their SFR, and vice versa.
          Moreover, the satellite galaxies affected by
          this are likely to have only the inner
          non-star-forming core as the galaxy
          (see for example Fig.~\ref{fig:merger}).
          This will drastically reduce their estimated
          SFR, while for their hosts it will be enhanced
          by the incorrect assignment of the star-forming
          outskirts of the satellite.
          %

          \begin{figure}
            \centering
            \includegraphics[width=\columnwidth]
                    {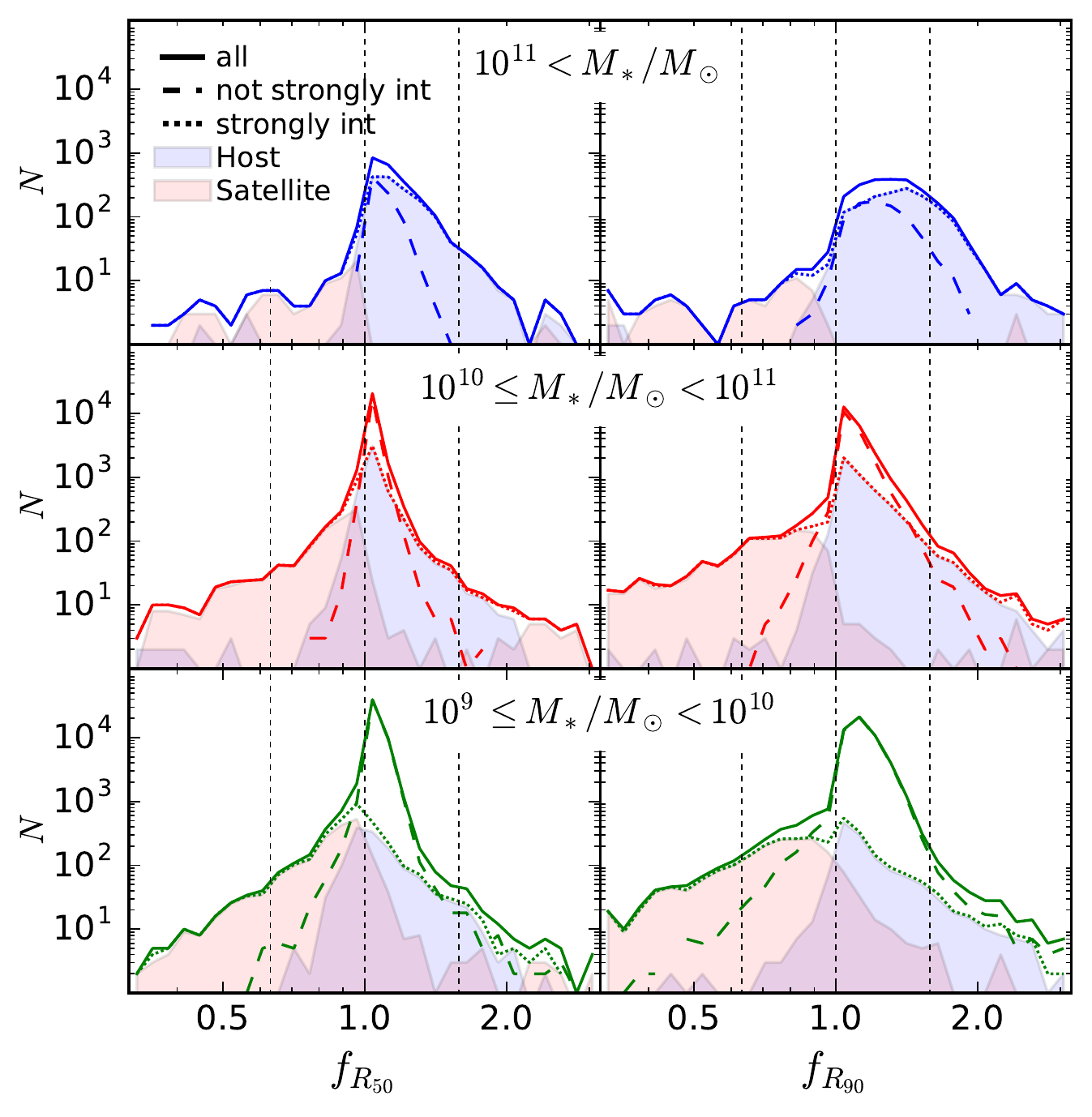}
            \caption{Distribution of the $R_{50}$, $f_{R_{50}}$,
                    and $R_{90}$, $f_{R_{90}}$, ratios for
                    different mass ranges at $z = 0$, 
                    as labelled.
                    The contribution from not strongly
                    interacting and strongly interacting host
                    and satellite galaxies is shown, as labelled.
                    For reference, vertical dashed lines 
                    show $\pm$0.2 dex from $f_{R_X} \equiv 1$.
                    }
            \label{fig:frdist}
          \end{figure}

        \subsubsection{Sizes - Enclosed Mass Radius}
        \label{sec:sizes}
          Accurate estimation of galaxy sizes is crucial
          as they are used not only to test how well galaxy
          formation models agree with observations, but have
          also been used for calibration of subgrid physics
          parameters by some of the present-day
          hydrodynamical simulations
          \citep[e.g][]{Crain2015}.
          We calculate spherical radii $R_{50}$ and $R_{90}$,
          which enclose 50\% and 90\%, respectively, of the 
          total stellar mass of a galaxy.
          We show in Fig.~\ref{fig:frdist} the 
          $f_{R_Y} = R_{Y,\mathrm{\sc HaloMaker}} 
          / R_{Y,\mathrm{\sc VELOCIraptor}}$ distribution, 
          for $R_{50}$ and $R_{90}$ at $z = 0$.
          Galaxy samples are colour coded as in
          Figs.~\ref{fig:fmdist} and~\ref{fig:fsfrdist}.
          %
          
          At all stellar masses the total $f_{R_{50}}$
          distribution peaks close to $f_{R_{50}} = 1$,
          and its tails extend beyond $\pm$ 0.3 dex from
          this value.
          At $10^{9} \leq M_* / \msuni \leq  10^{11}$ the
          peak of $f_{R_{50}}$ comes from not strongly
          interacting galaxies, while at 
          $M_* > 10^{11}\ \msuni$ isolated and
          interacting host galaxies contribute equally.
          Isolated and loosely interacting galaxies display
          a narrow distribution close to $f_{R_{50}} = 1$
          at all stellar masses, with its peak being slightly
          shifted towards $f_{R_{50}} > 1$.
          Both of these behaviours are similar to those
          seen for $f_M$ and obey to the same reasons
          (see Section~\ref{sec:totmass}).
          Not strongly interacting galaxies, however, at
          $M_* < 10^{10}\ \msuni$ show a larger spread on
          $f_{R_{50}}$, as galaxies can have values from
          $f_{R_{50}} \lesssim 0.5$ up to $f_{R_{50}} \gtrsim 2.0$; 
          while the former does not contribute largely to
          the low-$f_{R_{50}}$ tail, isolated and loosely
          interacting galaxies do contribute to the spread
          at the high-$f_{R_{50}}$ end.
          %
    
          Similarly to $f_M$ and $f_\mathrm{SFR}$ 
          (Figs.~\ref{fig:fmdist} and~\ref{fig:fsfrdist}, respectively),
          strongly interacting galaxies contribute the most to the
          spread in the $f_{R_{50}}$ distribution.
          Satellite galaxies comprise the majority of the
          low-$f_{R_{50}}$ population at all stellar masses,
          whereas the high-$f_{R_{50}}$ population tail arises
          from host galaxies, consistent with previous
          comparisons. 
          This is most evident for galaxies with
          $M_* > 10^{11}\ \msuni$ (top panel) where strongly 
          interacting host galaxies comprise the bulk of the
          total $f_{R_{50}}$ distribution, which is skewed
          towards $f_{R_{50}} > 1$.
          The latter can be seen in the bottom row of 
          Fig.~\ref{fig:clusters}, where the zoomed-out insets
          clearly show that the extension of the galaxy is
          expanded by {\sc HaloMaker} as it also considers other
          galaxies' outskirts as part of the central.
          %
    
          \begin{figure*}
            \centering
            \includegraphics[width=\textwidth]
                    {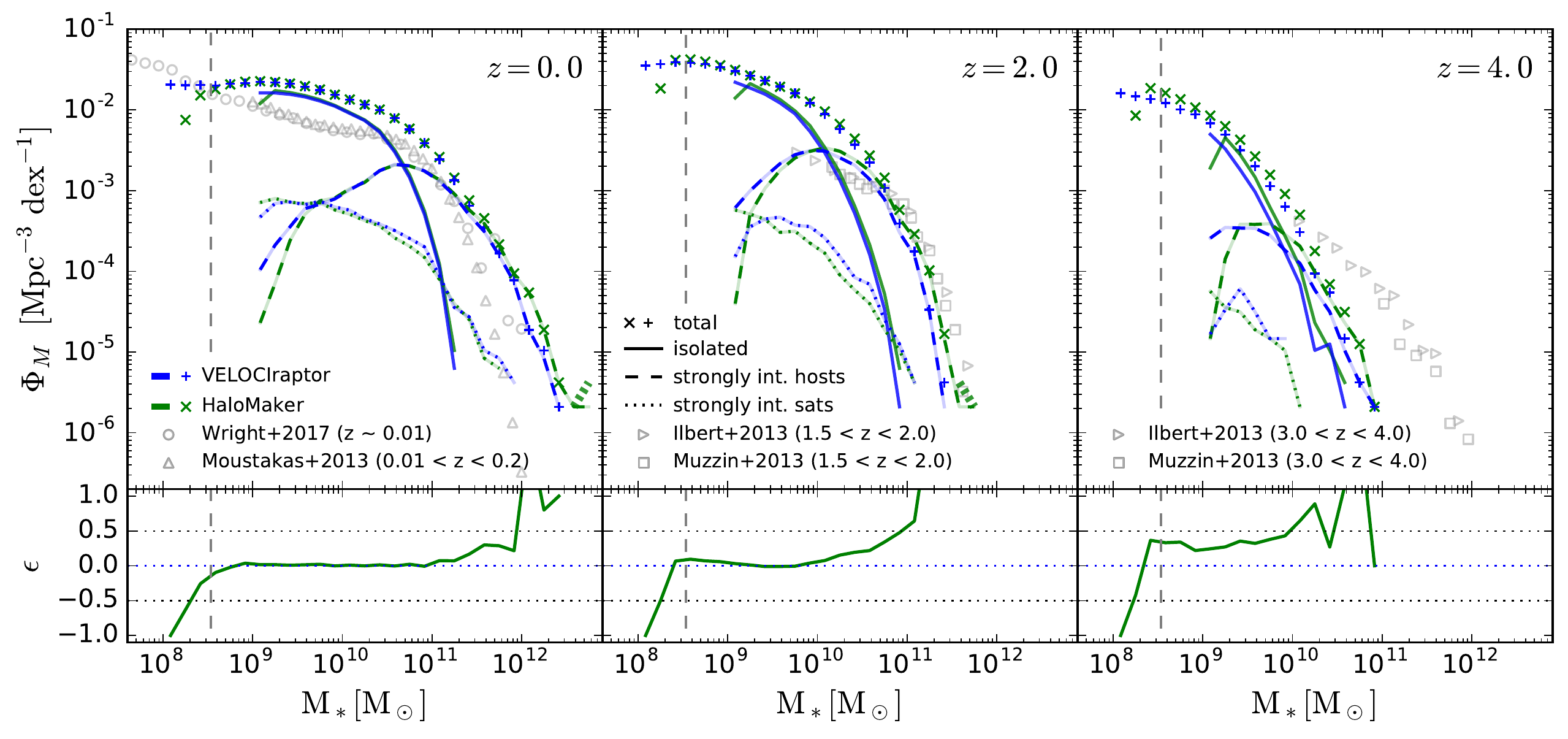}
            \caption{Galaxy Stellar Mass Function
                     (GSMF) of the Horizon-AGN simulation
                     calculated using
                     {\sc VELOCIraptor} (blue) and
                     {\sc HaloMaker} (green), measured
                     at redshifts $z=0,\ 2,\ 4$, as labelled
                     in each panel.
                     The GSMF of all galaxies is shown as
                     plus and cross symbols for 
                     {\sc VELOCIraptor} and {\sc HaloMaker}, 
                     respectively.
                     The contribution from not strongly interacting
                     galaxies is shown as a solid thin line;
                     and from strongly interacting hosts and
                     satellites as dashed and dotted lines,
                     respectively.
                     Bottom panels show the relative difference,  
                     $\epsilon = (\Phi_\mathrm{HaloMaker})/
                     \Phi_\mathrm{VELOCI} - 1$ of the total GSMF.
                     A dotted thick line is shown for 
                     {\sc HaloMaker}'s total GSMF at stellar
                     masses where {\sc VELOCIraptor} does not
                     find any structures.
                     A vertical dashed grey line is shown at the
                     mass equivalent to structures composed of 100 
                     particles for reference.
                     Observations of the GSMF from
                     \citet{Wright2017}, \citet{Moustakas2013},
                     \citet{Muzzin2013} and \citet{Ilbert2013}
                     are shown as symbols, as labelled.
                    }
            \label{fig:GSMF}
          \end{figure*}
          
          The overall behaviour of $f_{R_{90}}$ (right panels
          in Fig.~\ref{fig:frdist})
          is similar to $f_{R_{50}}$, however the spread is
          much larger at all stellar masses, for all the 
          galaxy samples.
          For not strongly interacting galaxies a narrow peak
          is no longer visible and $f_{R_{90}}$ extends well
          above 0.2 dex from $f_{R_{90}} = 1$.
          Although $R_{90}$ encloses almost all the mass of
          the galaxy, the distributions do not resemble to
          those of $f_M$ at the same redshift (see 
          Fig.~\ref{fig:fmdist}), especially for 
          $M_* > 10^{11}\ \msuni$ galaxies, suggesting that
          sizes are more sensitive to finder systematics than
          the stellar mass is.
          Interacting host galaxies peak at $f_{R_{90}} = 1.41$,
          showing that the size of central galaxies of groups
          and clusters are greatly increased by {\sc HaloMaker},
          which is consistent with the examples shown in 
          Fig.~\ref{fig:clusters}.
          %

          For completeness purposes we repeated the above 
          analysis for radius enclosing 20\% and 100\% of total
          stellar mass, and results 
          are consistent to those for $R_{50}$ and $R_{90}$, 
          respectively.
          %


      \subsection{Galaxy Population Statistics}
      \label{sec:popstatistics}
      In this section we study the impact of the identification 
      method on the statistical properties of the galaxy
      population of Horizon-AGN.
      We measure standard galaxy properties and compare
      {\sc VELOCIraptor} and {\sc HaloMaker} catalogues.
      Our main objective here is not to test how well 
      Horizon-AGN reproduces the observed galaxy population,
      but to compare how statistical measurements of
      galaxy population can be affected by identification
      and the resulting consequences of the biases
      discussed in Section~\ref{sec:galaxytogalaxy}.
      %
      
                    
        \subsubsection{Galaxy Stellar Mass Function}
        \label{sec:gsmf}
          We start with the simplest measurement, 
          the Galaxy Stellar Mass Function (GSMF).
          Simulations are often tuned to reproduce this
          quantity \citep[see review of][]{Somerville2015}.
          Moreover, it has been demonstrated that a
          GSMF consistent with observations can be
          obtained by tuning subgrid physical 
          model parameters
          \citep[see e.g.][]{Crain2015}.
          Therefore, it is essential that we understand
          and control for all the systematic effects
          behind measuring the GSMF in our simulations.
          %

          We show in Fig.~\ref{fig:GSMF} the GSMF,
          $\Phi_M = dN/d\log M_*$, of Horizon-AGN 
          measured with {\sc VELOCIraptor}
          (solid blue line), and {\sc HaloMaker}
          (solid green line) at redshifts 
          $z = 0,\ 2,\ 4$.
          To quantify the agreement between the catalogues,
          we compute the relative difference

          \begin{equation}
            \epsilon = (\Phi_{M_*,\mathrm{\sc HaloMaker}}/
                        \Phi_{M_*,\mathrm{\sc VELOCIraptor}})
                        - 1 \ .    
          \end{equation}
          
          \noindent We also show the contribution to the GSMF
          from isolated (solid thin line) and strongly interacting
          hosts and satellites (dashed and dotted lines,
          respectively) from the matched galaxies as described
          in Section~\ref{sec:galaxytogalaxy}.
          %

          At $z = 0$ the overall shape of the GSMF measured 
          by both catalogues is similar.
          However, differences
          can be seen at both the low and high mass ends.
          At $M_* < 10^{9}\ \msuni$
          {\sc HaloMaker} finds fewer galaxies than 
          {\sc VELOCIraptor}.
          The GSMF predicted by {\sc HaloMaker} displays a
          declining curve towards lower stellar masses, while
          {\sc VELOCIraptor}'s GSMF shows a plateau.
          This can be attributed to two factors.
          First at low number of particles, the density
          field used by {\sc HaloMaker} is likely to be poorly
          sampled, making it possible that structures are not
          dense enough in configuration space to be identified.
          {\sc VELOCIraptor} is better at picking up these
          structures as they are dense in velocity-space as well.
          This is consistent with comparison studies which
          have found that in general 6D-based finders tend
          to perform better at identifying structures with
          low number of particles
          \citep[][]{Knebe2011,Knebe2013b}.
          The second reason is attributed to the specific
          particle and density thresholds used by
          {\sc HaloMaker} to define relevant structures.
          At this mass range, galaxies are composed of 
          $\lesssim 300$ particles, close to the resolution
          limit of the simulation, making the identification
          of its peaks and saddle points challenging.
          %

          It has been pointed out by several studies that
          structures need to be composed by at least hundreds
          or thousands of particles in order to have reliable
          measurements of their internal properties, as well as
          resolved merger histories, 
          \citep[e.g][]{Knebe2013b,vandenBosch2018,Chisari2015,
          vandenBosch2017, Elahi2018}.
          We argue that {\sc VELOCIraptor} is capable of
          robustly identifying structures at very low particle
          numbers, as has been shown in other studies
          \citep{Elahi2018}.
          However, we leave further analysis on structures
          with small number of particles for an upcoming
          study (Elahi et al. in prep).
          Finally, it is worth mentioning that for Horizon-AGN,
          only galaxies with $M_* > 10^{9}\ \msuni$ are
          considered as resolved structures due to resolution.
          %

          At $M_* > 10^{11} \msuni$, the GSMF of
          {\sc HaloMaker} predicts between 20\% up
          to 100\% more galaxies than {\sc VELOCIraptor}. 
          This difference is a result of the IHSC
          being assigned to the central galaxy in 
          {\sc HaloMaker}.
          Therefore it is not that {\sc VELOCIraptor}
          is unable to find such big galaxies, but the
          fact that the mass of central 
          galaxies in {\sc HaloMaker} is systematically
          increased by the finder.
          Note that {\sc HaloMaker}'s GSMF extends to
          higher masses than the most
          massive galaxy obtained by {\sc VELOCIraptor},
          shown as a dotted thick green line in
          Fig~\ref{fig:GSMF}.
          %

          Despite the wide $f_M$ distributions shown in 
          Fig.~\ref{fig:fmdist}, the GSMFs practically
          overlap in the mass range between 
          $10^9 \leq \ \msuni \leq\ 10^{11}$.
          This is partially explained by the peak of
          the total $f_M$ distribution (Fig.~\ref{fig:fmdist}),
          located close to $f_M = 1$, which mostly
          comes from isolated galaxies (see 
          Fig.~\ref{fig:fmdistnotint}).
          The latter are the galaxies that contribute
          the most to the GSMF at $M_* < 10^{11}\msuni$.
          Another factor is that the over- and
          under-estimation of the stellar mass by 
          {\sc HaloMaker} is compensated between
          systems of different masses.
          This effect can be seen at 
          $M_* \gtrsim 10^9\msuni$ from the mass functions
          of different galaxy populations, 
          where {\sc VELOCIraptor} predicts more
          isolated and strongly interacting host galaxies
          than {\sc HaloMaker}.
          The opposite happens for strongly interacting
          satellites, giving a total GSMF that agrees at
          those stellar masses.
          This is even more evident at $z = 2$
          at $10^9 \leq M_*/\msuni \leq 10^{10.5}$,
          where catalogues predict different numbers of 
          galaxies for different populations, and still
          the total GSMFs agree relatively well.
          The observed difference in the estimation of
          $M_*$ (as seen in Section~\ref{sec:totmass}) leads
          to a shift in mass for the GSMF.
          Such difference can only be distinguished beyond the
          break of the GSMF, as the flat slope at lower masses
          has the effect of making the shift of the GSMF
          indistinguishable.
          %

          At higher redshifts, the GSMF has a different
          behaviour than at $z = 0$.
          Although the overall shape of the GSMF is roughly
          similar, there is a clear offset in the normalisation.
          At $z = 2$ (middle panel of Fig.~\ref{fig:GSMF})
          the GSMFs at $M_* < 10^9\ \msuni$ behave similarly
          to the $z = 0$ ones, except for a slightly higher
          number density obtained by {\sc HaloMaker} compared
          to {\sc VELOCIraptor} at 
          $2.5\ \times\ 10^{8}\ \lesssim M_*/\msuni \lesssim\ 9\ \times\ 10^{8}$.
          At $M_* \sim 10^9\ \msuni$, both GSMF agree well,
          however as we go to higher masses, {\sc HaloMaker}'s
          GSMF starts to deviate from the one measured by
          {\sc VELOCIraptor}, with up to $\sim 50$\% more
          galaxies at $M_* \sim 10^{11} \ \msuni$.
          At $z = 4.0$ (right panel in Fig.~\ref{fig:GSMF}),
          the GSMF of {\sc VELOCIraptor} and {\sc HaloMaker}
          are completely offset at all masses.
          For galaxies with 
          $10^{8.5} \lesssim M_*/\msuni \lesssim 10^{9.5}$,
          {\sc HaloMaker} predicts between 30\% and 40\% more
          galaxies than {\sc VELOCIraptor}.
          This difference increases to 50\% up to more than 
          100\% at $M_* \gtrsim 10^{10}\ \msuni$.
          We can see from the mass function of isolated
          galaxies and strongly interacting hosts that
          {\sc HaloMaker} predicts more galaxies at all
          stellar masses.
          This difference, however, to some degree can be
          explained by the IHSC that {\sc VELOCIraptor}
          is able to separate, but {\sc HaloMaker} includes
          as part of the galaxy.
          By adding the IHSC mass to their respective
          central galaxy we could in principle shift 
          these mass functions to the right, matching
          those obtained by {\sc HaloMaker}.
          As discussed for Fig.~\ref{fig:fmdist}, differences
          between the catalogues at high redshift can
          also be attributed to (i) higher merger rates, 
          (ii) bursty star formation (iii) AMR
          resolution implementation.
          %
          

        \subsubsection{Star Formation Rate Function}
        \label{sec:sfrf}
          \begin{figure}
            \centering
            \includegraphics[width=\columnwidth]{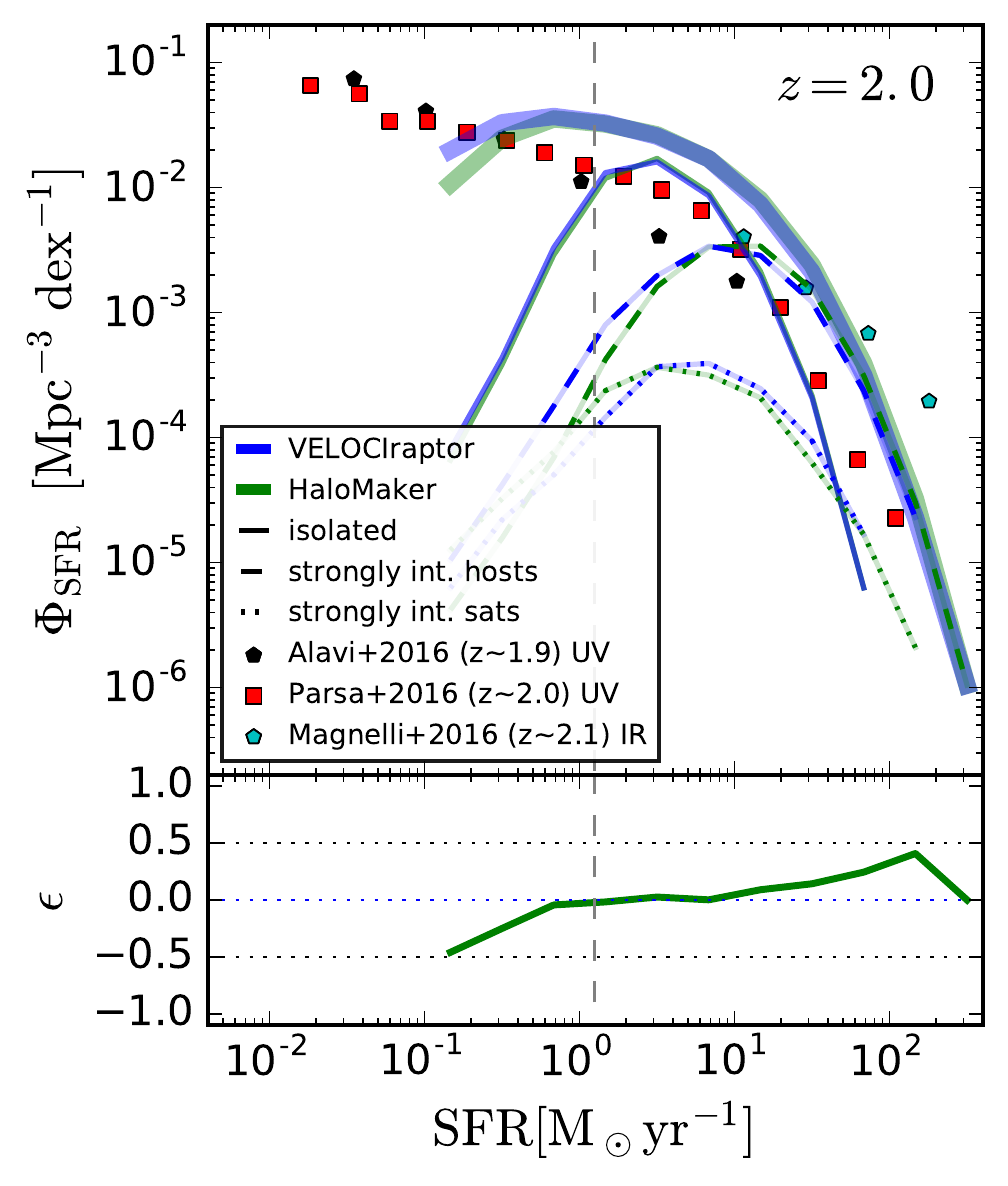}
            \caption{Star Formation Rate Function
                     (SFRF) of Horizon-AGN using {\sc VELOCIraptor}
                     (blue line) and {\sc HaloMaker} (green line)
                     at $z = 2$.
                     The SFRF from all galaxies is shown as a solid
                     thick line; the contribution from isolated
                     galaxies is shown as a solid thin
                     line; and from strongly interacting hosts and
                     satellites as a dashed and dotted lines,
                     respectively.
                     Bottom panels show the 
                     $\epsilon = \Phi_\mathrm{SFR,{\sc HaloMaker}} 
                     / \Phi_\mathrm{SFR,{\sc VELOCIraptor}}$.
                     The vertical dashed grey line shows a
                     SFR equivalent to $\sim$10 star particles 
                     formed in the last 50 Myr for reference.
                     From the compilation of observations presented
                     by \citet{Katsianis2017}, we show estimations
                     of the SFRF derived from IR LF from \citet{Magnelli2011}, 
                     and UV LF from \citet{Alavi2016} and
                     \citet{Parsa2016},
                     are shown as symbols, as labelled.
                     }
            \label{fig:SFRF}
          \end{figure}
          
          \begin{figure*}
            \centering
            \includegraphics[width=\textwidth]{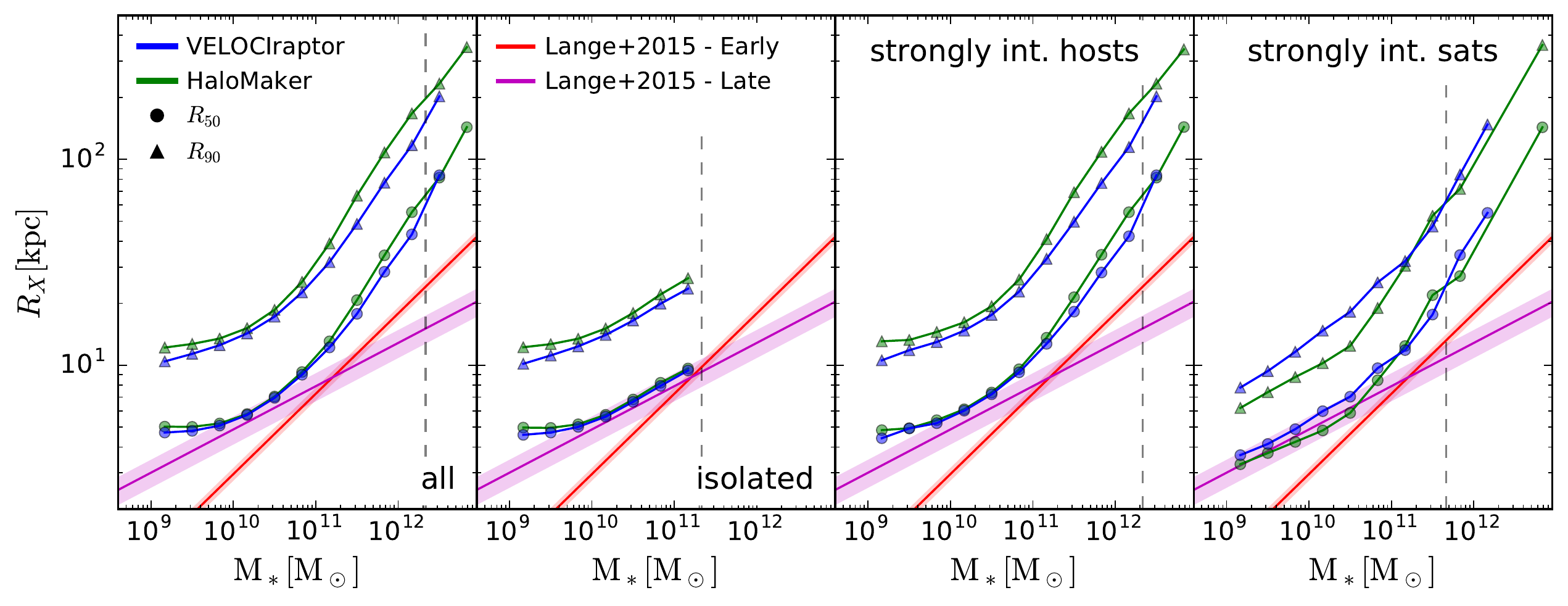}
            \caption{Galaxy size-mass relation as measured by 
                    {\sc VELOCIraptor} (green) and {\sc HaloMaker} (blue). 
                    The relation for $R_{50}$ and $R_{90}$
                    is shown as circles and triangles, respectively.
                    The mass-size relations are shown for all the
                    matched galaxies (first panel), isolated galaxies
                    (second panel), and for strongly interacting host 
                    and satellite
                    galaxies (third and fourth panel, respectively).
                    Dashed grey vertical lines delimit the left edge of
                    mass bins that have less than 10 galaxies.
                    $J$-band linear fit from \citet{Lange2015} 
                    (corrected for the observational 2D projection)
                    for early (late) type galaxies shown in red (magenta).
                    }
            \label{fig:sizemass}
          \end{figure*}          
          
          In Fig.~\ref{fig:SFRF} we show the estimated SFRF,
          $\Phi_\mathrm{SFR}$, of Horizon-AGN using 
          {\sc VELOCIraptor} (blue line) and {\sc HaloMaker} 
          (green line) at $z = 2$.
          SFRFs from isolated galaxies, as well as strongly
          interacting hosts and satellites are shown, 
          as labelled.
          Bottom panel shows the relative difference 
          $\epsilon = (\Phi_\mathrm{HaloMaker}
          /\Phi_\mathrm{VELOCI}) - 1$, for the total
          SFRF.
          The dashed vertical line shows 
          a SFR equivalent to $\sim$10 new star 
          particles formed in the last 50 Myr
          for reference.
          %
          
          The overall shape of the total SFRF is in good
          agreement between the finders, as well as with
          the estimated from observations.
          At $0.1\myri$, {\sc HaloMaker} predicts 50\%
          less galaxies than {\sc VELOCIraptor}.
          These are, however, SFR values close to 
          $\gtrsim1$ star particle formed in the 
          last 50 Myr.
          At higher SFRs, the values SFRFs start to
          become more similar, reaching a negligible
          difference at $0.6 \leq \mathrm{SFR}/ \myri
          \leq 6$.
          At these SFRs the total SFRF is principally
          dominated by isolated galaxies whose SFRF agree
          between the catalogues; 
          however, similarly to the GSMF, there is also
          `compensation' from different galaxy samples,
          as {\sc VELOCIraptor}
          predicts more strongly interacting hosts
          than {\sc HaloMaker}, but less strongly interacting
          satellites at SFR $\lesssim 3\myri$, and 
          vice versa at $3 \lesssim$ SFR $/\myri \lesssim 30$.
          At SFRs $> 30\,\myri$, {\sc HaloMaker}
          predicts more galaxies than 
          {\sc VELOCIraptor}, reaching a maximum difference
          of 50\% at $\sim 100 \myri$.
          This excess on number density predicted by
          {\sc HaloMaker} compared to {\sc VELOCIraptor}
          is caused by the addition of mass from 
          other galaxies to the central (see 
          Fig.~\ref{fig:clusters} and the IHSC that is
          separated in {\sc VELOCIraptor}).
          As we showed in Section~\ref{sec:totmass},
          systematics affect most of the systems composed
          by multiple galaxies, and not only contact mergers,
          hence the differences observed in the estimated
          SFRFs for different galaxy samples.
          %


        \subsubsection{Galaxy Mass-Size Relation}
        \label{sec:msize}
          We show in Fig.~\ref{fig:sizemass} the estimated
          galaxy mass-size relation using $R_{50}$ (circles) 
          and $R_{90}$ (triangles) for the galaxies found by
          {\sc VELOCIraptor} (blue) and {\sc HaloMaker} (green)
          in the sample of matched galaxies (as described in
          Section~\ref{sec:galaxytogalaxy}), at $z = 0$.
          Symbols show the median calculated in equal size bins
          in logarithmic scale, using the same bins for both
          catalogues.
          The four panels show the mass-size relation for all
          galaxies (first panel), isolated galaxies (second panel),
          and strongly interacting host and satellite galaxies
          (third and fourth panel, respectively).
          Vertical dashed lines are shown at the high-mass end
          where bins have less than 10 galaxies.
          For reference, the $J$-band mass-size relation linear fit
          from \citet{Lange2015} is shown as a solid
          red (magenta) line for early (late) type galaxies.
          We apply an average correction to the observations
          due to the fact that we are measuring 3D sizes in the
          simulation, while observations measure projected sizes.
          The latter is a simple scaling of $1.35$ applied to
          the observations (which comes from the fact that galaxies
          have minor to major axis ratios of $\approx 2$ and are
          inclined by $60$~degrees, on average). 
          %
    
          The overall shape for $R_{50}$ and $R_{90}$
          mass-size relation of the whole sample of matched
          galaxies is roughly similar for both finders.
          At $M_* \gtrsim 10^{9}\ \msuni$, $R_{50}$ ($R_{90}$)
          of all the galaxies in the sample are on average 10\% (20\%)
          larger in {\sc HaloMaker} than in {\sc VELOCIraptor}.
          At $M_*\sim 10^{10}\ \msuni$ this difference reaches a
          minimum for both radii, being almost negligible
          for $R_{50}$, and $\gtrsim$ 5\% for $R_{90}$.
          At $M_* > 10^{10.5}\ \msuni$, the difference in the sizes
          of galaxies between the catalogs starts to increase.
          The difference on the estimated radii of galaxies peaks
          at $M_* \sim 10^{12}\ \msuni$ where on average {\sc HaloMaker}
          galaxies have $R_{50}$ ($R_{90}$) values up to $\gtrsim$ 20\% 
          ($\sim50$\%) larger than {\sc VELOCIraptor}.
          Although there is agreement between both finders at 
          $M_* > 2\ \times\ 10^{12}\ \msuni$, the number of
          galaxies is very low.
          As discussed in Sections~\ref{sec:totmass} and
          ~\ref{sec:gsmf}, high-mass galaxies are more
          massive in {\sc HaloMaker} than in {\sc VELOCIraptor},
          extending the mass-size relation
          to larger values.
          It is interesting though, that despite the individual
          differences seen in Figs.~\ref{fig:fmdist} and 
          ~\ref{fig:frdist}, a simple extrapolation of the
          {\sc VELOCIraptor}'s relation would agree with the
          one described by {\sc HaloMaker}.
          %
    
          The mass-size relation for isolated and loosely
          interacting galaxies, as well as for strongly 
          interacting host galaxies (second and third panel of 
          Fig.~\ref{fig:sizemass}, respectively)
          has a similar behaviour as the complete sample at all
          stellar masses.
          Smaller $R_{50}$ and $R_{90}$ in {\sc VELOCIraptor} 
          are expected for isolated galaxies due to its 6DOF
          implementation, which, as discussed in
          Section~\ref{sec:totmass}, reduces the
          stellar particle budget for galaxies and is kept as
          the IHSC.
          Although the latter also affects sizes of interacting
          host galaxies, their sizes are again artificially
          increased because of how particles in a common 3DFOF
          object are distributed as was shown in
          Fig.~\ref{fig:clusters}.
          Both effects are evident at $M_* > 10^{11}\ \msuni$
          for $R_{50}$, and to a greater extent for $R_{90}$.
          Although at the very high-mass end there seems to be
          agreement, as discussed above, the number of
          galaxies for both finders is very low preventing us
          from reaching any conclusion.
          %

          The difference in the mass-size relation for
          strongly interacting satellites, however, has a
          different behaviour compared to other samples.
          At $M_* > 10^{11}\ \msuni$, interacting satellite
          galaxies are on average more compact in the
          {\sc HaloMaker} catalog;
          both $R_{50}$ and $R_{90}$
          have on average lower values than their
          {\sc VELOCIraptor} counterparts.
          Similarly to other samples, differences are larger
          for $R_{90}$ than for $R_{50}$ in the same stellar
          mass bin.
          At $M_* \gtrsim 10^9\ \msuni$, $R_{50}$ ($R_{90}$)
          is on average $\sim$ 10\% ($\sim$ 20\%) smaller
          in {\sc HaloMaker} than in {\sc VELOCIraptor}.
          At higher stellar masses the difference increases
          reaching a maximum at $\sim 10^{10.5}\ \msuni$ with
          galaxies being on average $\sim$ 20\% and $\sim$ 30\%
          smaller for $R_{50}$ and $R_{90}$, respectively,
          in {\sc HaloMaker} compared to {\sc VELOCIraptor}.
          At $10^{11} \leq\ M_* / \msuni \leq 5\ \times 10^{11}\ \msuni$,
          there are two mass bins where $R_{50}$ and $R_{90}$
          of interacting satellites are on average similar and
          even larger in {\sc HaloMaker}, contrary to what would
          be expected.
          This is likely to be caused by host-satellite
          swapping, and can be seen in Fig.~\ref{fig:frdist}
          as a small bump at $f_{R_X} > 2.$
          At $M_* > 10^{12}\ \msuni$ $R_{50}$ ($R_{90}$) is 
          on average $\sim$15\% ($\sim$ 20\%) smaller for 
          for {\sc HaloMaker} satellites.
          %


  \section{Discussion}
  \label{sec:discussion}
    We have presented an improved algorithm for 
    identifying galaxies in simulations and showed
    how galaxy properties are affected by the
    finding algorithm.
    In this section we discuss implications of
    our algorithm, as well as possible consequences
    that non-robust identification of galaxies can
    have in cosmological hydrodynamical simulations.

    \subsection{Identifying galaxies vs. dark matter halos}
      Many structure finding codes are capable of
      finding galaxies in simulations 
      \citep[see for reference][]{Knebe2013a}.
      The vast majority of them are generally
      limited to either taking all bound baryons
      inside a dark matter (sub)halo and label
      them as the galaxy, or use the same algorithm
      and parameters adopted for dark matter haloes
      to identify galaxies.
      Although both approaches are valid for the
      identification of galaxies, there are important
      differences between dark matter halos and
      galaxies to keep in mind:
      (i) Stars, that make up galaxies, are formed
      from gas elements at the bottom of potential
      wells, hence galaxies are expected to be more
      compact than dark matter haloes; such gas elements
      can also cool into discs, very different than
      the geometry of dark matter halos.
      Consequently, for a large fraction of the
      galaxies, the stellar density profiles do not
      resemble those of their dark matter counterpart.
      (ii) Taking into account this variety of
      shapes and distributions is extremely important
      for the identification of merging and
      interacting galaxies, as such morphologies
      can distort and become quite complex, making
      their identification a non-trivial task.
      (iii) During mergers, the outer dark matter
      component will at some point phase-mix, but
      the stars in its centre do that on a different
      timescale, with some features being
      long living (such as streams and shells).
      This makes it important
      to analyse them separately.
      Even if stars and dark matter are both
      collisionless in simulations and interact
      solely through gravity, we should not use the
      same approach if codes were designed under
      assumptions that are valid only for dark matter
      haloes.
      Although for some galaxies the above
      approaches might work, that is not expected
      to be the case for the entire galaxy
      population in cosmological simulations, as
      we have shown in this study.

      The algorithm presented here is \emph{a}
      solution to tackle this problem.
      It is particularly powerful as it was
      designed to work without any \emph{a priori}
      assumption on shape or
      distribution, which is
      capable of handling the large dynamical range
      covered in cosmological simulations.
      It is therefore a generalised solution
      that can be also easily applied to other
      components in simulations, which we
      explore in upcoming studies (Elahi et al. in prep).

    \subsection{Impact of identification}
      \subsubsection*{Simulation results}
        We have shown in this study how the 
        total mass, size and star formation
        rate of galaxies can be affected by 
        the assumptions and sometimes 
        oversimplification of the finder.
        Additionally, as seen in the case 
        studies (Section~\ref{sec:casestudy}),
        misestimation of masses of merging
        galaxies impact the estimated merger
        ratio.
        This has several consequences as
        galaxy mergers are essential for
        the growth of massive galaxies
        \citep{Robotham2014}.
        Inability to resolve galaxies in
        interactions can affect estimated 
        merger ratios and therefore estimated
        minor and major merger rates, and 
        the impact they have on the build up
        of galaxies.
        A related area of great interest is
        whether interactions enhance/suppress 
        the star formation activity in galaxies
        \citep[][Davies et al. submitted]
        {Ellison2008,Davies2015,Kaviraj2015,Martin2017}.
        As shown in this work the SFRs of galaxies
        that are strongly interacting have their
        SFRs affected within a fraction of 2 to 3
        by mostly due to how particles in the
        outskirts of galaxies are assigned to 
        density peaks.
        The latter is comparable to the enhancements
        inferred observationally, showing that it is
        critical to robustly measure SFRs in 
        simulated galaxies if we want to use them
        to offer physical interpretations at these
        environmental trends.
        %
        Misestimation of masses and sizes can impact the
        interpretation that we can give to galaxies in 
        dense environments such as groups and clusters,
        where both central and satellites can be largely
        affected.
        This implies that we could get misleading results
        when studying environmental effects on galaxy 
        quenching in hydrodynamical simulations.
        %
        
        These are not the only possible consequences. Our 
        case studies also showed that radial profiles can be affected,
        such as angular momentum or inertia tensors, both used
        for alignment studies.
        Regarding angular momentum, it has been recently shown
        by e.g. \citet[]{Cortese2016} in observations and
        \citet{Lagos17} in simulations, that the estimated specific
        angular momentum can be up to $\sim$2.5 times
        ($\sim$0.4 dex) higher if measured at two effective
        radius rather than one.
        It is therefore important for related studies in simulations
        to account for systematic effects that can severely
        affect the estimated sizes (e.g $R_{50}$) of interacting galaxies.
        Taking into account the offset we found when comparing
        {\sc HaloMaker} with our new algorithm in Section~\ref{sec:sizes},
        we would expect 3D finders to bias the specific angular momentum
        of satellites (centrals) towards low (high) values.

      \subsubsection*{The effects on our understanding of Galaxy Formation}
        We showed in this study that despite the large differences
        seen in individual galaxies, especially the interacting ones, the
        overall galaxy population statistics are not severely
        affected by finder systematics.
        This has important implications for the way the 
        galaxy formation is modelled and understood.
        We have already stated that population statistics
        are used to tune free parameters of subgrid physics
        models in simulations.
        To a certain extent, through tuning we can learn how
        recipes affect galaxies as well as the impact of 
        different models, e.g. star formation, stellar and 
        AGN feedback.
        However, we have shown that we can obtain the right
        amplitude of a relation or function (i.e. stellar mass
        function, or mass-size relation) using vastly different
        finders but for different reasons. We argue that the
        study of subsamples of the galaxy population
        (e.g. satellite galaxies, galaxy groups/clusters)
        can unveil such differences, and therefore provide
        key information to estimate the systematic effects
        introduced by the choice of finder.        
        Subgrid physics often model unresolved
        and generally not-so-well understood physical processes
        that can affect the large-scale properties of galaxies.
        This is the case of BH growth and its
        corresponding AGN feedback.
        A major growth channel of BHs are mergers, and we
        have shown that the choice of finder affects the
        derived merger ratio.
        This in turn affects our estimates of BH merging
        timescales, possibly causing the existence of multiple
        BHs in merger remnants, and thus changing the
        associated effect of AGN feedback on the galaxy properties.
        %

        Overall, there are many unknowns in simulations, and the exact way in which
        one decides to compare with observations or even among simulations is
        a non-trivial task.
        In this paper, we focus on the effect the galaxy identification has
        on the derived galaxy properties, and in many cases those differences
        will be smaller than other uncertainties, such as the exact way one
        measures a property \citep{Stevens2014}, or the systematic effects the
        physical modelling itself has on the predicted population.
        However, in some cases (such as galaxy mergers and satellite galaxies
        in dense environments), the bias introduced by the chosen algorithm 
        could be a dominant effect.
        %

        This all shows that perfecting our ability to identify
        galaxies and measure their properties in simulations
        is a key task that cannot be overlooked. Our new
        algorithm offers a new, robust and accurate way of doing
        this, yielding smoother stellar profiles
        (see Figs.~\ref{fig:merger}~and~\ref{fig:clusters}) and
        more robust stellar mass estimates than widely used 3D
        finders.
        This implementation of our algorithm in {\sc VELOCIraptor}
        will be made public in the next release of the code
        (Elahi et al. in prep).
        %
        

    \section{Summary and Conclusions}
    \label{sec:conclusions}
      We have extended the halo-finder code {\sc VELOCIraptor}
      \citep[][Elahi et al. in prep]{Elahi2011} to robustly 
      identify galaxies in state-of-the-art simulations of galaxy formation.
      This new implementation overcomes many common problems that even
      state-of-the-art structure finding codes struggle with, such as 
      particle assignment and accurate identification of strongly
      interacting systems.
      We have paid special attention to the appropriate selection and
      iterative adjustment of search parameters, to account for the wide
      dynamical range that simulations can have.
      Particle assignment (core growth) was improved by using the full
      phase-space dispersion tensor, allowing us not only to recover
      arbitrary galaxy shapes, but also to obtain smooth density profiles
      even for galaxies with satellites embedded within it.
      %

      With our improved code, we built an additional galaxy catalogue for
      the state-of-the-art cosmological hydrodynamical simulation 
      Horizon-AGN, and compared its outcomes with those of the complex
      configuration-space based finder, {\sc HaloMaker}.
      Case studies confirmed the versatility and robustness of our 
      algorithm, and provided insight into how identification tools
      can affect galaxy properties (e.g. mass and sizes), as well as
      the estimates of merger ratios.
      Below we summarize our main results. 
      %
      
      \begin{description}
        \item[\textbf{Galaxy-to-galaxy comparison.}] We matched the
              galaxy catalogues to quantify how the total
              $M_*$, SFR and sizes
              ($R_{50}$ and $R_{90}$) can be affected by the chosen finder.
              We built distributions 
              of $f_Y = Y_\mathrm{\sc HaloMaker} / Y_\mathrm{\sc VELOCIraptor}$,
              where $Y$ corresponds to each of the properties above, 
              and separate the contribution from isolated and interacting 
              galaxies.
              Interacting galaxies are those hosted by halos with more than
              1 substructure, otherwise galaxies are considered as isolated.
              \begin{itemize}
                \item Isolated galaxies are in general
                      narrowly distributed close to $f_Y = 1$ for $M_*$, SFR, 
                      and $R_{50}$.
                      Such similarities between catalogues are expected as the
                      identification of isolated galaxies should be
                      straightforward.
                      For $R_{90}$, however, the peak is not narrow and a 
                      considerable amount of isolated galaxies have $f_{R_{90}}$
                      values around $\pm$~0.3~dex from $f_{R_{90}} = 1$. This
                      suggests that $R_{90}$ is highly dependent on the finding
                      algorithm.
                \item Interacting galaxies show a very wide $f_Y$ distribution 
                      for all quantities studied.
                      There is an evident difference between host and satellite
                      galaxies, which peak at $f_Y > 1$ and $f_Y < 1$, respectively.
                      These differences are mainly caused by inadequate particle 
                      assignment in {\sc HaloMaker}, which we show our improved
                      version of {\sc VELOCIraptor} handles better.
                      {\sc HaloMaker} artificially increases 
                      (decreases) the estimated values of $M_*$, SFR and $R_{50,90}$
                      for host (satellite) galaxies in interacting systems.
                \item Differences between the catalogues are amplified at higher 
                      redshifts, where the $f_Y$ distributions of interacting and
                      isolated galaxies widen.
              \end {itemize}
              
        \item[\textbf{Galaxy population statistics.}] We investigate how
              the choice of finder affects the overall galaxy population
              statistics.
              We explore the GSMF, SFRF, as well as the 
              Mass-Size relation for $R_{50}$ and $R_{90}$. 
              \begin{itemize}
                \item At $M_* < 10^{11}\ \msuni$, the $z=0$ GSMFs of
                      {\sc HaloMaker} and {\sc VELOCIraptor} agree well,
                      while at higher stellar masses the former predicts
                      from 20\% to 100\% more galaxies than the latter.
                      At higher redshifts differences are amplified. At
                      $z = 4$, {\sc HaloMaker}'s GSMF predicts a number
                      density of galaxies at least 30\% higher than
                      {\sc VELOCIraptor} over the whole mass range,
                      increasing to 100\% at $M_* \gtrsim 10^{10}\ \msuni$.
                \item The SFRF at $z=0$ is also fairly similar between the
                      finders, with differences increasing with redshift.
                      At $z=2$, the peak of the cosmic star formation history,
                      {\sc HaloMaker} predicts up to 50\% more galaxies of
                      $3 \leq \mathrm{SFR} / \myri \leq 100$ than
                      {\sc VELOCIraptor}. This is important as these galaxies
                      are expected to dominate the cosmic SFR.
                \item We compare the $R_{50}$ and $R_{90}$ size-mass relation
                      predicted by both finders. 
                      We find that the $R_{50}$ mass-size 
                      relation resulting from the two finders are similar,
                      except at the high mass end,
                      $M_* \simeq 10^{12}\ \msuni$, where {\sc HaloMaker's}
                      galaxies are 20\% larger than {\sc VELOCIraptor's}. 
                      These differences increase by 30\% when we study $R_{90}$.
                      This results from the fact that the stellar content and
                      structure in the outskirts of galaxies is very sensitive
                      to the choice of finder.
              \end {itemize}      
      \end{description}

      Although we see that the overall $z=0$ galaxy statistics are not
      greatly impacted by the choice of finder, individual galaxies can
      display differences in mass and size of more than a factor of $3$
      between the two finders studied here. 
      We suggest that the tuning of simulations of galaxy formation is
      relatively robust as it has consistently focused on population
      statistics.
      However, comparisons of galaxy sub-populations with observations,
      specifically in the context of pairs, groups and clusters, can be
      greatly affected by the choice of finder. We showed that our new
      algorithm outperforms 3D finders and provided extensive evidence
      of this. 
      %
      
      One of our key findings is that the stellar outskirts of galaxies
      is greatly affected by the choice of finder.
      In upcoming studies we will explore in detail the diffuse Intra
      Halo Stellar Component, stellar streams and the outer stellar
      profiles of galaxies.
      Another important area of investigation will be comparing our 
      theoretical measurements of the diffuse stellar halo with
      observations, by mimicking the observational effects, such as
      selection, surface brightness biases, among others.
      %

     
    \section*{Acknowledgements}
    The authors thank Aaron Robotham, Matthieu Schaller and the theory and computing group at 
    ICRAR for helpful discussions. 
    RC is supported by the MERAC foundation postdoctoral
    grant awarded to CL and by the Consejo Nacional de Ciencia y Tecnolog\'ia CONACYT 
    CVU 520137 Scholar 290609 Overseas Scholarship 438594.
    CL is funded by a Discovery Early Career Researcher Award (DE150100618). 
    Parts of this research were conducted by the Australian Research
    Council  Centre  of  Excellence  for  All  Sky  Astrophysics  in  
    3  Dimensions (ASTRO 3D), through project number CE170100013.
    This research is part of Spin(e) (ANR-13- BS05-0005, http://cosmicorigin.org).
    This work has made use of the Horizon Cluster hosted by Institut 
    d'Astrophysique de Paris. We thank Stephane Rouberol for running 
    smoothly this cluster for us.
    %




\bibliographystyle{mnras}
\bibliography{references}



\appendix
  \section{6DFOF linking length}
  \label{appndx:ll}
\begin{figure*}
              \centering
              \includegraphics[width=\textwidth]{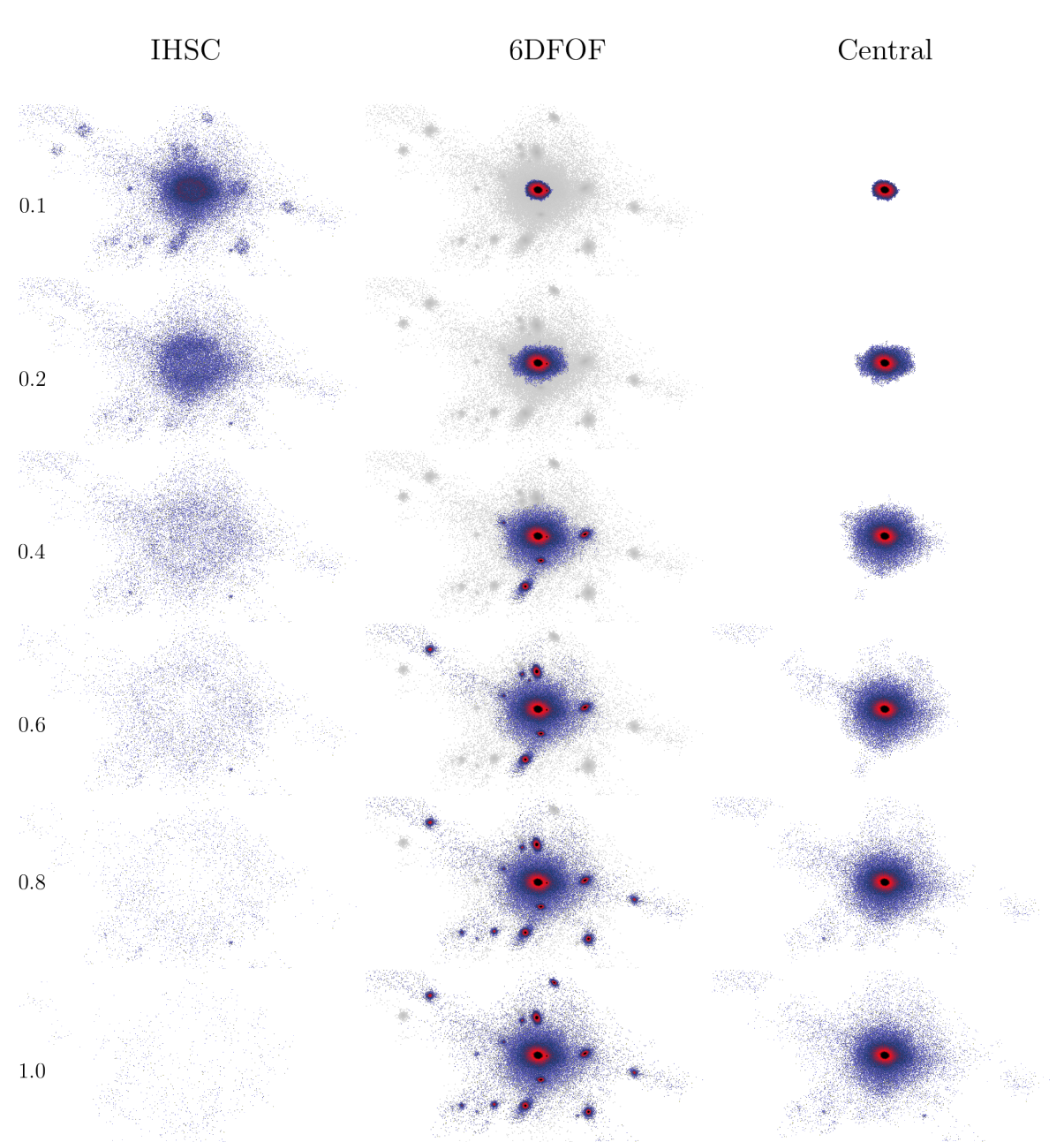}
              \caption{Projected stellar density of the IHSC, 6DFOF object
                      (step 2), and the final central galaxy, for 
                      different choices of $f_{x,\mathrm{(6D)}}$ (row labels).
                      The 3DFOF object (step 1) is shown in grey in the 
                      middle panels.
                      For this work our standard choice is
                      $f_{x,\mathrm{(6D)}} = 0.2$
                      }
              \label{fig:ll}
            \end{figure*}

      As described in Section~\ref{sec:step2}, we use
      a fraction $f_{x,\mathrm{(6D)}}$ of $b$ for the
      configuration space linking length, described in
      equation~(\ref{eq:lx6d}).
      We show in Fig.~\ref{fig:ll}, how the particular
      choice of $f_{x,\mathrm{(6D)}}$ can affect how galaxies
      are delimited in {\sc VELOCIraptor}, and how 
      its IHSC changes with it.

      Values of $0.6 \lesssim f_{x,\mathrm{(6D)}} \lesssim 1$
      link a large number of substructures to the
      6DFOF object.
      Though smaller values get rid of some of the
      substructures in the outskirts of the central/group,
      they can also shrink the size of the central galaxy
      to only the inner parts as is seen for
      $f_{x,\mathrm{(6D)}} = 0.1$.
      The latter is appreciated in the IHSC, where bubble-like
      features can be seen at the position of galaxies, as if
      the central parts of the galaxies were carved leaving
      the outer parts unassigned to any galaxy.
      Mid range values of 
      $0.2 \lesssim f_{x,\mathrm{(6D)}} \lesssim 0.4$
      leave out some of the outskirts of the group
      without leaving bubble-like structures in the IHSC.
      For this study our preferred choice is
      $f_{x,\mathrm{(6D)}} = 0.2$ as it leaves out most
      of the satellites, while preserving structures
      that can only be separated using the iterative 
      search in the 6DFOF object\footnote{It is important
      to notice that even 
      $f_{x,\mathrm{(6D)}} = 0.1$ a small satellite still 
      remains as part of the 6DFOF object (this can be seen
      by zooming the figure in the electronic version), and
      that independently of the $f_{x,\mathrm{(6D)}}$ value
      used, the inner profile of the galaxy is smooth 
      showing the power of the iterative search and 
      core growth.}.
      %

      Though, it can be seen that the estimated IHSC
      depends on the $f_{x,\mathrm{(6D)}}$ used, this example
      is illustrative of upcoming studies focused on the IHSC
      (Ca\~nas et al, in prep).

  \section{Core growth weighting}
    \label{appndx:weight}
    
    %
    %
    \begin{figure*}
      \centering
      \includegraphics[width=\textwidth]{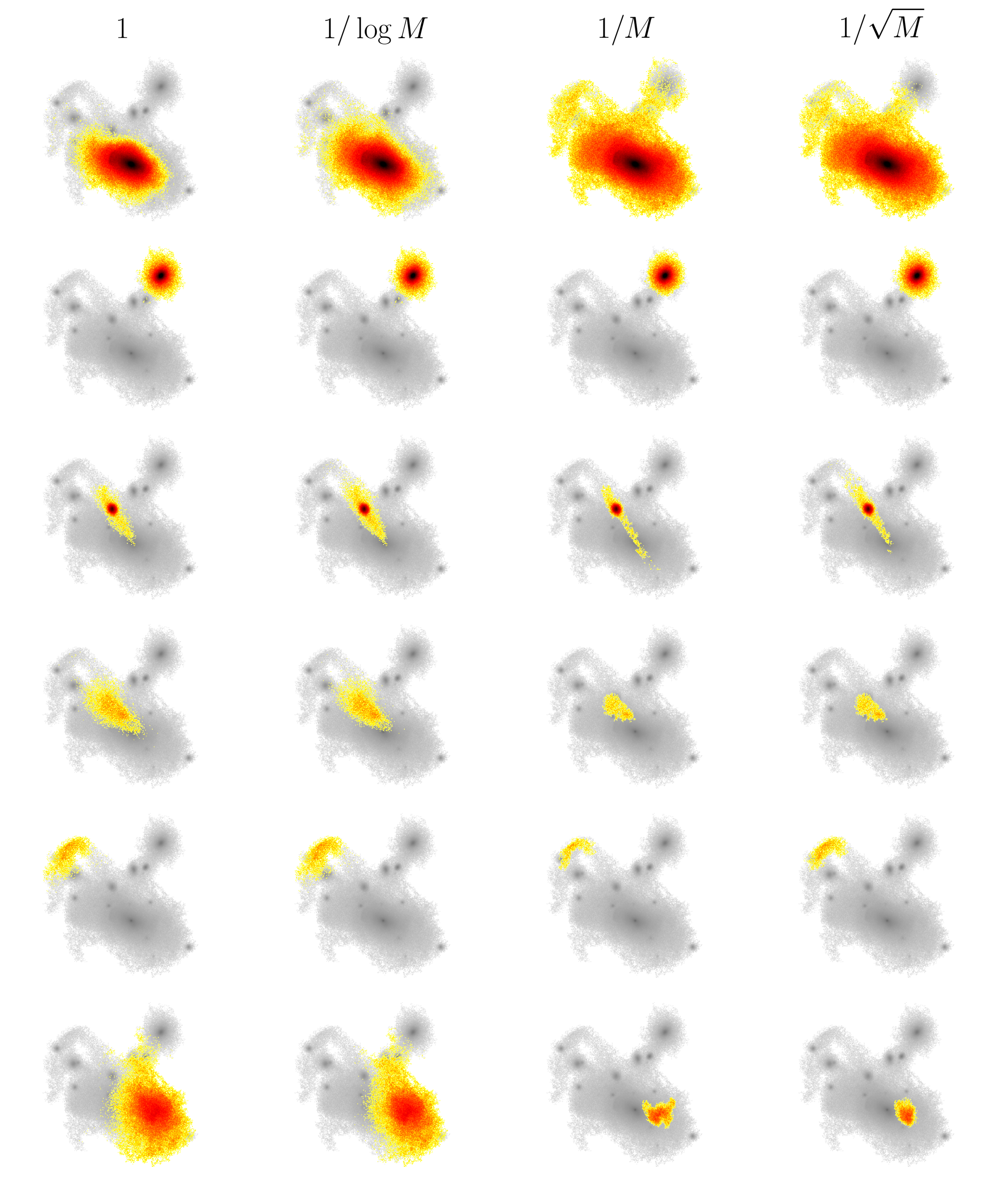}
      \caption{Projected stellar density of structures found
               by {\sc VELOCIraptor} inside a galaxy cluster
               in Horizon-AGN using different weights for the
               core growth.
              }
      \label{fig:weighting}
    \end{figure*}

    %
    %
    \begin{figure*}
      \centering
      \includegraphics[width=\textwidth]
                      {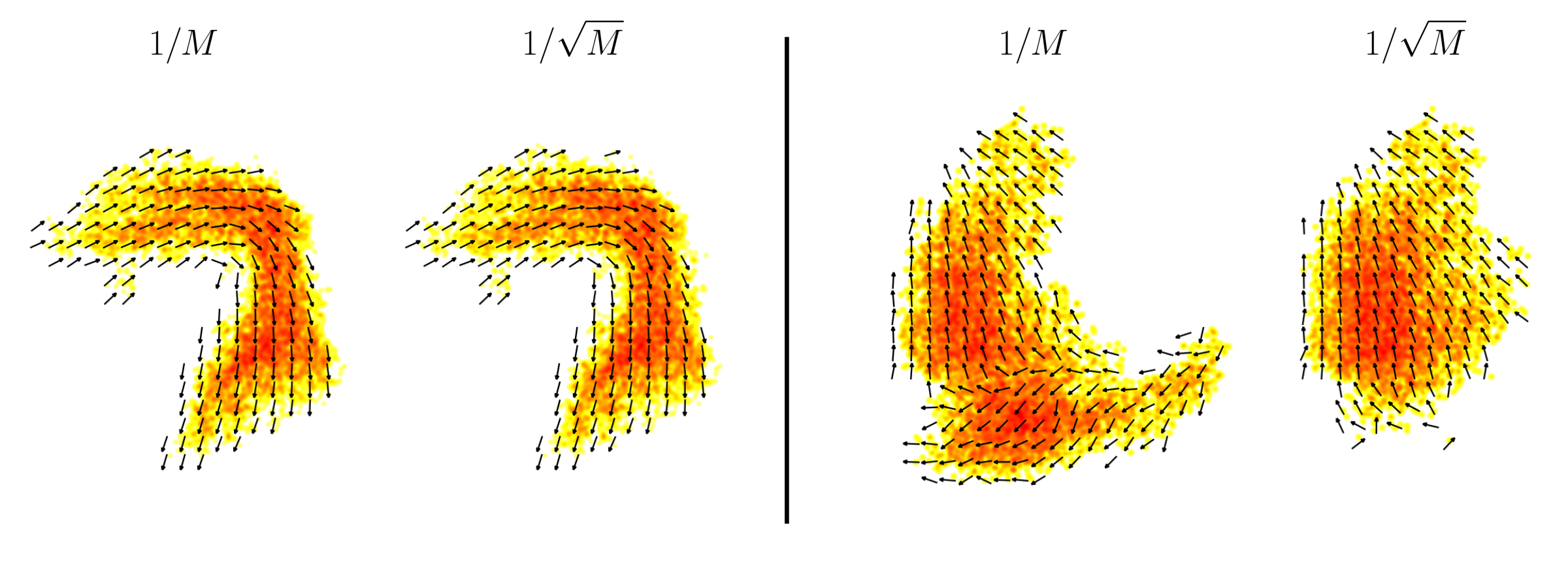}
      \caption{Projected stellar density (color) and velocity
               field (arrows), for two streams identified by
               {\sc VELOCIraptor} and grown with weights 
               proportional to $1/M$ and $1/\sqrt{M}$, 
               as labelled.
               For the left one it can be seen that the 
               velocity field changes smoothly along the
               structure for both weights.
               For the right one it can be seen that 
               $1/M$ has assigned particles of a distinct
               structure to the stream, while for $1/\sqrt{M}$
               the velocity field is much smoother.
              }
      \label{fig:stream}
    \end{figure*}

    In Section~\ref{sec:step4} we stated that when
    calculating phase-space distances from 
    cores, equation~(\ref{eq:coredistance}), 
    a weight $w_k$, was needed to compensate for the
    compactness of galaxies and extension of
    streams and shells.
    For this we use a mass-dependent weight, as this
    would reduced the phase-space distance from particles
    to phase-space compact high-mass objects (i.e. galaxy
    candidates), compared to low-density extended objects
    (streams and shells).
    We show in Fig.~\ref{fig:weighting} surface density
    projections of how the growth of phase-space cores
    changes for $w = \{1, 1/\log M, 1/M, 1/\sqrt{M}\}$,
    where $M$ is the total mass of the core at a
    given iteration level.
    %

    We show how the central galaxy is affected by 
    $w$ in the first row.
    For $w=1$ and $w = 1/\log M$ it can be seen 
    that some of the outer parts of the galaxy
    are missing.
    The reason behind this is the fact that 
    some streams can have very large dispersion
    so that `weak' weights do not compensate, 
    as is seen in the last row.
    On the other hand `strong' weights can in
    fact make that the central galaxy absorbs the
    vast majority of the particles, as they would
    have to be close to any structure by the same
    orders of magnitude difference of their masses.
    A weight of $1/\sqrt{M}$ does a 
    slightly better job than $1/{M}$, as for the
    largest satellite (second row) it is able to
    return correctly its outskirts.
    For satellites with leading or trailing tails,
    and some streams (rows 3 to 5), it can be 
    seen that weak weights recover a little better
    outer material than strong ones, as for the 
    latter streams seem to look a little fragmented.
    The last row shows an example of a tidal structure
    which grows drastically when weak weights are
    used, due to their large dispersions and the 
    metric used to assign particles.
    %
        
    {\sc VELOCIraptor} can also 
    identify tidal features with the same algorithm.
    Though the ability of identifying streams is
    important for many studies, our priority is first
    to identify galaxies as cleanly as possible.
    In order to avoid the extreme growth of streams, 
    we incline for stronger weights.
    Both $1/M$ and $1/\sqrt{M}$ give similar
    results, however the latter does a better job 
    at preventing that all particles are assigned
    to the central, and recovers better outskirts
    for satellites.
    We further compare these two weights by taking
    a closer look to the streams from the last rows
    of Fig.~\ref{fig:weighting}.
    In Fig.~\ref{fig:stream} we show zoomed-in
    projected surface densities and velocity
    field for both structures.
    The left panel shows how for an arc stream the 
    velocity field smoothly changes along it.
    The same is recovered for both weights.
    However, for the last stream it can be seen
    that for a weight of $1/M$, the velocity field
    is not continuous along the structure, and are
    in fact distinct structures, while for a weight
    of $1/\sqrt{M}$ the stream (or tidal feature)
    has a smooth the velocity field.
    %

    Weights proportional to other powers of $M$, or
    to any arbitrary quantity can be easily applied.
    However, a weight of $1/\sqrt{M}$ is used as it
    gives the desired results for the purposes of this study.
    %

    Physically, $M^\alpha$ for $\alpha < 1$ is the
    proportional to a galaxy's tidal the radius, $r_\mathrm{t}$.
    In the case of point masses, the tidal radii can
    be approximated as the Roche lobe radii, i.e.
    $r_\mathrm{t} \propto M^{1/3}$.
    For a King's profile the tidal radii is 
    approximately $\propto M^{0.4}$
    \citep[see e.g.][]{BinneyTremaine}.
    For more realistic profiles 
    $r_\mathrm{t} \propto R \ M^{1/3}$, which for
    the spherical collapse model gives
    $r_\mathrm{t} \propto M^{2/3}$.
    Therefore, $w \propto 1/r_\mathrm{t}$, give
    us $w \propto 1 / M^\alpha$, with 
    $1/3 \leq \alpha \leq 2/3$.
    Our choice of $\alpha = 0.5$ is then
    appropriate to properly account for
    the size of galaxies for particle
    assignment.

  \section{Evolution of massive galaxies in galaxy cluster}
  \label{appndx:evol}
\begin{figure*}
      \centering
        \includegraphics[width=\textwidth, clip=true, trim=0cm 0cm 0cm 0cm]{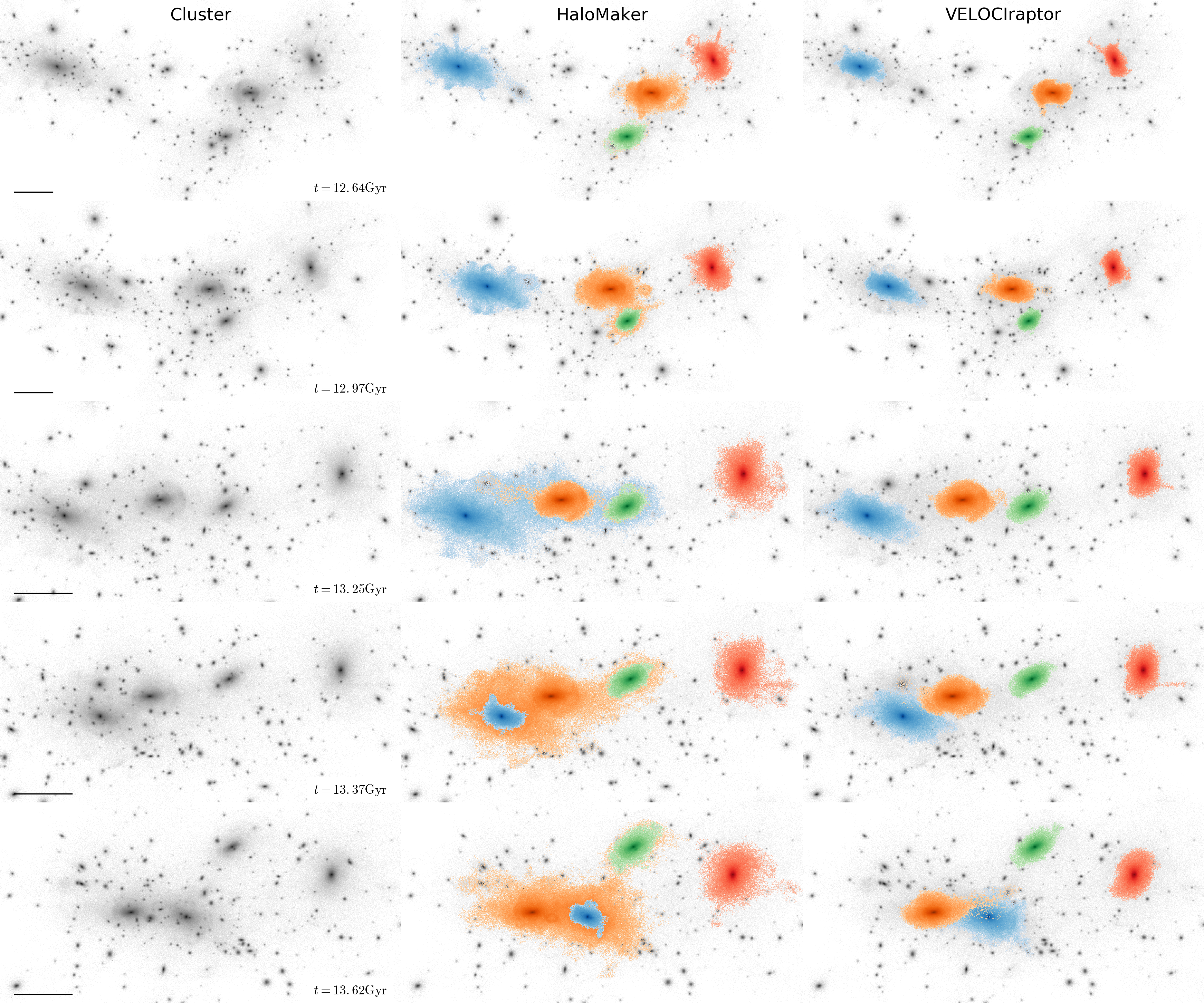}
        \caption{Evolution of the most massive galaxy cluster at $z=0$ in the
                 Horizon-AGN simulation (left), and how the four most massive
                 galaxies (\texttt{Galaxy 1-4}, coloured blue, orange, green
                 and red, respectively) are identified by {\sc HaloMaker} (centre)
                 and {\sc VELOCIraptor} (right).
                 Lower right corner of the first column displays the age of the
                 Universe at that snapshot, and a horizontal bar on the lower
                 left corner corresponds to a length of 400 kpc.
        }
      \label{fig:gal1234evol}
    \end{figure*}

    In Section~\ref{sec:evol} we show how the properties of
    the most massive galaxy in the most massive cluster in
    {\sc Horizon-AGN} evolve over the last Gyr.
    In this Appendix, we show the evolution of
    the subsequent three most massive galaxies in the same cluster.
    In Fig.~\ref{fig:gal1234evol} we show the projected
    mass density of the cluster (left column) at
    different times across the 40 snapshots analysed.
    Galaxies identified by {\sc HaloMaker} (central column)
    and {\sc VELOCIraptor} are coloured as   blue, orange, 
    green and red, for \texttt{Galaxy 1-4}, respectively.
    When galaxies are well separated (first row), both
    finders are able to identify them independently 
    without any contamination of other galaxies' outskirts.
    At subsequent snapshots, as galaxies get closer,
    {\sc HaloMaker} starts to assign the outskirts of other
    galaxies to what is considered the central, as seen
    in the second row for \texttt{Galaxy 2} (orange) and
    \texttt{3} (green).
    This problem gets worse at later times as the two most
    massive galaxies experience a flyby (rows three to five):
    first, \texttt{Galaxy 1} (blue) gets assigned the outskirts
    of galaxies \texttt{2} (orange) and \texttt{3} (green).
    Later, they are assigned to \texttt{Galaxy 2} (orange).
    It can be seen in rows three to
    five that the central galaxy can extend to very far regions.
    Due to the 6DFOF implementation of {\sc VELOCIraptor}
    and its phase-space dispersion tensor based particle
    assignment, the time-independent identification of
    galaxies is much more consistent over time even during the
    flyby of \texttt{Galaxy 1} and \texttt{2}.
    %
 
\begin{figure*}
      \centering
        \includegraphics[width=\textwidth, clip=true, trim=0cm 0cm 0cm 0cm]{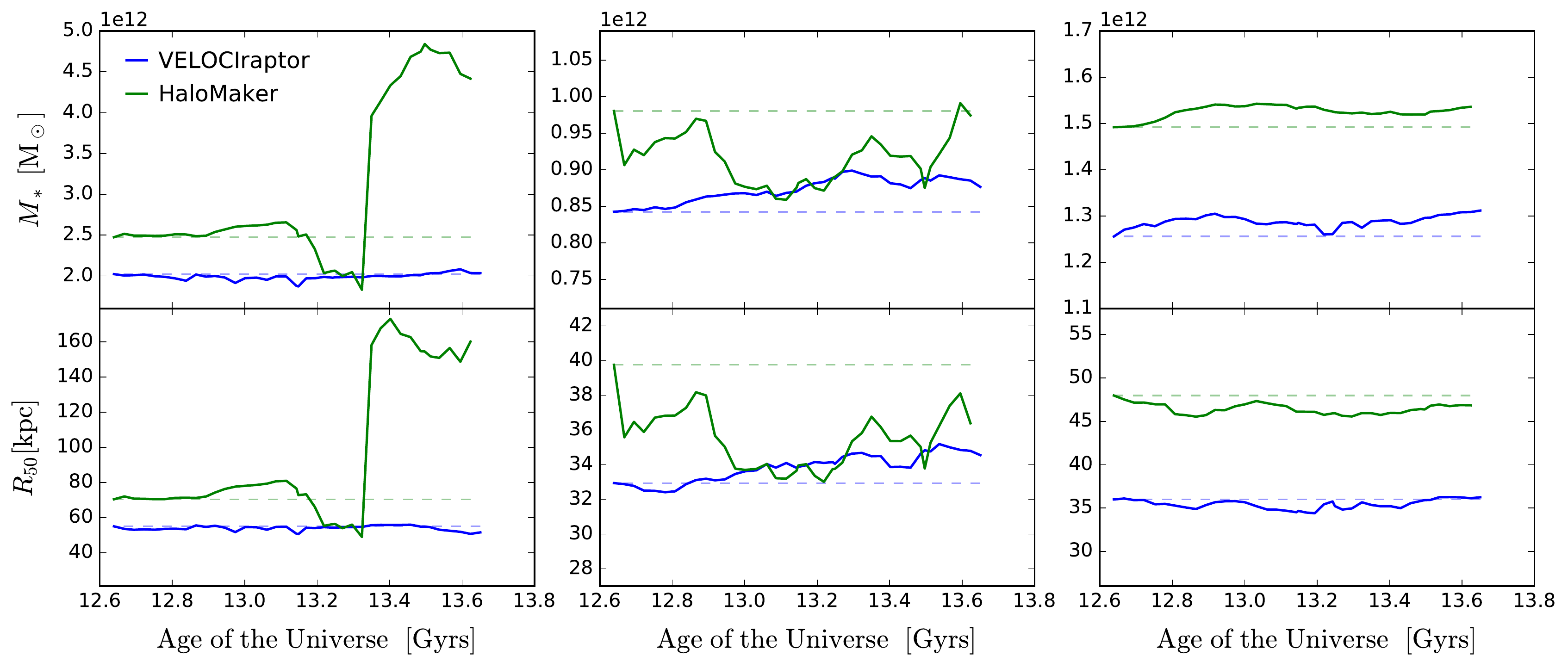}
        \includegraphics[width=\textwidth, clip=true, trim=0cm 0cm 0cm 0cm]{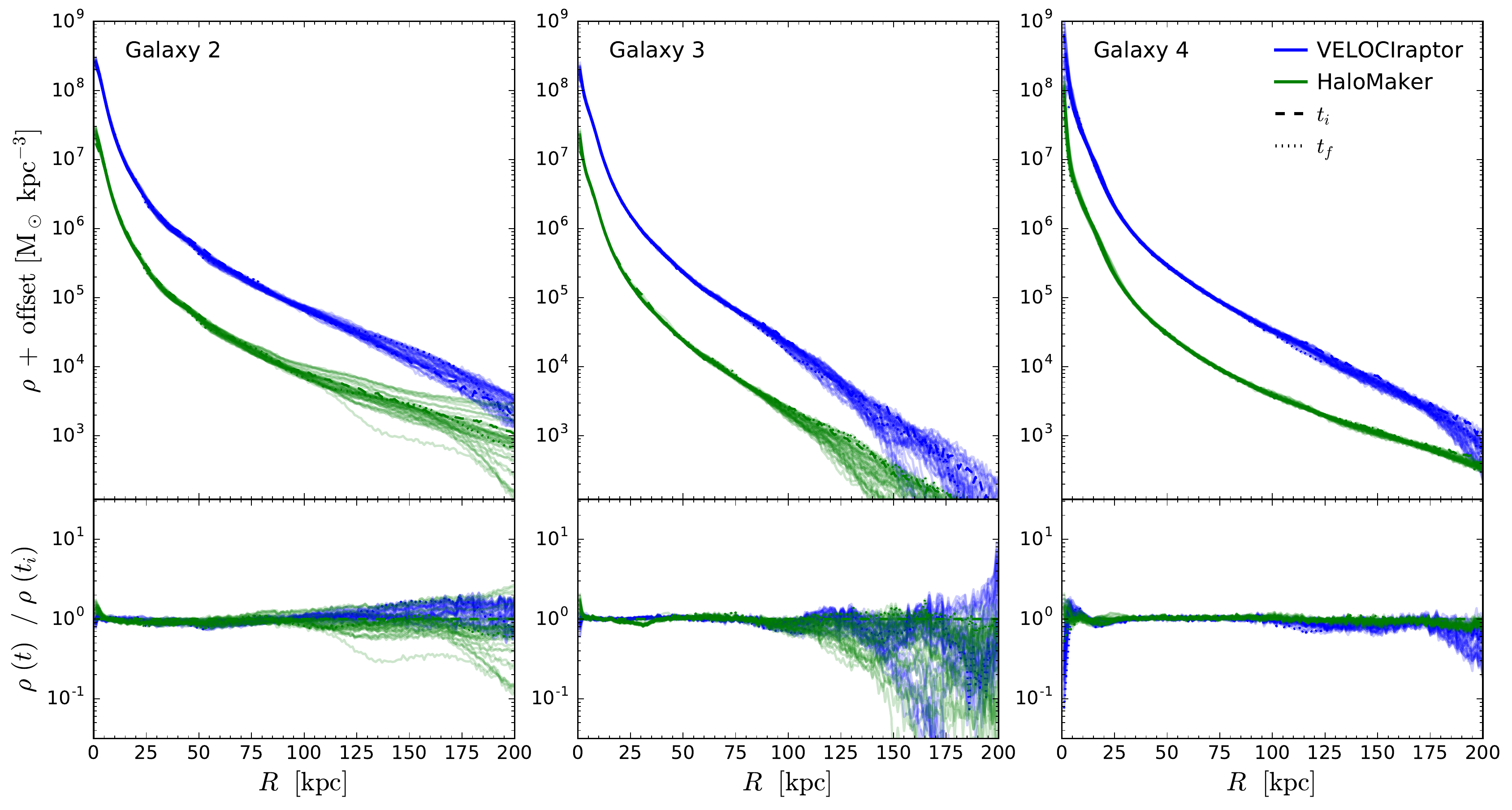}
        \caption{Evolution of $M_*$ and $R_{50}$ (top panels), and the stellar volume density profiles,  
                 $\rho$ (bottom panels), of \texttt{Galaxy 2-4}
                (left, centre and right columns, as labelled).
                Profiles of {\sc HaloMaker}'s galaxies are shifted by -1 dex for clarity.
        }
      \label{fig:gal234props}
    \end{figure*}

    We further test the consistency of the properties of 
    these galaxies.
    In Fig.~\ref{fig:gal234props}, we show the evolution of 
    $M_*$ and $R_{50}$ (top panels), and the stellar volume density
    profile, $\rho$ (bottom panels), of \texttt{Galaxy 2-4} 
    (left, centre and right column, respectively).
    As seen in Fig.~\ref{fig:gal1234evol} and consistent
    with Figs.~\ref{fig:gal1props} and~\ref{fig:gal1profile},
    the properties of \texttt{Galaxy 2}  change abruptly in time
    due to how particles are assigned to what is considered
    the central galaxy.
    A dip is seen when \texttt{Galaxy 1} is the central and 
    an abrupt increment when \texttt{Galaxy 2} is.
    For \texttt{Galaxy 3} and \texttt{4} the evolution of
    $M_*$, $R_{50}$ is more consistent over time, and is
    similar to what {\sc VELOCIraptor} estimates.
    The properties of {\sc HaloMaker}'s 
    \texttt{Galaxy 3} fluctuate significantly compared to 
    its counterpart in {\sc VELOCIraptor}, due to its outskirts
    being assigned to the central.
    These are, however, quite stable compared to the 
    evolution of those of \texttt{Galaxy 1} and 
    \texttt{2}.
    Consistent with what is shown throughout the paper, 
    strongly interacting galaxies can be affected 
    significantly due to the finder.

\bsp	
\label{lastpage}
\end{document}